%% file: main_doc.tex
\documentclass[12pt,defended,final,twoside]{cit_thesis}
\pdfoutput=1
\usepackage{graphicx}
\usepackage{amsmath}
\usepackage{amssymb}
\usepackage{latexsym}
\usepackage[sort]{cite}
\usepackage{rotating}

\usepackage{amsfonts}
\usepackage{bbm}
\usepackage[T1]{fontenc}
\usepackage[latin1]{inputenc}
\usepackage{booktabs}
\usepackage{amsthm}
\usepackage{color}

\newcommand{\be}{\begin{eqnarray}}
\newcommand{\ee}{\end{eqnarray}}
\newcommand{\bc}{\begin{center}}
\newcommand{\ec}{\end{center}}
\newcommand{\bea}{\begin{eqnarray}}
\newcommand{\eea}{\end{eqnarray}}
\newcommand{\ben}{\begin{equation}}
\newcommand{\een}{\end{equation}}

\newcommand{\eq}{\begin{equation}}
\newcommand{\eqe}{\end{equation}}
\newcommand{\g}{\gamma}

\newcommand{\eqa}{\begin{eqnarray}}
\newcommand{\eqae}{\end{eqnarray}}

\newcommand{\e}{\epsilon}

\def\IC{\mathbb{C}}

\def\CA{{\cal A}}

\def\CL{{\cal L}}

\def\a{\alpha}
\def\b{\beta}
\def\g{\gamma}
\def\d{\delta}
\def\e{\epsilon}

\def\i{\iota}

\def\l{\lambda}

\def\L{\Lambda}

\def\half{\frac{1}{2}}
\def\thalf{{\textstyle \frac{1}{2}}}

\def\goto{\rightarrow}

\def\det{{\rm det}}

\def\jmath{{j}}

\begin{document}

\include{front}


\numberwithin{equation}{section}

\include{intro}

\include{2pt}
\include{amplitude}
\include{conclusion}

\setcounter{equation}{0}
\numberwithin{equation}{chapter}
\appendix

\include{abjm}
\include{IIA}

\include{twistorgeom}
\include{bib}

\end{document}

%% file: front.tex
\degreeaward{Doctor of Philosophy }             
\university{California Institute of Technology}    
\address{Pasadena, California}                     
\unilogo{cit_logo}                                 
\copyyear{\the\year} 
\dedication{To my family.}
\title{Integrability of $\mathcal{N}=6$ Chern-Simons Theory}
\author{Arthur E. Lipstein}                                      
\maketitle

   \begin{frontmatter}
      \makecopyright            
      \makededication           
      \begin{acknowledgements}  
		\input{acknowledgements}

      \end{acknowledgements}
      \begin{abstract}
		\input{abstract}     
      \end{abstract}
	\tableofcontents
\clearpage
   \end{frontmatter}

%% file: acknowledgements.tex
I would like to thank my advisor, John Schwarz, for his guidance and support, for teaching me about string theory and M-theory, and for introducing me to the study of superconformal Chern-Simons theories. I would also like to thank Yu-tin Huang for teaching me about on-shell scattering amplitudes and Sakura Schafer-Nameki for teaching me about integrability and for all of her help in general. This thesis would not be possible without the help of these people.

I thank Miguel Bandres, Dongmin Gang, Yu-tin Huang, Eunkyung Koh, Sangmin Lee, and John Schwarz for their collaboration. I have also benefited from interactions with Nima Arkani-Hamed, Jonathan Bagger, Simon Caron-Huot, Hee-Joong Chung, Ori Ganor, David Gross, Christoph Keller, Juan Maldacena, Joseph Marsano, Donald Marolf, Tristan McLoughlin, Victor Mikhaylov, Karapet Mkrtchyan, Hiroshi Ooguri, Ari Pakman, Chang-Soon Park, Joseph Polchinski, Warren Siegel, Jaewon Song, Marcus Spradlin, Heywood Tam, Jaroslav Trnka, Ketan Vyas, Congkao Wen, and Xi Yin. I also thank Carol Silberstein for all of her help.

I would also like to thank Sergei Gukov and John Schwarz for giving me the opportunity to be the TA for the Relativistic Quantum Mechanics course and the String Theory course at Caltech, as well as Rana Adhikari, Yanbei Chen, Lee Lindblom, and Kip Thorne for giving me the opportunity to be a TA for the General Relativity course at Caltech. I am grateful to Sergei Gukov, Frank Porter, and Mark Wise for serving on my my thesis defense committee. 

I thank the organizers, lecturers, and participants of the 2010 PiTP and Cargese summer schools, the 2009 Mathematica Summer School, and the 2008 summer school on Particles, Fields, and Strings at UBC.

My work is supported in part by a Richard P. Feynman Fellowship, a James Albert Cullen Memorial Fellowship, and US DOE grant DE-FG02-92ER40701.

%% file: abstract.tex
In 2008, Aharony, Bergman, Jafferis, and Maldacena (ABJM) discovered a three-dimensional Chern-Simons theory with $\mathcal{N}=6$ supersymmetry and conjectured that in a certain limit, this theory is dual to type IIA string theory on $AdS_4 \times CP^3$. Since then, a great deal of evidence has been accumulated which suggests that the ABJM theory is integrable in the planar limit. Integrability is a very useful property that allows many physical observables, such as anomalous dimensions and scattering amplitudes, to be computed efficiently.  In the first half of this thesis, we will explain how to use integrabilty to compute the anomalous dimensions of long, single-trace operators in the ABJM theory. In particular, we will describe how to compute them at weak coupling using a Bethe Ansatz, and how to compute them at strong coupling using string theory. The latter approach involves using algebraic curve and world-sheet techniques to compute the energies of string states dual to gauge theory operators. In the second half of this thesis, we will discuss integrability from the point of view of on-shell scattering amplitudes in the ABJM theory. In particular, we will describe how to parameterize the amplitudes in terms of supertwistors and how to relate higher-point tree-level amplitudes to lower-point tree-level amplitudes using a recursion relation. We will also explain how this recursion relation can be used to show that all tree-level amplitudes of the ABJM theory are invariant under dual superconformal symmetry. This symmetry is hidden from the point of the action and implies that the theory has Yangian symmetry, which is a key feature of integrability. This thesis is mainly based on the material in \cite{Bandres:2009kw}, \cite{Huang:2010qy}, and \cite{Gang:2010gy}.         

%% file: intro.tex
\chapter{Introduction}    		
\label{intro}

What does it mean for a theory to be integrable? In practice, this means that the calculation of physical observables can be reduced to a well-posed finite mathematical problem. One necessary requirement for a theory to be integrable is that the number of symmetries matches the number of degrees of freedom. In order for a quantum field theory to be integrable, it must therefore have an infinite number of symmetries. In general, integrability is restricted to two-dimensional field theories. On the other hand, there are several examples where a higher dimensional theory is dual to an integrable system such as a spin chain. The best-known example of this phenomenon is the duality between $\mathcal{N}=4$ super Yang-Mills (sYM) theory \cite{Brink:1976bc} and type IIB string theory \cite{Green:1981yb} on the background geometry $AdS_{5}\times S^{5}$ \cite{Schwarz:1983qr}. The former is a four-dimensional gauge theory and the latter is a two-dimensional sigma model. This duality is an example of the AdS/CFT correspondence, which relates string theory on a background geometry consisting of $AdS_{d+1}$ times some compact space to a conformal field theory living on $d$-dimensional Minkowski space \cite{Maldacena:1997re}.

$\mathcal{N}=4$ sYM admits an expansion in $1/N$ at fixed 't Hooft coupling $\lambda=g_{YM}^{2}N$, where $N$ is the rank of the gauge group and $g_{YM}^{2}$ is the Yang-Mills coupling. The leading order in this expansion is known as the planar approximation and higher orders are suppressed as $N$ becomes large. The gauge theory parameters are related to the string theory parameters as follows
\begin{equation}
R^{2}/\alpha'\sim\sqrt{\lambda},\,\,\, g_{s}=\lambda/N
\end{equation}
where $R$ is the AdS radius, $1/(2 \pi \alpha')$ is the string tension, and $g_{s}$ is the coupling for string interactions. Since the planar approximation corresponds to taking $N$ to infinity while holding $\lambda$ fixed, this corresponds to taking $g_{s}$ to zero or equivalently taking the string theory to be non-interacting. Integrability has mainly been explored in this regime. Furthermore, if one takes $\lambda$ to be large, then quantum corrections in the string theory sigma model (which are suppressed by powers of $\alpha'/R^{2}$) become small and the string theory can be described using supergravity.

At small 't Hooft coupling, the anomalous dimensions of gauge invariant operators can be computed by solving a Bethe Ansatz \cite{Minahan:2002ve ,Beisert:2003tq}. At strong coupling, the equations of motion of the string theory sigma model can be recast as a flatness condition for a certain one-form known as the Lax connection \cite{sigma}. Using this property, any classical solution of the sigma model can be encoded in an algebraic curve \cite{Kazakov:2004qf,SchaferNameki:2004ik,Beisert:2005bm}. Ultimately, it is possible to define an all-loop Bethe Ansatz which interpolates from the gauge theory Bethe Ansatz to the string theory algebraic curve \cite{Beisert:2005fw}. This allows one to compute anomalous dimensions at strong 't Hooft coupling using string theory and interpolate these results to weak coupling. Although the all-loop Bethe Ansatz breaks down for finite-size operators after a certain order in the 't Hooft coupling, it is possible to generalize this formalism to compute the anomalous dimensions of finite-size operators to arbitrary order using a so-called Thermodynamic Bethe Ansatz (TBA) \cite{Arutyunov:2007tc,Arutyunov:2009zu,Gromov:2009tv, Bombardelli:2009ns, Gromov:2009bc,Arutyunov:2009ur,Arutyunov:2009kf,Cavaglia:2010nm}.

Integrable structure can also be found in the on-shell planar scattering amplitudes of $\mathcal{N}=4$ sYM. In fact, the scattering amplitudes of $\mathcal{N}=4$ sYM theory exhibit many hidden structures which are related to type IIB string theory on  $AdS_5 \times S^5$ through the AdS$_5$/CFT$_4$ correspondence. For example, if one uses the momenta of a maximal-helicity-violating (MHV) amplitude to define points in a dual space via $p_{i}=x_{i}-x_{i+1}$, then it turns out that the scattering amplitude is related to a light-like polygonal Wilson loop whose cusps are located at the dual points $x_i$. This duality was first proposed at strong coupling~\cite{Alday:2007hr}. Remarkably, this duality also holds at weak coupling~\cite{Drummond:2007aua,Brandhuber:2007yx,Drummond:2007cf}. Recently, there has also been progress in extending the amplitude/Wilson loop duality to non-MHV amplitudes \cite{Mason:2010yk,CaronHuot:2010ek}. Furthermore,there is a lot of evidence that this can be extended to an amplitude/Wilson-loop/correlator duality \cite{Alday:2010zy, Eden:2010zz,Eden:2010ce,Eden:2011yp,Eden:2011ku,Adamo:2011dq}.

Since the null-polygonal Wilson loops in $\mathcal{N}=4$ super Yang-Mills enjoy conformal symmetry, the amplitude/Wilson loop duality implies a hidden dual conformal symmetry of the scattering amplitudes which is inequivalent to the original conformal symmetry. Furthermore, it is possible to extend the dual conformal symmetry to dual superconformal symmetry, which becomes manifest once the amplitudes are written in a dual superspace \cite{Dualc,bht,Drummond:2008vq}.  The origin of dual superconformal symmetry can be understood using string theory. In particular, it is a consequence of the fact that type IIB string theory on $AdS_5 \times S^5$ is self-dual under a certain combination of T-duality transformations \cite{Berkovits:2008ic,Beisert:2008iq}. These T-duality transformation exchange superconformal symmetry with dual superconformal symmetry on the gauge theory side \cite{Drummond:2010qh}.

The presence of dual superconformal symmetry in $\mathcal{N}=4$ sYM is intimately related to its integrability since combining superconformal symmetry with dual superconformal symmetry at tree-level gives an infinite tower of symmetries known as the Yangian \cite{Dolan:2003uh,Dolan:2004ps,Drummond:2009fd}. Yangian symmetry also appears in the context of the $\mathcal{N}=4$ sYM spin chain and the dual string theory. The superconformal symmetry generators correspond to the level-0 Yangian generators and the dual superconformal symmetry generators provide part of the level-1 Yangian generators. Although higher-level generators do not impose new constraints on the amplitudes, the combination of superconformal symmetry and dual superconformal symmetry is sufficient to fix all tree-level amplitudes ~\cite{Korchemsky:2010ut}. It is also possible to extend dual superconformal symmetry to loop amplitudes \cite{Alday:2009zm, henn1,henn2,Henn:2011xk}. Ultimately, a generating function with manifest Yangian invariance has been proposed which is conjectured to capture the leading singularities of all scattering amplitudes in $\mathcal{N}=4$ sYM to all orders in perturbation theory \cite{ArkaniHamed:2009dn}. For a superamplitude with $n$ external legs and MHV-degree $k$, this generating function takes the form of an integral over the Grassmannian $G(k,n)$.

Using the amplitude/Wilson loop duality, it is possible to compute MHV amplitudes at strong coupling by computing null polygonal Wilson loops via string theory \cite{Alday:2007hr}. Furthermore, the string theory calculation can be reduced to solving a Thermodynamic Bethe Ansatz \cite{Alday:2009dv,Alday:2010vh}. Since a similar structure appears when computing the anomalous dimensions of finite-size operators, this suggests a deep connection between the computation of scattering amplitudes and anomalous dimensions.

The progress in computing scattering amplitudes in $\mathcal{N}=4$ sYM  is largely due to the spinor-helicity formalism, which uses (super)twistors to covariantly parameterize on-shell momenta, polarization vectors, and (for supersymmetric theories) on-shell multiplets. When written in terms of supertwistors, on-shell amplitudes look substantially simpler. Furthermore, twistor-inspired methods \cite{Witten:2003nn} have led to very efficient techniques for computing amplitudes. For example, the MHV expansion~\cite{Cachazo:2004kj} and the Britto, Feng, Cachazo, and Witten (BCFW)~\cite{Britto:2004ap,Britto:2005fq} recursion relations give a systematic procedure for  constructing higher-point tree-level amplitudes from lower-point tree-level amplitudes. This procedure was recently extended to the loop integrands of planar scattering amplitudes in $\mathcal{N}=4$ sYM \cite{ArkaniHamed:2010kv,Boels:2010nw}.

In this thesis, we will focus on integrability in a new example of the AdS/CFT correspondence which relates superconformal Chern-Simons theories in three dimensions to type IIA string theory in $AdS_{4}\times CP^{3}$. The study of superconformal Chern-Simons theories was originally motivated by the need to better understand M-theory, whose fundamental objects are referred to as M2-branes and M5-branes \cite{Schwarz:2004yj}. A superconformal Chern-Simons theory with maximal supersymmetry was discovered by Bagger, Lambert, and Gustavsson (BLG) \cite{BLG,Bagger:2007vi,Gustavsson:2007vu}, and is thought to describe two interacting M2-branes \cite{Mukhi:2008ux,Distler:2008mk}. Subsequent studies showed that the BLG theory is the only theory which has maximal supersymmetry at the classical level and is unitary \cite{Bandres:2008vf,Gauntlett:2008uf,Benvenuti:2008bt,Ho:2008ei,Bandres:2008kj}. It was then shown by Aharony, Bergman, Jafferis, and Maldacena (ABJM) that in order to generalize the BLG theory to describe an arbitrary number of M2-branes, one must sacrifice maximal supersymmetry \cite{Aharony:2008ug}. The ABJM theory has $OSp(6|4)$ superconformal symmetry and $U(N)_k\times U(N)_{-k}$ gauge symmetry, where $k$ is the Chern-Simons level. When $k=1,2$, the supersymmetry is enhanced to $\mathcal{N}=8$ due to quantum effects \cite{Gustavsson:2009pm,Kapustin:2010xq}. When $k\ll N\ll k^{5}$, the ABJM theory is dual to type IIA string theory on $AdS_{4}\times CP^3$.

Like $\mathcal{N}=4$ sYM, the ABJM theory admits an expansion in $1/N$ at fixed 't Hooft parameter $\lambda=N/k$. The parameters of the string theory are related to the gauge theory parameters as follows:
\begin{equation}
R^{2}/\alpha'\sim\sqrt{\lambda},\,\,\, g_{s}=\lambda^{5/4}/N
\end{equation}
where $R$ is proportional to the AdS radius. Since the discovery of the ABJM theory, a lot of evidence has been found which suggests that it is integrable in the planar limit. In particular, the planar dilatation operator in the gauge theory was shown to be integrable up to six loops ~\cite{Minahan:2008hf,Zwiebel:2009vb,Minahan:2009te,Bak:2009mq}, classical integrability was demonstrated in various subsectors of the dual string theory \cite{Stefanski:2008ik,Arutyunov:2008if,Sorokin:2010wn}, an $AdS_4/CFT_3$ algebraic curve was formulated in \cite{Gromov:2008bz}, an all-loop Bethe Ansatz was conjectured in \cite{AllLoop}, and a Thermodynamic Bethe Ansatz for the ABJM theory was proposed in \cite{Bombardelli:2009xz,Gromov:2009at}. Moreover, a spinor-helicity formalism was developed for three-dimensional superconformal Chern-Simons theories \cite{Bargheer:2010hn,Huang:2010rn} and used to show that the four- and six-point tree-level amplitudes of the ABJM theory have Yangian symmetry \cite{Bargheer:2010hn} and dual superconformal symmetry \cite{Huang:2010qy}. A recursion relation for tree-level amplitudes was then constructed and used to show that all tree-level amplitudes have dual superconformal symmetry \cite{Gang:2010gy}. A Grassmannian integral formula similar to the integral formula of $\mathcal{N}=4$ sYM has also been proposed for the ABJM theory \cite{Lee:2010du}.

While the $AdS_{4}/CFT_{3}$ correspondence shares certain features with the $AdS_{5}/CFT_{4}$ correspondence, it also exhibits several new features. First of all, the $AdS_{4}/CFT_{3}$ magnon dispersion relation was found to be  $\epsilon=\frac{1}{2}\sqrt{1+8h(\lambda)\sin^{2}\frac{p}{2}}$, where  \cite{Nishioka:2008gz,Gaiotto:2008cg}
\[
h(\lambda)=\lambda,\,\,\,\lambda \gg 1,
\]
\[
h(\lambda)=2\lambda^{2},\,\,\, \lambda \ll 1.
\]
This is in contrast to the magnon dispersion relation for $AdS_{5}/CFT_{4}$, where $h(\lambda)=\frac{\sqrt{\lambda}}{4\pi}$ for all values of $\lambda$ \cite{Beisert:2005tm,Berenstein:2009qd}. The interpolating function $h(\lambda)$ appears in the all-loop Bethe Ansatz and its nontrivial structure in the $AdS_{4}/CFT_{3}$ correspondence can be attributed to the fact that the theory only has 3/4 maximal supersymmetry. Another consequence of the less-than-maximal supersymmetry is that the radius of $AdS_4 \times CP^3$ receives $\mathcal{O}(1/\sqrt{\lambda})$ corrections, although this only becomes relevant at two loops in the string theory sigma model ~\cite{Bergman:2009zh}.

Furthermore, unlike the string theory dual to $\mathcal{N}=4$ sYM, the string theory dual to the ABJM theory is not fully described by a coset sigma model since the coset sigma model description breaks down for string configurations which have no support in $CP^3$ \cite{Arutyunov:2008if}. In order to go beyond one-loop in the string theory sigma model, one must use a Green-Schwarz sigma model whose target space is the full $AdS_4 \times CP^3$ superspace \cite{Gomis:2008jt,Grassi:2009yj}. New techniques have to be developed to prove integrability in the full superspace \cite{Sorokin:2011mj}.

Another new feature of type IIA string theory on $AdS_{4}\times CP^3$ is that in the Penrose limit, half of the excitations are twice as massive as the other half \cite{Nishioka:2008gz,Gaiotto:2008cg,Grignani:2008is}. The latter are subsequently referred to as {}``light'' and the former are referred to as {}``heavy''. This is in contrast to what was found when looking at the Penrose limit of type IIB string theory on $AdS_{5}\times S^{5}$, where all the excitations have the same mass \cite{Berenstein:2002jq}. The origin of these two types of fluctuations can be understood from the ABJM spin chain. In particular, the light fluctuations correspond to the elementary excitations of the spin chain while heavy fluctuations correspond to composite excitations of the spin chain. Whereas the ABJM spin chain has only eight elementary excitations, the $\mathcal{N}=4$ sYM spin chain has sixteen elementary excitations, which are in one-to-one correspondence with the fluctuations in the Penrose limit of type IIB string theory on $AdS_5 \times S^5$.

Since the heavy fluctuations in type IIA string theory on $AdS_4 \times CP^3$ are not dual to elementary excitations of the ABJM spin chain, care must be taken when computing quantum corrections to the world-sheet theory, which correspond to $\alpha'$ corrections to the classical string theory. In particular, several groups found a disagreement with the all-loop Bethe Ansatz after computing the one-loop correction to the energy of the folded spinning string in $AdS_{4}\times CP^3$ \cite{Krishnan:2008zs, McLoughlin:2008ms, Alday:2008ut}. In computing the one-loop correction, these groups used the same prescription for adding up fluctuation frequencies that was used in $AdS_{5}\times S^{5}$ . Several ways to resolve this discrepancy were proposed in \cite{Gromov:2008fy,McLoughlin:2008he,Bandres:2009kw}. In particular two alternative summation prescriptions were proposed which achieve agreement with the all-loop Bethe Ansatz by treating the frequencies of heavy and light fluctuations on unequal footing. Furthermore, the analysis of \cite{Bandres:2009kw} indicates that the standard summation prescription for computing one-loop corrections leads to divergent results when applied to fluctuation frequencies computed using algebraic curve techniques. Reference \cite{McLoughlin:2008he} pointed out that the discrepancy can also be resolved if the interpolating function has the form $\sqrt{h(\lambda)}=\sqrt{\lambda}+a_1+\mathcal{O}\left(1/\sqrt{\lambda}\right)$ with $a_1\neq 0$.

The ABJM theory also exhibits many new features in terms of its on-shell scattering amplitudes. First of all, since there is no chirality in three dimensions, there is only one type of spinor with which one can parameterize the scattering amplitudes of massless particles. More concretely,  in four dimensions, a null momentum can be written in terms of two types of spinors as follows:
\begin{equation}
 p_{\alpha\dot{\alpha}}=\lambda_{\alpha}\tilde{\lambda}_{\dot{\alpha}}
\end{equation}
where $\alpha$ and $\dot{\alpha}$ are $SU(2)$ indices of opposite chirality. On the other hand, a null momentum in three dimensions
can be written in bi-spinor form as follows:
\begin{equation}
p_{\alpha\beta}=\lambda_{\alpha}\lambda_{\beta}
\end{equation}
where $\alpha$ is an $SL(2,R)$ index. Furthermore, the BCFW recursion relations are not applicable to three-dimensional conformal field theories. In the usual BCFW approach, one linearly deforms two external momenta of an on-shell amplitude by a complex parameter. In order to define a similar recursion relation in three dimensions which preserves on-shell properties and conformal symmetry, one must allow the deformation to be nonlinear \cite{Gang:2010gy}.

Another implication of the lack of chirality in three dimensions is that half of the dual supersymmetry generators fail to commute with the equations that relate the dual superspace coordinates to the on-shell superspace coordinates. This problem can be remedied by augmenting the dual superspace by three Grassmann-even coordinates which carry only R-symmetry indices \cite{Huang:2010qy}. Such coordinates are not required in order to define a dual superspace in $\mathcal{N}=4$ sYM, although they do appear if one formulates $\mathcal{N}=4$ sYM using a non-chiral superspace \cite{Huang:2011um}.  Furthermore, the inclusion of these coordinates allows one to match the nontrivial dual superconformal generators with level-1 Yangian generators, which implies that the amplitudes of the ABJM theory have Yangian symmetry. Although the ABJM theory  is only $3/4$ maximal, this is precisely the amount of supersymmetry for which dual superconformal symmetry is possible in three dimensions.

The structure of this thesis is as follows. Chapter 2 focuses on anomalous dimensions of long gauge-invariant operators in the ABJM theory. First, we describe how to compute them at weak coupling using an asymptotic Bethe Ansatz. In section 2.1.3, we present the two-loop Bethe Ansatz for the $SU(2) \times SU(2)$ sector of the ABJM spin chain and use it to compute the anomalous dimensions of non-BPS operators. In section 2.2, we describe how to compute anomalous dimensions at strong coupling using string theory. The material in section 2.2 is taken from \cite{Bandres:2009kw}. First we describe two formalisms for computing the spectrum of fluctuation frequencies about classical string solutions, which are known as the world-sheet approach and the algebraic curve approach. In section 2.2.3, we apply these techniques to two classical solutions which are dual to the gauge theory operators analyzed in section 2.1.3, notably a point particle and a circular string which are spinning in $CP^3$. In section 2.2.4, we analyze different prescriptions for computing one-loop corrections to the energies of these classical solutions and match the energies with the anomalous dimensions of the dual gauge theory operators. Chapter 3 discusses on-shell amplitudes of the ABJM theory. The material in Chapter 3 is mostly taken from \cite{Huang:2010qy} and \cite{Gang:2010gy}. In section 3.1, we describe supertwistors which can be used to parameterize the on-shell amplitudes of three-dimensional superconformal field theories. Section 3.2.1 describes how to construct a dual superspace for the ABJM theory. In particular, we will show that in order to define dual supersymmetry generators in a consistent way, this space must contain three Grassmann-even coordinates in addition to three bosonic and six fermionic coordinates. In section 3.2.3, we define the dual conformal boost generator and demonstrate that the four-point amplitude of the ABJM theory has dual superconformal symmetry. Section 3.3 then demonstrates that the nontrivial dual superconformal generators can be matched with level-one Yangian generators, which establishes that combining ordinary supersymmetry with dual superconformal symmetry gives Yangian symmetry. In section 3.4, we describe the difficulties of extending the BCFW recursion relation to three dimensional theories and present an alternative recursion relation for the ABJM theory. In section 3.5, we demonstrate that this recursion relation preserves dual superconformal symmetry, which implies that all tree-level amplitudes of the ABJM theory have Yangian symmetry. In section 3.6, we briefly describe the Grassmannian integral formula which is conjectured to generate all tree-level amplitudes of the ABJM theory. In Chapter 4, we present some conclusions. There are also several appendices. Appendix A reviews the ABJM theory, Appendix B reviews type IIA string theory on $AdS_4 \times CP^3$, and Appendix C describes the twistor geometry of three-dimensional Minkowski space.

%% file: 2pt.tex
\chapter{Anomalous Dimensions}    		
\label{2pt}

\section{Weak Coupling}

In this section, we will describe how to compute anomalous dimensions of long gauge invariant operators in the planar limit of the ABJM theory. We begin with a general discussion of two-point correlators, and then describe the spin chain of the ABJM theory. We then specialize to the $SU(2) \times SU(2)$ sector, which is the simplest closed subsector of the spin chain, and compute the dispersion relation of elementary and composite excitations as well as the anomalous dimension of a non-BPS operator which is dual to a circular spinning string in the dual string theory.

\subsection{Two-Point Correlators}

A Euclidean conformal field theory (CFT) is completely characterized by its two- and three-point correlators. It is possible to choose a basis of local gauge invariant operators such that the two-point correlators of these operators takes the form \footnote{This is a formula for scalar operators, but it can be generalized to describe operators with spin.}

\[
\left\langle \mathcal{O}_{A}(x)\mathcal{O}_{B}(y)\right\rangle =\frac{\delta_{AB}}{|x-y|^{2D_{A}}}\]
where $D_{A}$ is the scaling dimension of $O_{A}$, which can be
split into two pieces

\[
D_{A}=D_{A}^{(0)}+\delta D_{A}\]
where $D_{A}^{(0)}$ is the classical scaling dimension and $\delta D_{A}$
is the anomalous dimension coming from quantum corrections.
The operators $\mathcal{O}_{A}$ and their scaling dimensions can be thought
of as eigen-operators and eigenvalues of an anomalous dimension
matrix, or dilatation operator, $\delta D$. If the CFT has a large-$N$ expansion, where $N$
is related to the rank of the gauge-group, then $\delta D$ has the
following topological expansion:
\eq
\delta D\left(\lambda,1/N\right)=\sum_{g=0}^{\infty}\frac{1}{N^{2g}}\sum_{l=1}\lambda^{l}\delta D_{l,g}
\label{texp}
\eqe
where $\lambda$ is the 't Hooft coupling. If $\lambda\ll1$, then
$\delta D$ can be computed perturbatively by computing the renormalization
matrix $Z$ which cancels the divergences of the two-point correlators,
and then taking the logarithmic derivative of $Z$ with respect to
the renormalization scale $\mu$. If the CFT is
dual to a string theory, then it is also possible to compute the anomalous
dimensions when $\lambda\gg1$ using the $AdS/CFT$ correspondence. In particular,
\[
D_{A}\left(\lambda,1/N\right)=E_{A}\left(R^{2}/\alpha',g_{s}\right)\]
where $E_{A}$ is the energy of the string state dual to the operator
$\mathcal{O}_{A}$, $R$ is the radius of the dual supergravity background, $1/(2 \pi \alpha')$ is the string tension, and $g_s$ is the coupling for string interactions \cite{Gubser:1998bc,Witten:1998qj}. If $D_{A}$ scales like $\sqrt{\lambda}$, then
$\mathcal{O}_{A}$ is dual to a classical solution of string
theory. In practice, one matches the scaling dimension computed in
the gauge theory with the energy computed in the string theory by
first expanding in $\lambda$, and then expanding in $1/J$ in the gauge theory (where $J$ is R-charge of the gauge theory operator),
while first expanding in $R^{2}/\alpha'$, and then expanding in $\lambda/J^{2}$
in the string theory \cite{Berenstein:2002jq}. We will describe how to compute string
energies in section 2.2.

In the large-$N$ limit, only the $g=0$ term in eq. \ref{texp} survives. This
is known as the planar limit. In this limit, multi-trace operators
decouple so we only need to consider single-trace operators. Furthermore,
the dilatation operator can be thought of as the Hamiltonian
of a spin chain. If the spin chain is integrable, then the Hamiltonian can be diagonalized via a Bethe
Ansatz. In the planar limit, the string theory becomes non-interacting,
since only world-sheets zero genus contribute.

\subsection{The ABJM Spin Chain}

A nice review of the ABJM spin chain can be found in \cite{Klose:2010ki}. Since the matter
fields in the ABJM theory are in the bifundamental representation
of the gauge group $U(N)\times U(N)$, gauge invariant operators are
constructed by taking the trace of an even number of fields which alternate
between the $\left(N,\bar{N}\right)$ and $\left(\bar{N},N\right)$
representation. They can therefore be thought of as alternating spin
chains. Schematically, they have the form
\[
\mathcal{O}\sim tr\left(\phi^{I_{1}}\phi_{J_{1}}...\phi^{I_{L}}\phi_{J_{L}}\right)\]
where the scalar fields $\phi^I$, $I=1,...,4$, are in the fundamental representation of the R-symmetry group $SU(4)$, and their adjoints $\phi_I$ are in the anti-fundamental representation (see Appendix A for more details).
One can also insert fermionic fields $\psi_{I}$ on the odd sites,
$\psi^I$ on the even sites, and covariant derivatives $D_{\mu}$
which do not introduce extra sites. Field strength insertions can
be replaced with currents using the quantum equations of motion. One choice of vacuum
for the ABJM spin chain is \[
\mathcal{O}=tr\left[\left(\phi^{1}\phi_{3}\right)^{J}\right]{\normalcolor .}\]
Note that the vacuum is protected by supersymmetry and therefore has
a zero anomalous dimension. For this choice of vacuum, the elementary
bosonic excitations correspond to inserting $\phi^{2}$ or $\phi^{4}$
on the odd sites and $\phi_{2}$ or $\phi_{4}$
on the even sites. By supersymmetry, there are also four elementary fermionic excitations. All other insertions correspond to composite
excitations of the spin chain. In particular, there are eight independent
composite excitations which correspond to one elementary excitation
on the even sites and one elementary excitation on the odd sites.

\begin{figure}
\center
\includegraphics[scale=0.6]{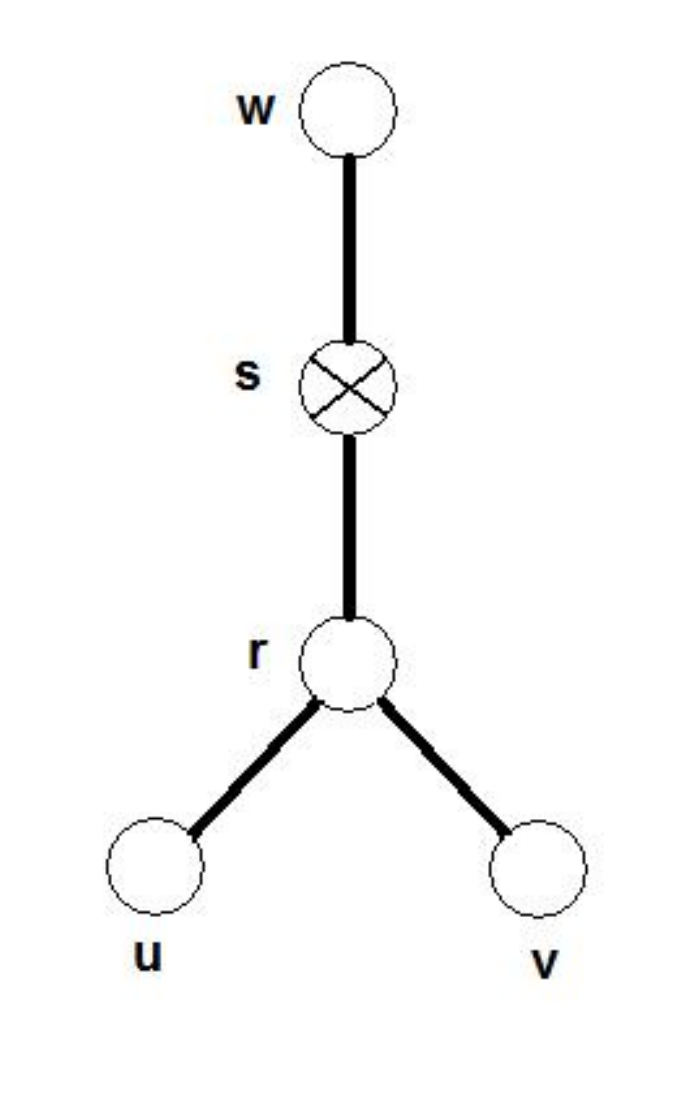}
\caption{Dynkin diagram of $OSp(6|4)$. The $SU(2)\times SU(2)$ sector corresponds to exciting only the momentum carrying roots ($u$ and $v$). To go beyond the $SU(2)\times SU(2)$ sector, one must excite the auxiliary roots in the $SU(2|2)$ tail ($r$, $s$, and  $w$).  Note that the root $s$ is fermionic.}
\label{dynkin}
\end{figure}

The vacuum of the spin chain breaks the superconformal group from
$OSp(6|4)$ to $SU(2|2)$ and the eight elementary excitations transform
in the representation $(2|2)_{even}\oplus(2|2)_{odd}$, where $even/odd$
refers to the even/odd sites of the spin chain. This can also be understood
by looking at the Dynkin diagram of $OSp(6|4)$, which is depicted
in Fig. \ref{dynkin}. Note that the Dynkin diagram has an $SU(2|2)$ tail, corresponding
to the nodes $r,s,w$, and two wings corresponding to the nodes $u,v$.
As we mentioned in the previous subsection, if the ABJM spin chain
is integrable, this implies that the dilatation operator can be diagonalized using a Bethe Ansatz. In this case, the roots of
the Bethe equations are in one-to-one correspondence with the nodes
of the Dynkin diagram. In particular, $u$ and $v$ are referred to
as the momentum-carrying Bethe roots because they are related to the
momenta of excitations on the odd/even sites of the spin chain.
Once the momentum-carrying Bethe roots are determined, the anomalous
dimension and all higher charges of the spin chain can be computed.
The other roots are auxiliary in the sense that the spectrum of the spin chain does not depend on them explicitly. The elementary excitations
on the odd sites correspond to exciting one $u$ root and various combinations
of roots in the $SU(2|2)$ tail, notably $\left\{ K_{r},K_{s},K_{w}\right\} =\left\{ 0,0,0\right\} ,\left\{ 1,0,0\right\} ,\left\{ 1,1,0\right\},$ and $\left\{ 1,1,1\right\} $, where $K_r,K_s,K_w$ denote the number of $r,s,w$ roots which are excited. Note that the first two combinations are bosonic and the second two are fermionic
since $s$ is a fermionic root. The elementary excitations
on the even sites correspond to replacing $u$ with $v$.

In the next subsection, we will specialize to the $SU(2)\times SU(2)$
subsector of the ABJM spin chain, which corresponds to exciting only
the momentum-carrying roots $u$ and $v$. We will show that the two-loop
planar dilatation operator in this sector is integrable by writing
down a Bethe Ansatz which can be used to diagonalize it. We will also
describe some simple solutions to the Bethe Ansatz.

\subsection{$SU(2) \times SU(2)$ Sector}
In the $SU(2)\times SU(2)$ sector, the odd sites of the spin chain
are $\phi^{I}\in\left(\phi^{1},\phi^{2}\right)$ and the even sites
of the spin chain are $\phi_{I}\in\phi_{3},\phi_{4}$.
For an operator of length $2L$, the planar two-loop dilatation operator
is given by

\eq
\delta D=\lambda^{2}\sum_{i=1}^{L}\left(1_{2i-1,2i+1}-P_{2i-1,2i+1}+1_{2i,2i+2}-P_{2i,2i+2}\right)
\label{dila}
\eqe
where $P_{ij}$ permutes the fields on sites $i$ and $j$ and the
indices are periodic, i.e., $2L+1\sim1$. It is easy to see that
\begin{equation}
\mathcal{O}=tr\left[\left(\phi^{1}\phi_{3}\right)^{J}\right]
\label{eq:vac}
\end{equation}
is annihilated by the dilatation operator, which is expected since this
operator is the vacuum of the spin chain. Furthermore, this operator
is protected by supersymmetry since its scaling dimension is equal to its
R-charge, i.e., it is a BPS operator. Note that there are two types of excitations
in the $SU(2)\times SU(2)$ sector: $\phi^{2}$ and $\phi_{4}$.
If we think of $\phi^{1}$ and $\phi_{3}$ as being down
spins and the excitations $\phi^{2}$ and $\phi_{4}$ as being up spins, the dilatation operator
in eq. \ref{dila} is just the Hamiltonian for two Heisenberg spin chains, one
on the even sites and one on the odd sites. The spin chains are only
coupled by the constraint that the sum of the momenta of all the excitations
is zero (modulo $2\pi$). This constraint arises from the fact that the spin chain
corresponds to a trace of fields, which is invariant under cyclic
permutations.

The Heisenberg spin chain was solved by Hans Bethe \cite{Bethe}. The key property of this model which makes it solvable is that the scattering of an arbitrary number
of excitations factorizes into $2\rightarrow2$ scatterings.
In this sense, once one solves the Schrodinger equation for two excitations, it is easy to generalize the solution to an arbitrary
number of excitations. A nice review of the coordinate Bethe Ansatz for integrable spin chains in $\mathcal{N}=4$ sYM can be found in  \cite{Plefka:2005bk}, for example. Coming back to the $SU(2)\times SU(2)$ spin chain,
suppose we have an operator of length $2L$ with $K_{u}$ excitations on the odd sites and $K_{v}$
excitations on the even sites. Then the Bethe Ansatz equations are
given by
\begin{equation}
\left(\frac{u_{j}+i/2}{u_{j}-i/2}\right)^{L}=\Pi_{k\neq j}^{K_{u}}\frac{u_{j}-u_{k}+i}{u_{j}-u_{k}-i},
\label{bethe1}
\end{equation}
\begin{equation}
\left(\frac{v_{j}+i/2}{v_{j}-i/2}\right)^{L}=\Pi_{k\neq j}^{K_{v}}\frac{v_{j}-v_{k}+i}{v_{j}-v_{k}-i},
\label{bethe2}
\end{equation}
\begin{equation}
\Pi_{j=1}^{K_{u}} \left(\frac{u_{j}+i/2}{u_{j}-i/2} \right) \Pi_{k=1}^{K_{v}} \left(\frac{v_{k}+i/2}{v_{k}-i/2} \right)=1,
\label{trace}
\end{equation}
\begin{equation}
\delta D=\lambda^{2}\left(\sum_{j=1}^{K_{u}}\frac{1}{u_{j}^{2}+1/4}+\sum_{k=1}^{K_{v}}\frac{1}{v_{k}^{2}+1/4}\right),
\label{energy}
\end{equation}
where the variables $u$ and $v$ are related to the momenta of the
excitations on the odd/even sites by $u=\frac{1}{2}\cot\frac{p}{2}$
and $v=\frac{1}{2}\cot\frac{p}{2}$. Eqs. \ref{bethe1} and \ref{bethe2} contain
dynamical information. In particular, the right-hand side of each
of these equations is just a product of two-body scattering matrices. Note that the excitations on the odd sites do not scatter with excitations on the even sites since there is no coupling between them in the Hamiltonian. Eq. \ref{trace} simply states that the sum of all the momenta should
be zero (modulo $2\pi$). After solving eqs. \ref{bethe1}, \ref{bethe2}, and \ref{trace} for the $u$'s and $v$'s,
one just plugs them into eq. \ref{energy} to obtain the anomalous dimension.

Let's work out a simple example. In particular, let's consider inserting
one excitation on the even sites and one excitation on the odd sites
of the operator in eq. \ref{eq:vac}. If we think of the resulting non-BPS operator as two Heisenberg spin chains, then each spin chain has
$J+1$ sites. Furthermore, we can think of the two elementary excitations as a single composite excitation
of the ABJM spin chain. In this case, eqs. \ref{bethe1}, \ref{bethe2}, and \ref{trace}
reduce to
\[
\left(\frac{u+i/2}{u-i/2}\right)^{J+1}=\left(\frac{v+i/2}{v-i/2}\right)^{J+1}=\frac{u+i/2}{u-i/2}\frac{v+i/2}{v-i/2}=1{\normalcolor .}\]
Suppose the odd excitation has momentum $p_{1}$ and the even excitation has momentum $p_{2}$. Then $u=\frac{1}{2}\cot\frac{p_{1}}{2}$
and $v=\frac{1}{2}\cot\frac{p_{2}}{2}$ and the equations above can
be written as \[
e^{ip_{1}\left(J+1\right)}=e^{ip_{2}\left(J+1\right)}=e^{i\left(p_{1}+p_{2}\right)}=1{\normalcolor .}\]
These are solved by $p_{1}=-p_{2}=\frac{2\pi n}{J+1}$. Plugging
this into eq. \ref{energy} then gives\[
\delta D_{composite}=\lambda^{2}\left(4\sin^{2}\frac{p_{1}}{2}+4\sin^{2}\frac{p_{2}}{2}\right)=8\lambda^{2}\sin^{2}\left(\frac{\pi n}{J+1}\right){\normalcolor .}\]
Hence, after we add one impurity to the even sites and one to the odd sites of the operator
in eq. \ref{eq:vac}, the the quantum-corrected
scaling dimension of the resulting operator is given by \[
D=J+1+8\lambda^{2}\sin^{2}\left(\frac{\pi n}{J+1}\right) + \mathcal{O}(\lambda^4){\normalcolor .}\]
Note that the scaling dimension of the vacuum operator is $J$ (since
the scalar fields have mass dimension $1/2$). It follows that the
energy of a composite excitation of the spin chain is given by $1+8\lambda^{2}\sin^{2}\left(\frac{\pi n}{J+1}\right)$.
In the large-$J$ limit, this reduces to
\eq
E_{composite}=1+\frac{8\pi^{2}\lambda^{2}n^{2}}{J^{2}}+\mathcal{O}\left(\frac{1}{J^{6}}\right){\normalcolor .}
\label{comp}
\eqe
Since the composite excitation is made up of two elementary excitations,
we find that in the large-$J$ limit the energy of an elementary
excitation is
\eq
E_{elementary}=\frac{1}{2}+\frac{4\pi^{2}\lambda^{2}n^{2}}{J^{2}}+\mathcal{O}\left(\frac{1}{J^{6}}\right){\normalcolor .}
\label{elem}
\eqe

As one more example, let's consider adding $J$ excitations to the
even sites and $J$ excitations to the odd sites of the operator in
eq. \ref{eq:vac}. We then obtain the following non-BPS operator:
\eq
\mathcal{O}=tr\left[\left(\phi^{1}\phi_{3}\right)^{J}\left(\phi^{2}\phi_{4}\right)^{J}+...\right]
\label{NB}
\eqe
where the dots stand for permutations of $\left(\phi^{1}\phi_{3}\right)$
and $\left(\phi^{2}\phi_{4}\right)$. Once again, this operator
can be thought of as two decoupled spin chains except in this case,
each spin chain has $J$ up spins and $J$ down spins. Since the two
spin chains are identical, we can set $u_{j}=v_{j}$ and eqs. \ref{bethe1} -- \ref{energy} reduce
to

\begin{eqnarray}
\left( \frac{u_{j}+i/2}{u_{j}-i/2}\right) ^{2J} &=&\prod_{k\neq j}^{J}\frac{%
u_{j}-u_{k}+i}{u_{j}-u_{k}-i}, \\
\left( \prod_{j=1}^{J}\left( \frac{u_{j}+i/2}{u_{j}-i/2}\right) \right)
^{2} &=&1\Longrightarrow \sum_{j=1}^{J}\ln \left( \frac{u_{j}+i/2}{u_{j}-i/2%
}\right) =-\pi mi, \label{BA1} \\
\delta D &=&2\lambda ^{2}\sum_{j=1}^{J}\frac{1}{u_{j}^{2}+1/4},
\end{eqnarray}%
where $m$ is an integer which is introduced after taking the log of both sides of eq. \ref{BA1}. This integer corresponds to the winding number of the classical string which is dual to the operator in eq. \ref{NB}.
In the large-$J$ limit, the Bethe equations simplify and can be solved using the methods described in \cite{Kazakov:2004qf,Beisert:2005mq}. In particular, following the manipulations in  section 3 of \cite{Beisert:2005mq}, one finds that
\eqa
\delta D&=&\left(\frac{\pi^{2}\lambda^{2}m^{2}}{J}+...\right)+\frac{1}{J}\left(\frac{2a\pi^{2}\lambda^{2}}{J}+...\right),\\
a&=&m^{2}/4+\sum_{n=1}^{\infty}\left(n\left(\sqrt{n^{2}-m^{2}}-n\right)+m^{2}/2\right).
\label{AD}
\eqae

\section{Strong Coupling}

In this section, we will describe how to compute the one-loop energies of the classical string theory solutions dual to the gauge theory operators analyzed in section 2.1, notably a rotating point particle and a circular string which are spinning in $CP^3$ and have trivial support in $AdS_4$. The latter solution is the $AdS_{4}/CFT_{3}$ analogue of the $SU(2)$ circular
string which was discovered in~\cite{Frolov:2003qc} and studied extensively in the $AdS_{5}/CFT_{4}$ correspondence \cite{Frolov:2003tu,Frolov:2004bh,Beisert:2005mq}. The point-particle and spinning string solutions are especially interesting to study in the $AdS_{4}/CFT_{3}$ context because $\kappa$ symmetry breaks down in the coset sigma model for solutions with trivial support in $CP^3$ \cite{Arutyunov:2008if}.

In order to compute the one-loop correction to the energy of a classical solution, we must first compute the spectrum of fluctuations about the solution. This can be computed by expanding the Green-Schwarz (GS) action to quadratic
order in the fluctuations and finding the normal modes of the resulting equations of motion. We refer to this method
as the world-sheet (WS) approach.
Alternatively, the spectrum can be computed
from the algebraic curve corresponding to this solution using semi-classical
techniques. We refer to this as the algebraic curve (AC) approach. This approach was developed for type IIB string theory in $AdS_{5} \times S^{5}$ in~\cite{Gromov:2007aq} and then adapted to type IIA string theory in $AdS_{4} \times CP^3$ in~\cite{Gromov:2008bz}. In the next two subsections, we describe the world-sheet and algebraic curve formalisms in greater detail.

\subsection{World-Sheet Formalism}
The world-sheet approach for computing the spectrum of fluctuations about a classical
solution in $AdS_5 \times S^5$ was developed in \cite{Frolov:2002av}. In this subsection, we will describe the analogous formalism for $AdS_4 \times CP^3$.

Using the metric $G_{MN}$ in eq.~\ref{IIAGeomM}, the bosonic part of the GS
Lagrangian in conformal gauge is given by
\begin{equation}
\ \mathcal{L}_{bose}=\frac{1}{4\pi }\eta ^{ab}G_{MN}\partial
_{a}X^{M}\partial _{b}X^{N},  \label{eq:bosaction}
\end{equation}%
where $a,b=\tau ,\sigma $ are world-sheet indices, $\eta ^{ab}={\normalcolor %
diag}\left[ -1,1\right] $, and we have set $\alpha ^{\prime }=1$. Any
solution to the bosonic equations of motion has at least five conserved charges. In
particular, there are two $AdS_{4}$ charges given by
\begin{equation}
E=\sqrt{{\lambda /2}}\int_{0}^{2\pi }d\sigma \cosh ^{2}\rho \dot{t}{,}
\label{eq:E}
\end{equation}%
\begin{equation}
S=\sqrt{{\lambda /2}}\int_{0}^{2\pi }d\sigma \sinh ^{2}\rho \sin ^{2}\theta
\dot{\phi},  \label{SpinS}
\end{equation}%
and there are three $CP^{3}$ charges given by
\begin{subequations}
\label{Js}
\begin{eqnarray}
J_{\psi } &=&2\sqrt{{2\lambda}}\int_{0}^{2\pi }d\sigma \cos ^{2}\xi \sin
^{2}\xi \left( \dot{\psi}+\frac{1}{2}\cos \theta _{1}\dot{\phi}_{1}-\frac{1}{%
2}\cos \theta _{2}\dot{\phi}_{2}\right) ,  \label{eq:Jpsi1} \\
J_{\phi _{1}} &=&\sqrt{{\lambda /2}}\int_{0}^{2\pi }d\sigma \cos ^{2}\xi
\sin ^{2}\theta _{1}\dot{\phi}_{1}  \label{eq:JPhi1} \\
&&+\sqrt{{2\lambda }}\int_{0}^{2\pi }d\sigma \cos ^{2}\xi \sin ^{2}\xi
\left( \dot{\psi}+\frac{1}{2}\cos \theta _{1}\dot{\phi}_{1}-\frac{1}{2}\cos
\theta _{2}\dot{\phi}_{2}\right) \cos \theta _{1},  \notag \\
J_{\phi _{2}} &=&\sqrt{{\lambda /2}}\int_{0}^{2\pi }d\sigma \sin ^{2}\xi
\sin ^{2}\theta _{2}\dot{\phi}_{2}  \label{eq:JPhi2} \\
&&+\sqrt{{2\lambda }}\int_{0}^{2\pi }d\sigma \cos ^{2}\xi \sin ^{2}\xi
\left( \dot{\psi}+\frac{1}{2}\cos \theta _{1}\dot{\phi}_{1}-\frac{1}{2}\cos
\theta _{2}\dot{\phi}_{2}\right) \cos \theta _{2,}  \notag
\end{eqnarray}%
where $E$ is the energy and $S$, $J_{\psi }$, $J_{\phi _{1}}$, and $J_{\phi
_{2}}$ are angular momenta.

A solution to the bosonic equations of motion is said to be a classical
solution if it also satisfies the Virasoro constraints
\end{subequations}
\begin{equation}
\ G_{MN}\left( \partial _{\tau }X^{M}\partial _{\tau }X^{N}+\partial
_{\sigma }X^{M}\partial _{\sigma }X^{N}\right) =0,\,\,\,G_{MN}\partial
_{\tau }X^{M}\partial _{\sigma }X^{N}=0{\normalcolor.}  \label{eq:Virasoro}
\end{equation}%
Note that these are the only constraints that relate motion in $AdS_{4}$ to
motion in $CP^{3}$.

The spectrum of bosonic fluctuations around a classical solution can be
computed by expanding the bosonic Lagrangian in eq.~\ref{eq:bosaction} to
quadratic order in the fluctuations and finding the normal modes of the
resulting equations of motion. In the examples we consider in section 2.2.3, two of the
bosonic modes are massless and the other eight are massive. While the eight
massive modes correspond to the physical transverse degrees of freedom, the
two massless modes can be discarded. One way to see that the massless modes
can be discarded is by expanding the Virasoro constraints to linear order in
the fluctuations \cite{Frolov:2002av}.

To compute the spectrum of fermionic fluctuations, we only need the
quadratic part of the fermionic GS action for type IIA string theory. This
action describes two 10-dimensional
Majorana-Weyl spinors of opposite chirality which can be combined into a
single non-chiral Majorana spinor $\Theta $. The quadratic GS action for type IIA string theory in a general background
can be found in \cite{Cvetic:1999zs}. For the supergravity background in
eqs.~\ref{IIAGeomM} -- \ref{F2}, the quadratic Lagrangian for the fermions is given by
\begin{equation}
\mathcal{L}_{fermi}=\bar{\Theta}\left( \eta ^{ab}-\epsilon ^{ab}\Gamma _{11}\right)
e_{a}\left[ (\partial _{b}+\frac{1}{4}\omega _{b})+\frac{1}{8}e^{\phi
}\left( -\Gamma _{11}\Gamma \cdot F_{2}+\Gamma \cdot F_{4}\right) e_{b}%
\right] \Theta {,}  \label{eq:GS}
\end{equation}%
where $\bar{\Theta}=\Theta ^{\dagger }\Gamma ^{0}$, $\epsilon ^{\tau \sigma
}=-\epsilon ^{\sigma \tau }=1$, $e_{a}=\partial _{a}X^{M}e_{M}^{A}\Gamma _{A}
$, $\omega _{a}=\partial _{a}X^{M}\omega _{M}^{AB}\Gamma _{AB}$, and $\Gamma
\cdot F_{(n)}=\frac{1}{n!}\Gamma ^{N_{1}...N_{n}}F_{N_{1}...N_{n}}$. Note
that $M$ is a base-space index while $A,B=0,...,9$ are tangent-space
indices.

We will now recast the fermionic Lagrangian in eq.~\ref{eq:GS} in a form that allows us to compute the fermionic fluctuation frequencies in a straightforward way. First we note that after rearranging terms, eq.~\ref{eq:GS} can be written as
\begin{equation}
\frac{\mathcal{L}_{fermi}}{2K}=-\bar{\Theta}_{+}\Gamma _{0}\left[ \partial
_{\tau}-\Gamma _{11}\partial _{\sigma}+\frac{1}{4}\left( \omega _{\tau}-\Gamma
_{11}\omega _{\sigma}\right)\right]\Theta-2K\bar{\Theta}_{+}\Gamma _{0}\Gamma \cdot F\Gamma _{0}\Theta_{+} ,  \label{GS2}
\end{equation}%
where we define $K = \partial _{\tau }X^{M}e_{M}^{0}$, $\Theta
_{+}=P_{+}\Theta $, and%
\begin{equation}
P_{+}= -\frac{1}{2K}\Gamma _{0}\left( e_{\tau}+e_{\sigma}\Gamma _{11}\right)
,  \label{Proj}
\end{equation}
\begin{equation}
\Gamma \cdot F =\frac{1}{8}e^{\phi }\left( -\Gamma _{11}\Gamma \cdot
F_{2}+\Gamma \cdot F_{4}\right) \label{GammaF}.
\end{equation}%
Note that $P_{+}=P_{+}^{\dag }$ and if
the classical solution satisfies
\begin{equation}
\partial _{\sigma }X^{M }e_{M}^{0}=0,  \label{ct1}
\end{equation}
then $P_{+}$ is a projection operator, i.e., $P_{+}^{2}=P_{+}$. In addition, if the classical solution satisfies
\begin{equation}
P_{+}\left[ P_{+},\omega _{\tau }-\Gamma _{11}\omega _{\sigma }\right] =0,
\label{ct2}
\end{equation}%
then the fermionic Lagrangian simplifies to
\begin{equation}
\frac{\mathcal{L}_{fermi}}{2K}=-\bar{\Theta}_{+}\Gamma _{0}\left[ \partial
_{\tau}-\Gamma _{11}\partial _{\sigma}+\frac{1}{4}\left( \omega _{\tau}-\Gamma
_{11}\omega _{\sigma}\right) +2K\left( \Gamma \cdot F\Gamma _{0}\right) %
\right] \Theta _{+}.  \label{GS22}
\end{equation}
Finally, if we consider the Fourier mode $\Theta \left( \sigma ,\tau \right) =%
\tilde{\Theta} \exp \left( -i\omega \tau +in\sigma \right) $, where $%
\tilde{\Theta}$ is a constant spinor, then the equations of motion for the
fermionic fluctuations are given by%
\begin{equation}
\left\{ P_{+}\left[ i\omega +in\Gamma _{11}-\frac{1}{4}\left( \omega
_{\tau}-\Gamma _{11}\omega _{\sigma}\right) -2K\left( \Gamma \cdot F\Gamma
_{0}\right) \right] P_{+}\right\} \tilde{\Theta}=0.  \label{Feom}
\end{equation}
One can choose a basis where $P_{+}$ has the form $\left(\begin{array}{cc}
1 & 0\\
0 & 0\end{array}\right)$ (where each element in the $2\times2$ matrix corresponds to a $16\times16$
matrix). In this basis, the matrix on the left-hand side of eq.~\ref{Feom} will
have the form $\left(\begin{array}{cc}
{\bf A} & 0\\
0 & 0\end{array}\right)$. The fermionic frequencies are determined by taking the determinant
of ${\bf A}$ and finding its roots.

Only half of the fermionic components appear in the
Lagrangian in eq.~\ref{GS22}. Hence, a natural choice for fixing kappa-symmetry is to set the
other components to zero by imposing the gauge condition $\Theta =\Theta _{+}
$. This gives the desired number of fermionic degrees of freedom.

\subsection{Algebraic Curve Formalism}
The procedure for computing the spectrum of excitations about a classical
string solution using the $AdS_{4}/CFT_{3}$ algebraic curve was first
presented in~\cite{Gromov:2008bz}. In this section, we reformulate this
procedure in terms of an off-shell formalism similar to the one that was
developed for the $AdS_{5}/CFT_{4}$ algebraic curve in~\cite{Gromov:2008ec}.
The off-shell formalism makes things much more efficient. First we describe
how to construct the classical algebraic curve. Then we describe how to
semi-classically quantize the curve and obtain the spectrum of excitations.

\subsubsection*{Classical Algebraic Curve}

For type IIA string theory in $AdS_{4}\times CP^{3}$, any classical solution
can be encoded in a 10-sheeted Riemann surface whose branches, called
quasimomenta, are denoted by
\begin{equation*}
\left\{q_{1},q_{2},q_{3},q_{4},q_{5},q_{6},q_{7},q_{8},q_{9},q_{10}\right\}.
\end{equation*}
This algebraic curve corresponds to the fundamental representation of $OSp(6|4)$, which is ten-dimensional.
Furthermore, the quasimomenta are not all independent. In particular
\begin{equation}
\left( q_{1}(x),q_{2}(x),q_{3}(x),q_{4}(x),q_{5}(x)\right) =-\left(
q_{10}(x),q_{9}(x),q_{8}(x),q_{7}(x),q_{6}(x)\right) ,
\label{nind}
\end{equation}%
where $x$ is a complex number called the spectral parameter. More precisely, $\left\{ e^{iq_{i}},e^{-iq_{i}}\right\}$ should be regarded as ten branches of the same analytical function. As we will see shortly, the algebraic curve arises by diagonalizing a monodromy matrix which is computed from the Lax connection of the string theory sigma model.

To compute the quasimomenta, it is useful to parameterize $AdS_{4}$ and $CP^3$ using the following embedding coordinates
\begin{eqnarray*}
n_{1}^{2}+n_{2}^{2}-n_{3}^{2}-n_{4}^{2}-n_{5}^{2} &=&1 \\
\sum_{I=1}^{4}\left\vert z^{I}\right\vert ^{2} &=&1,\qquad z^{I}\sim
e^{i\lambda }z^{I},
\end{eqnarray*}%
where $\lambda \in {\mathbb{R}}$.
One can then compute the following connection:
\begin{equation}
j_{a}(\tau ,\sigma )=2\left(
\begin{array}{cc}
n_{i}\partial _{a}n_{j}-n_{j}\partial _{a}n_{i} & 0 \\
0 & z_{I}^{\dagger }D _{a}z^{J}-z^{J}D_{a}z_{I}^{\dagger }%
\end{array}%
\right) {,}  \label{eq:curr}
\end{equation}%
%
where $a \in \left\{\tau ,\sigma \right\} $, $D_{a}=\partial_{a}+i A_{a}$, and $A_{a}=i \sum_{I=1}^{4} z_I^{\dagger} \partial_{a}z^{I}$~\cite{Gromov:2008bz}. This connection is a $9\times 9$ matrix and
transforms under the bosonic part of the supergroup $OSp(6|4)$, notably $SU(4)\times SO(3,2)\sim O(6)\times Sp(4)$. If the equations of motion of the string theory sigma model are satisfied, then $j$ is a conserved current:
\begin{equation}
d \star j =0.
\label{dj}
\end{equation}
Furthermore, $j$ is a flat connection:
\begin{equation}
d j+j \wedge j=0.
\label{flat}
\end{equation}
Eqs. \ref{dj} and \ref{flat} are equivalent to the flatness of the bosonic Lax connection
\begin{equation}
J(x)=\frac{j+ x \star j}{1-x^2},
\end{equation}
where $x$ is the spectral parameter introduced earlier.

Using the Lax connection, we can construct the following monodromy matrix:
\begin{equation}
\Lambda (x)=P\exp \frac{1}{x^{2}-1}\int_{0}^{2\pi }d\sigma \left[ j_{\sigma
}(\tau ,\sigma )+xj_{\tau }(\tau ,\sigma )\right],  \label{eq:mono}
\end{equation}%
where $P$ is the path-ordering symbol and the integral is over a loop of
constant world-sheet time $\tau $. Note that the eigenvalues of $\Lambda (x)$ are independent of the world-sheet time $\tau $. Hence, the monodromy matrix encodes an infinite set of conserved charges.

The quasimomenta are related to the eigenvalues of the monodromy matrix. In
particular, if we diagonalize the monodromy matrix we will find that the eigenvalues of the $AdS_{4}$ part are in general given by%
\begin{equation}
\left\{ e^{i\hat{p}_{1}(x)},e^{i\hat{p}_{2}(x)},e^{i\hat{p}_{3}(x)},e^{i\hat{%
p}_{4}(x)},1\right\} ,  \label{eq:hat}
\end{equation}%
where $\hat{p}_{1}(x)+\hat{p}_{4}(x)=\hat{p}_{2}(x)+\hat{p}_{3}(x)=0$, while
the eigenvalues from the $CP^{3}$ part are given by%
\begin{equation}
\left\{ e^{i\tilde{p}_{1}(x)},e^{i\tilde{p}_{2}(x)},e^{i\tilde{p}%
_{3}(x)},e^{i\tilde{p}_{4}(x)}\right\} ,  \label{eq:tilde}
\end{equation}%
where $\sum_{i=1}^{4}\tilde{p}_{i}(x)=0$. The classical quasimomenta are
then defined as
\begin{equation}
\left( q_{1},q_{2},q_{3},q_{4},q_{5}\right) =\left( \frac{\hat{p}_{1}+\hat{p}%
_{2}}{2},\frac{\hat{p}_{1}-\hat{p}_{2}}{2},\tilde{p}_{1}+\tilde{p}_{2},%
\tilde{p}_{1}+\tilde{p}_{3},\tilde{p}_{1}+\tilde{p}_{4}\right) ,
\label{eq:pq}
\end{equation}%
where we have suppressed the $x$-dependence. From this formula, we see that $%
q_{1}(x)$ and $q_{2}(x)$ correspond to the $AdS_{4}$ part of the algebraic
curve, while $q_{3}(x)$, $q_{4}(x)$, and $q_{5}(x)$ correspond to the $CP^{3}$
part of the algebraic curve.

\subsubsection*{Semi-Classical Quantization}

\begin{figure}[tb]
\center
\includegraphics{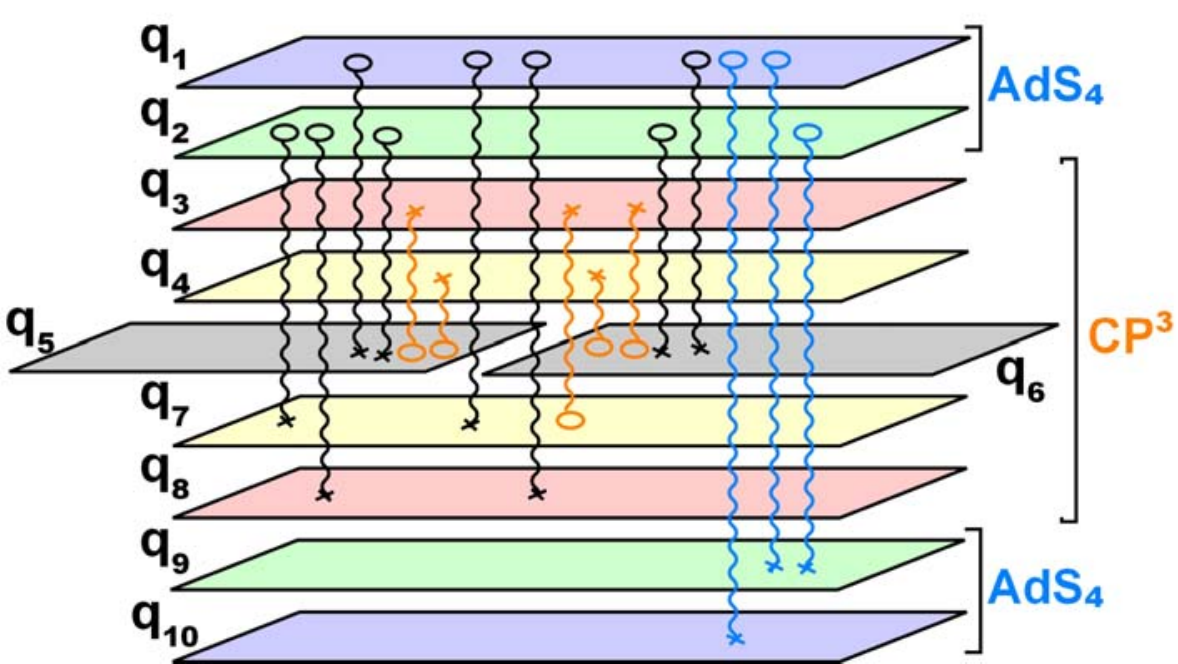}
\caption{Depiction of the fluctuations of the $AdS_{4}\times CP^{3}$ algebraic
curve. Each fluctuation corresponds to a pole which connects two sheets.}
\label{F1}
\end{figure}

The algebraic curve will generically have cuts connecting several pairs of
sheets. These cuts encode the classical physics. To perform semiclassical
quantization, we add poles to the algebraic curve which correspond to
quantum fluctuations. If we think of the poles as infinitesimal cuts, each pole connects two sheets. In particular, the bosonic
fluctuations connect two $AdS$ sheets or two $CP^3$ sheets and the fermionic
fluctuations connect an $AdS$ sheet to a $CP^3$ sheet. See Fig.~\ref{F1} for a depiction of the fluctuations. In total, there are eight
bosonic and eight fermionic fluctuations, and they are labeled by the pairs
of sheets that their poles connect. The labels are referred to as
polarizations and are summarized in Table~\ref{T1}.

\begin{table}[th]
\caption{Labels for the fluctuations (\textcolor{red}{\bf heavy}, \textcolor{blue}{\bf light}) of the $AdS_{4}\times CP^3$ algebraic curve  }
\label
{T1}
\begin{equation}
\begin{array}{c|c}
\toprule & \mathrm{\mathbf{Polarizations~(i,j)}} \\
\midrule {\rm \bf AdS} &
\begin{array}{l}
\textcolor{red}{\bf{(1,10/1,10);(2,9/2,9);(1,9/2,10)}}%
\end{array}
\\
\midrule {\rm \bf Fermions} &
\begin{array}{l}
\textcolor{red}{\bf{(1,7/4,10);(1,8/3,10);(2,7/4,9);(2,8/3,9)}} \\
\textcolor{blue}{\bf{(1,5/6,10);(1,6/5,10);(2,5/6,9);(2,6/5,9)}}%
\end{array}
\\
\midrule CP^{3} &
\begin{array}{l}
\textcolor{red}{\bf{(3,7/4,8)}} \\
\textcolor{blue}{\bf{(3,5/6,8);(3,6/5,8);(4,5/6,7);(4,6/5,7)}}%
\end{array}
\\
\bottomrule
\end{array}
\notag
\end{equation}%
\end{table}
Notice that every fluctuation can be labeled by two equivalent
polarizations because every pole
connects two equivalent pairs of sheets as a consequence of eq. \ref{nind}. Fluctuations connecting sheet 5 or 6 to any other sheet are defined to be light. Notice
that there are eight light excitations. All the others are defined to be
heavy excitations. The physical significance of this terminology will become
clear later on. When we compute the spectrum of fluctuations about the point-particle in $CP^3$, for example, we will find that the heavy excitations
are twice as massive as the light excitations.

When adding poles, we must take into account the level-matching condition
\begin{equation}
\sum_{n=-\infty }^{\infty }n\sum_{ij}N_{n}^{ij}=0,  \label{eq:11}
\end{equation}%
where $N_{n}^{ij}$ is the number of excitations with polarization $ij$ and
mode number $n$. Furthermore, the locations of the poles are not arbitrary;
they are determined by the following equation:
\begin{equation}
q_{i}\left( x_{n}^{ij}\right) -q_{j}\left( x_{n}^{ij}\right) =2\pi n,
\label{poloc}
\end{equation}
where $x_{n}^{ij}$ is the location of a pole corresponding to a fluctuation
with polarization $ij$ and mode number $n$. 

In addition to adding poles to the algebraic curve, we must also add fluctuations to the classical
quasimomenta. These fluctuations will depend on the spectral parameter $x$
as well as the locations of the poles, which we will denote by the
collective coordinate $y$. The functional form of the fluctuations is
determined by some general constraints:
\begin{itemize}
\item They are not all independent:%
\begin{equation*}
\left(%
\begin{array}{c}
\delta q_{1}(x,y) \\
\delta q_{2}(x,y) \\
\delta q_{3}(x,y) \\
\delta q_{4}(x,y) \\
\delta q_{5}(x,y)%
\end{array}%
\right)=-\left(%
\begin{array}{c}
\delta q_{10}(x,y) \\
\delta q_{9}(x,y) \\
\delta q_{8}(x,y) \\
\delta q_{7}(x,y) \\
\delta q_{6}(x,y)%
\end{array}%
\right){\normalcolor .}
\end{equation*}

\item They have poles near the points $x=\pm1$ and the residues of these
poles are synchronized as follows:
\begin{equation}
\lim_{x\rightarrow\pm1}\left(\delta q_{1}(x,y),\delta q_{2}(x,y),\delta
q_{3}(x,y),\delta q_{4}(x,y),\delta q_{5}(x,y)\right)\propto\frac{1}{x\pm1}%
\left(1,1,1,1,0\right){\normalcolor .}  \label{eq:synch}
\end{equation}
This encodes the Virasoro constraints.

\item There is an inversion symmetry:
\begin{equation}
\left(%
\begin{array}{c}
\delta q_{1}(1/x,y) \\
\delta q_{2}(1/x,y) \\
\delta q_{3}(1/x,y) \\
\delta q_{4}(1/x,y) \\
\delta q_{5}(1/x,y)%
\end{array}%
\right)=\left(%
\begin{array}{c}
-\delta q_{2}(x,y) \\
-\delta q_{1}(x,y) \\
-\delta q_{4}(x,y) \\
-\delta q_{3}(x,y) \\
\delta q_{5}(x,y)%
\end{array}%
\right){\normalcolor .}  \label{inversion}
\end{equation}

\item The fluctuations have the following large-$x$ behavior:
\begin{equation}
\lim_{x\rightarrow \infty }\left(
\begin{array}{c}
\delta q_{1}(x,y) \\
\delta q_{2}(x,y) \\
\delta q_{3}(x,y) \\
\delta q_{4}(x,y) \\
\delta q_{5}(x,y)%
\end{array}%
\right) \simeq \frac{1}{2gx}\left(
\begin{array}{c}
\Delta (y)+N_{19}+2N_{1\,10}+N_{15}+N_{16}+N_{17}+N_{18} \\
\Delta (y)+2N_{29}+N_{2\,10}+N_{25}+N_{26}+N_{27}+N_{28} \\
-N_{35}-N_{36}-N_{37}-N_{39}-N_{3\,10} \\
-N_{45}-N_{46}-N_{48}-N_{49}-N_{4\,10} \\
N_{35}+N_{45}-N_{57}-N_{58}-N_{15}-N_{25}+N_{59}+N_{5\,10}%
\end{array}%
\right) ,  \label{eq:asymptotics}
\end{equation}%
where $g=\sqrt{\lambda /8}$, $N_{ij}=\sum_{n=-\infty }^{\infty }N_{n}^{ij}$,
and $\Delta (y)$ is called the anomalous part of the energy shift. Whereas
the $N_{n}^{ij}$ are inputs of the calculation, $\Delta (y)$ will be
determined in the process of determining the fluctuations of the
quasimomenta. The factor of two that appears in front of $N_{1\,10}$ and $N_{29}$ is a consequence of the symmetry in eq. \ref{nind}.

\item Finally, when the spectral parameter approaches the location of one of
the poles, the fluctuations have the following form:
\begin{equation}
\lim_{x\rightarrow x_{n}^{ij}}\delta q_{k}\propto \frac{\alpha
(x_{n}^{ij})N_{n}^{ij}}{x-x_{n}^{ij}},\,\,\,\alpha (x)=\frac{1}{2g}\frac{%
x^{2}}{x^{2}-1},  \label{eq:alpha}
\end{equation}%
where the proportionality constants can be read off from the coefficient of
$N_{ij}$ in the $k$'th row of eq.~\ref{eq:asymptotics}.
\end{itemize}

After computing the anomalous part of the energy shift, the
total energy of the fluctuations is given by
\begin{equation}
\Omega (y)=\Delta (y)+\sum_{AdS_{4}}N^{ij}+\frac{1}{2}\sum_{ferm}N^{ij},
\label{omegaoff}
\end{equation}%
where the first sum is over polarizations $(i,j)$ which correspond to fluctuations in $AdS_4$, and the second sum is over polarizations $(i,j)$ which correspond to fermionic fluctuations. We will denote the energy (or frequency) of a fluctuation with polarization $(i,j)$ as  $\Omega_{ij}(y)$. It is useful to consider the fluctuation frequency without fixing the value
of $y$. In this case, the fluctuation frequency is said to be off-shell. 

Using arguments similar to those in \cite{Gromov:2008ec}, one can deduce the relations among the off-shell frequencies.
First of all, the light off-shell frequencies are related by
\begin{equation}
\label{LL}
\Omega_{i6}(y)=\Omega _{i5}(y)
\end{equation}
where $i=1,2,3,4$.

Second, all the heavy off-shell frequencies can be written as the sum of two light off-shell frequencies as summarized in Table \ref{tLH}.
\begin{table}[th]
\caption{Relations between heavy and light off-shell frequencies }
\label{tLH}
\begin{equation*}
\begin{array}{cc|c}
\toprule \textcolor{red}{\bf{Heavy}} & \textcolor{blue}{\bf{Light}} &  \\
\midrule
\begin{array}{r}
\Omega _{29} =  \\
\Omega _{1~10}=   \\
\Omega _{19}=     \\
\end{array}
&
\begin{array}{l}
2\Omega _{25}   \\
2\Omega _{15}  \\
\Omega _{15}+\Omega _{25}
\end{array}
&
{\rm \bf AdS}
\\
\midrule
\begin{array}{l}
\Omega _{27}= \\
\Omega _{17}= \\
\Omega _{28}= \\
\Omega _{18}=
\end{array}
&
\begin{array}{l}
\Omega _{25}+\Omega _{45}  \\
\Omega _{15}+\Omega _{45} \\
\Omega _{25}+\Omega _{35}   \\
\Omega _{15}+\Omega _{35}
\end{array}
&
{\rm \bf Fermions}
\\
\midrule
\begin{array}{l}
\Omega _{37}=\\
\end{array}
&
\begin{array}{l}
\Omega _{35}+\Omega _{45}
\end{array}
&
{\mathbb C \mathbb P}^3
\\
\bottomrule
\end{array}
\end{equation*}%
\end{table}

Finally, any off-shell frequency $\Omega _{ij}$ is related to its mirror
off-shell frequency $\Omega _{\overline{ij}}$ by
\[
\Omega _{ij}\left( y\right) =-\Omega _{\overline{ij}}\left( 1/y\right)
+\Omega _{\overline{ij}}\left( 0\right) +C,
\]%
where $C=1,1/2$, or $0$ for AdS, Fermionic, or $CP^{3}$ polarizations, respectively. The
mirror polarization $\left( \overline{i,j}\right) $ of the polarization $%
\left( i,j\right) $ can be readily found using eq.~\ref{inversion}, e.g.,  $\left( \overline{1,10}\right)=\left( 2,9\right) ,$ $
\left( \overline{2,5}\right)= \left( 1,5\right) ,$ $\left( \overline{4,5}\right)= \left( 3,5\right) ,$ $\left( \overline{3,7}\right)= \left( 3,7\right) ,$ etc.
Using these relations, only two of the
eight light off-shell frequencies are independent. For example,
\begin{subequations}
\label{M35}
\begin{eqnarray}
\Omega _{35}\left( y\right)  &=&-\Omega _{45}\left( 1/y\right) +\Omega
_{45}\left( 0\right) , \\
\Omega _{25}\left( y\right)  &=&-\Omega _{15}\left( 1/y\right) +\Omega
_{15}\left( 0\right) +1/2.
\end{eqnarray}%
\end{subequations}
In conclusion, if we compute the off-shell frequencies $\Omega _{15}$ and $%
\Omega _{45}$, then we can determine all the other off-shell frequencies
automatically from the relations in eqs.~\ref{LL},\ref{M35}, and Table \ref{tLH}.

The on-shell frequencies are then obtained by evaluating the off-shell frequencies at the location of the
poles, which are determined by solving eq.~\ref{poloc}, i.e. $\omega_{n}^{ij}=\Omega _{ij}\left( x_{n}^{ij}\right) $. It will be convenient to
organize them into the following linear combinations:
\begin{equation}
\omega _{L}(n)=\omega _{n}^{35}+\omega _{n}^{36}+\omega _{n}^{45}+\omega
_{n}^{46}-\omega _{n}^{15}-\omega _{n}^{16}-\omega _{n}^{25}-\omega _{n}^{26}
\label{eq:light}
\end{equation}%
\begin{equation}
\omega _{H}(n)=\omega _{n}^{19}+\omega _{n}^{29}+\omega _{n}^{1\,10}+\omega
_{n}^{37}-\omega _{n}^{17}-\omega _{n}^{18}-\omega _{n}^{27}-\omega _{n}^{28}
\label{eq:heavy}
\end{equation}%
where $L$ stands for light and $H$ stands for heavy. 

\subsection{Fluctuation Frequencies}

In this subsection, we use the algebraic curve formalism to compute the spectrum of fluctuations about two classical solutions corresponding to a point particle and a circular string which are spinning in $CP^3$. The corresponding world-sheet calculations can be found in \cite{Bandres:2009kw}. For each classical solution, we compare the algebraic curve spectrum to the world-sheet spectrum. We also compare the light/heavy fluctuation frequencies about the point-particle solution to the spectrum of elementary/composite excitations of the ABJM spin chain.
\subsubsection{Point-Particle}

In terms of the coordinates of eqs. \ref{adsmetric} and \ref{cpmett}, the
solution for a point-particle rotating with angular momentum $J$ in $CP^{3}$ is given
by
\begin{equation}
t=\kappa \tau ,\,\,\,\rho=0,\,\,\,\xi =\pi
/4,\,\,\,\theta _{1}=\theta _{2}=\pi /2,\,\,\,\psi =\mathcal{J}\tau
,\,\,\,\phi _{1}=\phi _{2}=0,  \label{eq:partglob}
\end{equation}%
where $\mathcal{J}=\frac{J}{4\pi g}$ and $g=\sqrt{\lambda /8}$. This version
of the solution is useful for doing calculations in the world-sheet
formalism. Alternatively, we can write this solution in embedding
coordinates by plugging eq.~\ref{eq:partglob} into eqs.~\ref{adsmap}
and \ref{mapping}:
\begin{equation}
n_{1}=\cos \kappa \tau ,\,\,\,n_{2}=\sin \kappa \tau
,\,\,\,n_{3}=n_{4}=n_{5}=0,\,\,\,z^{1}=z^{2}=z_{3}^{\dagger }=z_{4}^{\dagger
}=\frac{1}{2}e^{i\mathcal{J}\tau /2}{\normalcolor.}  \label{eq:partemb}
\end{equation}%
This version of the solution is useful for doing calculations in the
algebraic curve formalism. The energy and angular momenta of the particle can be read
off from eqs.~\ref{eq:E}--\ref{Js}: $E=4\pi g\kappa $, $S=0$, $J_{\psi }=J$, $%
J_{\phi _{1}}=J_{\phi _{2}}=0$. Furthermore, the Virasoro constraints in
eq.~\ref{eq:Virasoro} give $\kappa =\mathcal{J}$, or equivalently $E=J$.
Note that this is a BPS condition. We therefore expect that the dimension of
the dual gauge theory operator should be protected by supersymmetry. We claim that this classical solution is dual to the BPS operator in eq \ref{eq:vac}.

\subsubsection*{Classical Quasimomenta}

\begin{figure}[tb]
\center
\includegraphics{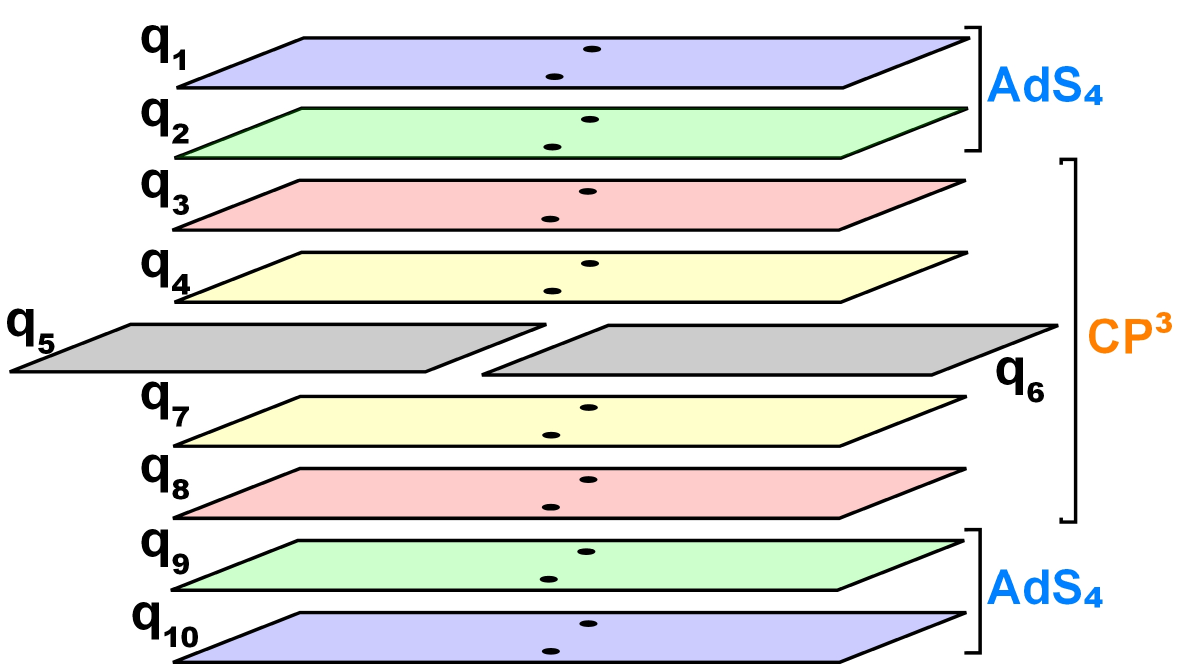}
\caption{Classical algebraic curve for the point-particle rotating in $CP^3$.}
\label{BMN}
\end{figure}

In this section, we compute the algebraic curve for the classical
solution given in eq.~\ref{eq:partemb}. First we plug this solution into eq.~\ref{eq:curr}:
\begin{equation*}
\left(j_{\tau}\right)_{AdS_{4}}=2\kappa\left(%
\begin{array}{ccccc}
0 & 1 & 0 & 0 & 0 \\
-1 & 0 & 0 & 0 & 0 \\
0 & 0 & 0 & 0 & 0 \\
0 & 0 & 0 & 0 & 0 \\
0 & 0 & 0 & 0 & 0%
\end{array}%
\right),\,\,\,\left(j_{\tau}\right)_{CP^{3}}=i\mathcal{J}\left(%
\begin{array}{cccc}
1 & 0 & 0 & 0 \\
0 & 0 & 0 & 0 \\
0 & 0 & -1 & 0 \\
0 & 0 & 0 & 0%
\end{array}%
\right),\,\,\, j_{\sigma}=0{\normalcolor .}
\end{equation*}
Note that this connection is independent of $\sigma$, so it is trivial to
compute the monodromy matrix in eq.~\ref{eq:mono} since path ordering is
not an issue. Diagonalizing the monodromy matrix and comparing the
eigenvalues to eqs.~\ref{eq:hat} and \ref{eq:tilde} then gives $\hat{p%
}_{1}=-\hat{p}_{4}=\frac{4\pi\kappa x}{x^{2}-1}$, $\tilde{p}_{1}=-\tilde{p}%
_{4}=\frac{2\pi\mathcal{J}x}{x^{2}-1}$, and $\hat{p}_{2}=\hat{p}_{3}=\tilde{p%
}_{2}=\tilde{p}_{3}=0$. Recalling that $\kappa=\mathcal{J}$ and plugging
these results into eq.~\ref{eq:pq}, we find that the classical
quasimomenta are
\begin{equation}
q_{1}=q_{2}=q_{3}=q_{4}=\frac{2\pi\mathcal{J}x}{x^{2}-1},\,\,\, q_{5}=0{%
\normalcolor .}  \label{eq:partquasi}
\end{equation}
The algebraic curve corresponding to these quasimomenta is depicted in Fig. \ref{BMN}. Note that all sheets except those corresponding to $q_{5}$ and $%
q_{6}$ have poles at $x=\pm1$.

\subsubsection*{Off-Shell Frequencies}

Recall from eqs.~\ref{LL} and \ref{M35} and Table \ref{tLH} that if we know the off-shell
frequencies $\Omega_{15}(y)$ and $\Omega_{45}(y)$, then
all the others are determined. Let's begin by computing $\Omega _{15}(y)$. Suppose we have two fluctuations
between $q_{1}$ and $q_{5}$. To satisfy level-matching, let's take one of
these fluctuations to have mode number $+n$ and the other to have mode
number $-n$. Each fluctuation corresponds to adding a pole to the classical algebraic curve. The locations of the poles are determined by solving
eq. \ref{poloc}. We will denote the pole locations by $x_{\pm n}^{15}$. We then make the following ansatz for the
fluctuations:
\begin{eqnarray*}
\delta q_{1}(x,y) &=&\sum_{\pm }\frac{\alpha \left( x_{n}^{15}\right) }{%
x-x_{n}^{15}},\,\,\,\delta q_{2}(x,y)=-\delta q_{1}(1/x,y),\,\,\, \\
\delta q_{5}(x,y) &=&-\sum_{\pm }\frac{\alpha \left( x\right) }{%
x-x_{n}^{15}}-\sum_{\pm }\frac{\alpha \left(1/x\right) }{%
1/x-x_{n}^{15}},
\end{eqnarray*}%
where $\alpha(x)$ is defined in eq.~\ref{eq:alpha}, $\pm $ stands for the sum over the positive and negative mode number, and $y$
is a collective coordinate for the positions of the two poles $x_{\pm
n}^{15} $. We have not made an ansatz for $\delta q_{3}$ and $\delta q_{4}$
because they are not needed to compute $\Omega _{15}(y)$. Notice that this
ansatz satisfies the inversion symmetry in eq.~\ref{inversion} and has pole structure
in agreement with eq.~\ref{eq:alpha}. In the large-$x$ limit, the fluctuations reduce
to
\begin{eqnarray*}
\lim_{x\rightarrow \infty }\delta q_{1}(x,y) &\sim &\frac{1}{x}\sum_{\pm
}\alpha \left( x_{n}^{15}\right) ,\,\,\, \\
\lim_{x\rightarrow \infty }\delta q_{2}(x,y) &\sim &\frac{1}{2gx}\sum_{\pm }%
\frac{1}{\left( x_{n}^{15}\right) ^{2}-1},\, \\
\,\,\lim_{x\rightarrow \infty }\delta q_{5}(x,y) &\sim &-\frac{1}{gx},
\end{eqnarray*}%
where we neglect $\mathcal{O}\left( x^{-2}\right) $ terms. Comparing these expressions to eq.~\ref{eq:asymptotics}
implies that the anomalous energy shift is given by
\begin{equation*}
\Delta (y)=\sum_{\pm }\frac{1}{\left( x_{n}^{15}\right) ^{2}-1}{\normalcolor.%
}
\end{equation*}%
The off-shell fluctuation frequency is then obtained by plugging this into eq. \ref{omegaoff}
and recalling that the (1,5) fluctuation is fermionic:
\begin{equation*}
\Omega _{15}(y)=\Delta (y)+\frac{1}{2}N^{15}=\Delta (y)+1=\sum_{\pm }\frac{1%
}{2}\frac{\left( x_{n}^{15}\right) ^{2}+1}{\left( x_{n}^{15}\right) ^{2}-1}.
\end{equation*}%
This implies that the off-shell frequency for a single fluctuation between $%
q_{1}$ and $q_{5}$ is given by%
\begin{equation*}
\Omega _{15}(y)=\frac{1}{2}\frac{y^{2}+1}{y^{2}-1}{\normalcolor.}
\end{equation*}

Now let's compute $\Omega _{45}(y)$. Once again, let's suppose that we have two
fluctuations between $q_{4}$ and $q_{5}$ which have opposite mode numbers $%
\pm n$. We make the following ansatz for the fluctuations:
\begin{eqnarray*}
\delta q_{1}(x,y) &=&\frac{\alpha _{+}(y)}{x+1}+\frac{\alpha _{-}(y)}{x-1}%
,\,\,\,\delta q_{2}(x,y)=-\delta q_{1}(1/x,y), \\
\delta q_{4}(x,y) &=&-\sum_{\pm }\frac{\alpha \left( x\right) }{x-x_{n}^{45}}%
,\,\,\,\delta q_{3}(x,y)=-\delta q_{4}(1/x,y),\,\,\, \\
\delta q_{5}(x,y) &=&\sum_{\pm }\frac{\alpha \left( x\right) }{x-x_{n}^{45}}%
+\sum_{\pm }\frac{\alpha \left( 1/x\right) }{1/x-x_{n}^{45}},
\end{eqnarray*}%
where $\alpha _{\pm }(y)$ are some functions to be determined. Note that
this ansatz satisfies the inversion symmetry in eq~\ref{inversion} and has pole structure
in agreement with eq.~\ref{eq:alpha}. Taking the large-$x$ limit gives%
\begin{eqnarray*}
\lim_{x\rightarrow \infty }\delta q_{1}(x,y) &\sim &\frac{\alpha
_{+}(y)+\alpha _{-}(y)}{x},\,\,\,\lim_{x\rightarrow \infty }\delta
q_{2}(x,y)\sim \alpha _{-}(y)-\alpha _{+}(y)+\frac{\alpha _{+}(y)+\alpha
_{-}(y)}{x}, \\
\lim_{x\rightarrow \infty }\delta q_{3}(x,y) &\sim
&0,\,\,\,\lim_{x\rightarrow \infty }\delta q_{4}(x,y)\sim -\frac{1}{gx}%
,\,\,\,\lim_{x\rightarrow \infty }\delta q_{5}(x,y)\sim \frac{1}{gx}{%
\normalcolor.}
\end{eqnarray*}%
Comparing these limits with eq.~\ref{eq:asymptotics} implies that%
\begin{equation}
\alpha _{+}(y)=\alpha _{-}(y)=\frac{\Delta (y)}{4g}{\normalcolor.}
\label{eq:apmd}
\end{equation}%
Furthermore, the residues of the poles at $x=\pm 1$ must be synchronized
according to eq.~\ref{eq:synch}. Near $x=+1$, $\delta q_1$ and $\delta q_4$ are given by
\[
\lim_{x\rightarrow +1}\delta q_{1}(x,y)\sim \frac{\alpha _{-}(y)}{x-1}%
\]
\[
\lim_{x\rightarrow +1}\delta q_{4}\sim \left( \frac{1}{4g}\sum_{\pm }\frac{1%
}{x_{n}^{45}-1}\right) \frac{1}{x-1}.
\]
Equating the residues of $\delta q_{1} $ and $\delta q_{4}$ near $x=+1$ then gives
\[
\alpha _{-}(y)=\frac{1}{4g}%
\sum_{\pm }\frac{1}{x_{n}^{45}-1}{\normalcolor.}
\]
Combining this with eq. \ref{eq:apmd} implies that
\begin{equation}
\Delta (y)=\sum_{\pm }\frac{1}{x_{n}^{45}-1}{\normalcolor.}
\label{eq:anom45}
\end{equation}%
At this point, it is useful to recall that $x_{n}^{45}$ is a root of the
following equation (which comes from plugging eq~\ref{eq:partquasi}
into eq.~\ref{poloc}):
\begin{equation*}
\frac{2\pi \mathcal{J}x_{n}^{45}}{\left( x_{n}^{45}\right) ^{2}-1}=2\pi n{%
\normalcolor.}
\end{equation*}%
Note that this equation has two roots. The convention that we will follow is
to assign the pole to the root with larger magnitude. Hence, if $n<0$ then $%
x_{n}^{45}=\frac{\mathcal{J}}{n}-\sqrt{1+\frac{\mathcal{J}^{2}}{n^{2}}}$ and
if $n>0$ then $x_{n}^{45}=\frac{\mathcal{J}}{n}+\sqrt{1+\frac{\mathcal{J}^{2}%
}{n^{2}}}$. The point to take away from this discussion is that
\begin{equation*}
x_{+n}^{45}=-x_{-n}^{45}{\normalcolor.}
\end{equation*}%
Using this fact, eq.~\ref{eq:anom45} can be
written as follows:
\begin{equation*}
\Delta (y)=\frac{1}{x_{+n}^{45}-1}-\frac{1}{x_{+n}^{45}+1}=\frac{2}{\left(
x_{+n}^{45}\right) ^{2}-1}=\sum_{\pm }\frac{1}{\left( x_{n}^{45}\right)
^{2}-1}{\normalcolor.}
\end{equation*}%
The off-shell fluctuation frequency is then obtained by plugging this into
eq.~\ref{omegaoff} and recalling that the $(4,5)$ fluctuation is a $CP^{3}$ fluctuation:
\begin{equation*}
\Omega _{45}(y)=\Delta (y)=\sum_{\pm }\frac{1}{\left( x_{n}^{45}\right)
^{2}-1}.
\end{equation*}%
It follows that the off-shell frequency for a single fluctuation between $%
q_{4}$ and $q_{5}$ is given by%
\begin{equation*}
\Omega _{45}(y)=\frac{1}{y^{2}-1}{\normalcolor.}
\end{equation*}

The remaining off-shell frequencies are now easily computed from eqs.~\ref{LL} and \ref{M35} and Table \ref{tLH}.
We summarize the off-shell frequencies in Table~\ref{offshell}.

\begin{table}[tb]
\caption{ Off-shell frequencies for fluctuations about the point-particle
solution }
\label{offshell}%
\begin{equation}
\begin{array}{c|c|l}
\toprule & \mathrm{\mathbf{\Omega(y)}} & \mathrm{\mathbf{Polarizations}} \\
\midrule {\rm \bf AdS} &
\begin{array}{l}
\frac{y^{2}+1}{y^{2}-1}%
\end{array}
&
\begin{array}{l}
\textcolor{red}{\bf{(1,10);(2,9);(1,9)}}%
\end{array}
\\
\midrule {\rm \bf Fermions} &
\begin{array}{l}
\frac{y^{2}+3}{2(y^{2}-1)} \\
\frac{y^{2}+1}{2(y^{2}-1)} \\
\end{array}
&
\begin{array}{l}
\textcolor{red}{\bf{(1,7);(1,8);(2,7);(2,8)}} \\
\textcolor{blue}{\bf{(1,5);(1,6);(2,5);(2,6)}}%
\end{array}
\\
\midrule CP^{3} &
\begin{array}{l}
\frac{2}{y^{2}-1} \\
\frac{1}{y^{2}-1}%
\end{array}
&
\begin{array}{l}
\textcolor{red}{\bf{(3,7)}} \\
\textcolor{blue}{\bf{(3,5);(3,6);(4,5);(4,6)}}%
\end{array}
\\
\bottomrule
\end{array}
\notag
\end{equation}%
\end{table}

\subsubsection*{On-Shell Frequencies}

To compute the on-shell frequencies, we must compute the locations of the
poles by solving eq.~\ref{poloc}. Recall that fluctuations that connect $q_{5}$ or $q_{6}$ to
any other sheets are referred to as light, and all the others are referred
to as heavy. A little
thought shows that for light fluctuations, eq.~\ref{poloc} reduces to
\begin{equation*}
\frac{\mathcal{J}x_{n}}{x_{n}^{2}-1}=n,
\end{equation*}%
and for heavy fluctuations it reduces to
\begin{equation*}
\frac{\mathcal{J}x_{n}}{x_{n}^{2}-1}=\frac{n}{2}.
\end{equation*}%
Each of these equations admits two solutions. We
will assign the location of the pole to the solution with greater magnitude.
Assuming $n>0$, the location of the pole for light excitations is then
given by
\begin{equation*}
x_{n}=\frac{\mathcal{J}}{2n}+\sqrt{\frac{\mathcal{J}^{2}}{4n^{2}}+1},
\end{equation*}%
and the location of the pole for heavy excitations is given by
\begin{equation*}
x_{n}=\frac{\mathcal{J}}{n}+\sqrt{\frac{\mathcal{J}^{2}}{n^{2}}+1}{%
\normalcolor.}
\end{equation*}%
Plugging these solutions into the off-shell frequencies in Table~\ref{offshell}
readily gives the on-shell algebraic curve frequencies in Table~\ref{tabBMN}.

We summarize the spectrum of fluctuations obtained with the algebraic curve
and world-sheet formalisms in Table~\ref{tabBMN}. The algebraic curve frequencies have been re-scaled by a factor of $\kappa$ in order to compare them to the world-sheet frequencies. Note that both sets of frequencies agree with the spectrum of
fluctuations that were found in the Penrose limit (up to constant shifts) \cite{Nishioka:2008gz,Gaiotto:2008cg,Grignani:2008is}.
\begin{table}[tb]
\caption{Spectrum of fluctuations about the point-particle solution computed
using the world-sheet (WS) and algebraic curve (AC) formalisms ($\protect\omega_n=\protect\sqrt{n^2+\protect\kappa^2}$).
Polarizations (\textcolor{red}{\bf{heavy}}/\textcolor{blue}{\bf{light}}) indicate which
pairs of sheets are connected by a fluctuation in the AC formalism, and $%
\pm$ indicates that half of the frequencies have a + and the other half have
a $-$. }
\label{tabBMN}%
\begin{equation}
\begin{array}{c|c|c|l}
\toprule & \mathrm{\mathbf{WS}} & \mathrm{\mathbf{AC}} & \mathrm{\mathbf{%
Polarizations}} \\
\midrule {\rm \bf AdS} &
\begin{array}{c}
\omega_n%
\end{array}
&
\begin{array}{c}
\omega_n%
\end{array}
&
\begin{array}{c}
\textcolor{red}{\bf{(1,10);(2,9);(1,9)}}%
\end{array}
\\
\midrule {\rm \bf Fermions} &
\begin{array}{c}
\omega_n\pm\frac{\kappa}{2} \\
\frac{1}{2}\omega_{2n}%
\end{array}
&
\begin{array}{c}
\omega_n-\frac{\kappa}{2} \\
\frac{1}{2}\omega_{2n}%
\end{array}
&
\begin{array}{l}
\textcolor{red}{\bf{(1,7);(1,8);(2,7);(2,8)}} \\
\textcolor{blue}{\bf{(1,5);(1,6);(2,5);(2,6)}}%
\end{array}
\\
\midrule CP^{3} &
\begin{array}{c}
\omega_n \\
\frac{1}{2}\omega_{2n}\pm\frac{\kappa}{2}%
\end{array}
&
\begin{array}{c}
\omega_n-\kappa \\
\frac{1}{2}\omega_{2n}-\frac{\kappa}{2}%
\end{array}
&
\begin{array}{l}
\textcolor{red}{\bf{(3,7)}} \\
\textcolor{blue}{\bf{(3,5);(3,6);(4,5);(4,6)}}%
\end{array}
\\
\bottomrule
\end{array}
\notag
\end{equation}%
\end{table}

While the constant shifts in the world-sheet spectrum occur with opposite
signs and can be removed by gauge transformations, this is not
the case for the algebraic curve frequencies. In fact, the constant shifts
in the algebraic curve frequencies have physical significance, which can be
seen by taking the mode number $n=0$. In this limit, the $AdS$ frequencies reduce to~$\kappa$, the $%
CP^{3}$ frequencies reduce to 0, and the fermionic frequencies reduce to~$\kappa/2$. In
this sense, the $n=0$ algebraic curve frequencies have {}``flat-space'' behavior. This property was also observed for algebraic curve frequencies computed about solutions in $AdS_{5} \times S^{5}$ \cite{Gromov:2007aq}.
On the other hand, the world-sheet frequencies do not have this property. In
the next subsection, we will see that the constant shifts in the algebraic
curve spectrum have important implications for the one-loop correction to
the classical energy.

Finally, let's compare the light/heavy fluctuations in the string theory to the elementary/composite excitations of the gauge theory spin chain. Recalling that $\mathcal{\kappa}=\frac{J}{\sqrt{2\pi^{2}\lambda}}$,
if we divide the light frequencies by $\kappa$ and expand in the
parameter $\lambda/J^{2}$ we find that \[
\frac{\omega_{light}}{\kappa}=\frac{1}{2}\sqrt{1+\frac{8\pi^{2}n^{2}\lambda}{J^{2}}}=\frac{1}{2}+\frac{2\pi^{2}n^{2}\lambda}{J^{2}}+\mathcal{O}\left(\frac{\lambda^2}{J^4}\right),\]
where we are neglecting constant shifts. From this, we see that the
spectrum of light fluctuations matches the spectrum of elementary
excitations of the spin chain given in eq. \ref{elem} if one makes the replacement
$\lambda\rightarrow2\lambda^{2}$. This confirms that the magnon dispersion
relation contains an interpolating function $h(\lambda)$ which is
equal to $\lambda$ when $\lambda\gg1$ and $2\lambda^{2}$ when $\lambda\ll1$,
as mentioned in the introduction. Similarly, for the heavy excitations
one finds that \[
\frac{\omega_{heavy}}{\kappa}=\sqrt{1+\frac{2\pi^{2}n^{2}\lambda}{J^{2}}}=1+\frac{\pi^{2}n^{2}\lambda}{J^{2}}+\mathcal{O}\left(\frac{\lambda^2}{J^4}\right)\]
which matches the spectrum of composite excitations given in eq. \ref{comp} if
one makes the replacements $\lambda\rightarrow2\lambda^{2}$ and $n\rightarrow2n$.
This suggests that only heavy fluctuations with even mode number are
dual to composite excitations of the spin chain.

\subsubsection{Spinning String}

In the global coordinates of eqs. \ref{adsmetric} and \ref{cpmett}, the
solution for a circular spinning string with two equal nonzero spins in $%
CP^{3}$ is
\begin{equation}
t=\kappa \tau ,\,\,\,\rho =0,\,\,\,\xi =\pi /4,\,\,\,\theta _{1}=\theta
_{2}=\pi /2,\,\,\,\psi =m\sigma ,\,\,\,\phi _{1}=\phi _{2}=2\mathcal{J}\tau
,  \label{eq:spinglob}
\end{equation}%
where $\mathcal{J}=\frac{J}{4\pi g}$, $g=\sqrt{\lambda/8}$, and $m$ is the winding number. Using eqs.~\ref{adsmap}
and \ref{mapping}, we can also
write this solution in embedding coordinates (which are useful for doing
algebraic curve calculations):
\[
n_{1}=\cos \kappa \tau ,\,\,\,n_{2}=\sin \kappa \tau
,\,\,\,n_{3}=n_{4}=n_{5}=0,
\]
\begin{equation}
z^{1}=z_{4}^{\dagger }=\frac{1}{2}e^{i\left( \mathcal{J}\tau +m\sigma/2
\right) },\,\,\,z^{3}=z_{2}^{\dagger }=\frac{1}{2}e^{i\left( \mathcal{J}\tau
-m\sigma/2 \right) }.
\label{eq:spinemb}
\end{equation}%
Eqs. \ref{eq:E}--\ref{Js} imply that $E=4\pi g\kappa $, $S=0$, $J_{\psi }=0$, and $J_{\phi _{1}}=J_{\phi _{2}}=J$. Furthermore, the Virasoro constraints
in eq.~\ref{eq:Virasoro} give $\kappa =\sqrt{m^{2}+4\mathcal{J}^{2}}$, or
equivalently $E=2J\sqrt{1+\frac{\pi ^{2}m^{2}\lambda }{2J^{2}}}$. In the
limit $\mathcal{J}\gg m$, this reduces to the BPS condition $E=2J$, so we
expect that the dual operator should have engineering dimension $2J$ and a nonzero but finite anomalous dimension. We claim that this classical solution is dual to the non-BPS operator in eq. \ref{NB}. Expanding the dispersion relation to first order in the parameter $\lambda /J^{2}$ gives
\begin{equation}
E=2J+\frac{\pi ^{2}m^{2}\lambda }{2J}+\mathcal{O}\left( \lambda
^{2}/J^{3}\right) {\normalcolor.}  \label{eq:circexpand}
\end{equation}%
In order to extrapolate this formula to the gauge theory, we must make the
replacement $\lambda \rightarrow 2\lambda ^{2}$. We then get
the following prediction for the anomalous dimension of the dual gauge
theory operator:
\begin{equation}
\delta D=\frac{\pi ^{2}\lambda ^{2}m^2}{J}+\mathcal{O}\left( \lambda^2 /J^{2}\right) .
\label{eq:spinad}
\end{equation}%
The higher-order terms in the expansion of the classical
string energy in eq.~\ref{eq:circexpand} correspond to $\mathcal{O}\left(
\lambda ^{4}/J^{3}\right) $ corrections to the anomalous dimension, but the
one-loop correction to the energy provides $\mathcal{O}%
\left( \lambda ^{2}/J^{2}\right) $ corrections to the anomalous dimension, as we will now demonstrate.

The calculation of the fluctuation frequencies about the spinning string using the algebraic curve formalism is somewhat involved, so we will just list the main results. The details of the algebraic curve and world sheet calculations can be found in \cite{Bandres:2009kw}.

\begin{figure}[tb]
\center
\includegraphics{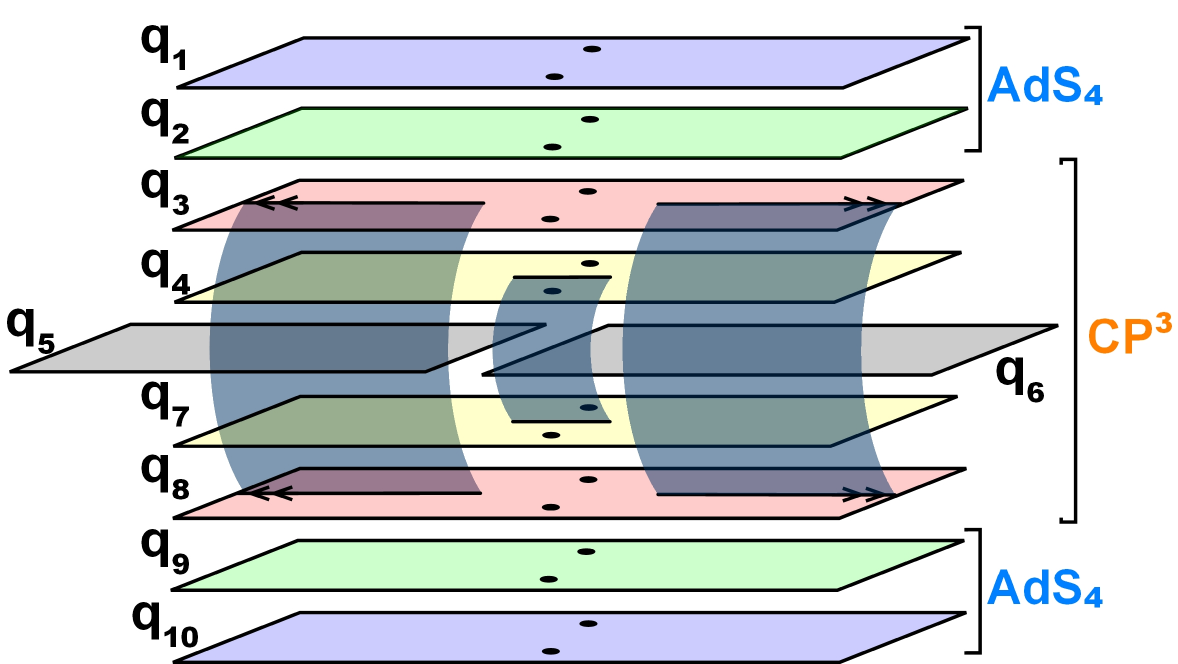}
\caption{Classical algebraic curve for the circular spinning string in $CP^3$}
\label{ACspin}
\end{figure}

Since the spinning string has the same motion in $AdS_{4}$ as the point-particle, the $AdS_{4}$ quasimomenta have the same structure and are given by
\begin{equation*}
q_{1}(x)=q_{2}(x)=\frac{2\pi \kappa x}{x^{2}-1},
\end{equation*}%
where $\kappa =\sqrt{4\mathcal{J}^{2}+m^{2}}$ for the spinning string.
On the other hand, the quasimomenta $q_{3}(x)$, $q_{4}(x)$, and $q_{5}(x)$ are nontrivial and
are given by
\begin{equation}  \label{spinquasi}
\begin{array}{l}
q_{3}(x) =\frac{4\pi x}{x^{2}-1}K(x)-2\pi m, \\
q_{4}(x) =-q_{3}(1/x)-2\pi m=\frac{4\pi x}{x^{2}-1}K(1/x), \\
q_{5}(x) =0%
\end{array}%
\end{equation}
where $K(x)=\sqrt{\mathcal{J}^{2}+m^{2}x^{2}/4}$. From these quasimomenta, we see that the spinning string algebraic curve has
a cut between $q_{3}$ and $q_{8}$ and between $q_{4}$ and $q_{7}$ (by inversion symmetry). The classical algebraic curve is depicted
in Fig.~\ref{ACspin}.

We summarize the spectrum of fluctuations about the spinning string in Tables~\ref{tabAC}~and~\ref{freqT}. The algebraic curve frequencies have been re-scaled by a factor of $\kappa$ in order to compare them to the world-sheet frequencies.
\begin{table}[tbp]
\caption{Spectrum of fluctuations about the spinning string solution
computed using the world-sheet (WS) and algebraic curve (AC) formalisms. The notation for the frequencies is given in Table~\protect\ref%
{freqT}. Polarizations (\textcolor{red}{\bf{heavy}}/\textcolor{blue}{\bf{light}})
indicate which pairs of sheets are connected by a fluctuation in the AC
formalism. }
\label{tabAC}%
\begin{equation}
\begin{array}{c|c|c|l}
\toprule & \mathrm{\mathbf{WS}} & \mathrm{\mathbf{AC}} & \mathrm{\mathbf{%
Polarizations}} \\
\midrule {\rm \bf AdS} &
\begin{array}{l}
\omega_n^{A}%
\end{array}
&
\begin{array}{l}
\omega_n^{A}%
\end{array}
&
\begin{array}{l}
\textcolor{red}{\bf{(1,10);(2,9);(1,9)}}%
\end{array}
\\
\midrule {\rm \bf Fermions} &
\begin{array}{l}
\omega_n^{F}+\frac{\kappa}{2} \\
\omega_n^{F}-\frac{\kappa}{2} \\
\omega_{2n}^{A}/2%
\end{array}
&
\begin{array}{l}
\omega_n^{F}+\frac{\kappa}{2}-2\mathcal{J} \\
\omega_{m+n}^{F}-\frac{\kappa}{2} \\
\omega_{2n}^{A}/2%
\end{array}
&
\begin{array}{l}
\textcolor{red}{\bf{(1,7);(2,7)}} \\
\textcolor{red}{\bf{(1,8);(2,8)}} \\
\textcolor{blue}{\bf{(1,5);(1,6);(2,5);(2,6)}}%
\end{array}
\\
\midrule CP^{3} &
\begin{array}{l}
\omega_n^{C} \\
\omega_n^{C_{-}} \\
\omega_n^{C_{+}}%
\end{array}
&
\begin{array}{l}
\omega_{m+n}^{C}-2\mathcal{J} \\
\omega_{m+n}^{C_{-}} \\
\omega_n^{C_{+}}-2\mathcal{J}%
\end{array}
&
\begin{array}{l}
\textcolor{red}{\bf{(3,7)}} \\
\textcolor{blue}{\bf{(3,5);(3,6)}} \\
\textcolor{blue}{\bf{(4,5);(4,6)}}%
\end{array}
\\
\bottomrule
\end{array}
\notag
\end{equation}%
\end{table}
\begin{table}[tb]
\caption{Notation for spinning string frequencies}
\label{freqT}%
\begin{equation}
\begin{array}{l|l}
\toprule \mathrm{\mathbf{\quad\quad\quad eigenmodes}} & \mathrm{\mathbf{%
notation}} \\
\midrule
\begin{array}{l}
\sqrt{2\mathcal{J}^2+n^2\pm \sqrt{4\mathcal{J}^4+n^2\kappa^2}} \\
\sqrt{4\mathcal{J}^2+n^2-m^2} \\
\end{array}
&
\begin{array}{l}
\omega_n^{C_\pm} \\
\omega_n^{C} \\
\end{array}
\\
\midrule \sqrt{n^2+\kappa^2} &
\begin{array}{l}
\omega_n^{A}%
\end{array}
\\
\midrule \sqrt{4\mathcal{J}^2+n^2} &
\begin{array}{l}
\omega_n^{F}%
\end{array}
\\
\bottomrule
\end{array}
\notag
\end{equation}%
\end{table}
We find
that the algebraic curve spectrum matches the world-sheet spectrum up to
constant shifts and shifts in mode number. Furthermore, if we set the
winding number $m=0$ and take $\mathcal{J}\rightarrow\mathcal{J}/2$, we find
that all the frequencies in Table~\ref{tabAC} reduce to the
corresponding frequencies in Table~\ref{tabBMN}, which is expected since
setting the winding number to zero reduces the string to a point-particle.
This is an important check of our results for the spinning string.
On the other hand, if we set the mode number $n=0$, we find that the
algebraic curve frequencies once again have flat-space behavior, i.e., the $AdS$
frequencies reduce to~$\kappa$, the $CP^{3}$ frequencies reduce to 0, and the fermionic
frequencies reduce to~$\kappa/2$.

Finally, we would like to point out that both the
algebraic curve and world-sheet spectra have instabilities when $|m|\geq2$. For example if we set $m=2$, then the algebraic curve frequencies
labeled by (3,5) and (3,6) become imaginary for $n=-3$ and $n=-1$ and the
corresponding world-sheet frequencies become imaginary for $n=\pm1$. \footnote{We would like to thank Victor Mikhaylov for showing us his unpublished notes on the spinning string algebraic curve \cite{notes}. In these notes, he also derives the algebraic curve for the spinning string and uses it to compute the fluctuation frequencies, however the asymptotics that he imposes on the algebraic curve are different from the asymptotics we use in eq.~\ref{eq:asymptotics}. The differences occur in the signs of several terms on the right-hand side of eq.~\ref{eq:asymptotics}. As a result, we obtain frequencies with different constant shifts.}

\subsection{One-Loop Energies}

In the previous subsection, we computed the spectrum of fluctuations about the point-particle and spinning string solutions using world-sheet and algebraic curve techniques and found that the algebraic curve frequencies agree with the world-sheet frequencies up to constant shifts and shifts in mode number. Although the algebraic curve and world-sheet frequencies look similar, they have very different properties. In the next section, we will describe how to compute the one-loop corrections to the energy of these classical solutions. Computing one-loop corrections is subtle because it involves evaluating an infinite sum of fluctuation frequencies. In particular, the algebraic curve spectrum gives a divergent one-loop correction if we use the same prescription for adding up the frequencies that was used for type IIB string theory on $AdS_{5}\times S^5$. Since the point-particle is a BPS solution, we expect that its one-loop correction should vanish. Furthermore, since the spinning string solution becomes near-BPS in a certain limit, we expect its one-loop correction to be nonzero but finite. Hence the algebraic curve does not give one-loop corrections which are compatible with supersymmetry if one uses the standard summation prescription. In the next section, we describe an alternative summation prescription which gives  a vanishing one-loop correction for the point-particle and a finite one-loop correction for the spinning string when used with both the algebraic curve spectrum and the world-sheet spectrum. In section 2.2.4.3, we will show that the one-loop energy of the spinning string can be matched with the anomalous dimension of the dual gauge theory operator in eq. \ref{NB}, which was computed using a Bethe Ansatz.

\subsubsection{Summation Prescriptions}

Given the spectrum of fluctuations about a classical string solution, we
compute the one-loop correction to the string energy by adding up the
spectrum. The standard formula, which we refer to as the old summation prescription, is
\begin{equation}
\delta E_{1-loop,old}=\lim_{N\rightarrow \infty }\frac{1}{2\kappa}\sum_{n=-N}^{N}%
\left( \sum_{i=1}^{8}\omega _{n,i}^{B}-\sum_{i=1}^{8}\omega
_{n,i}^{F}\right) ,  \label{eq:bad1}
\end{equation}%
where $\kappa$ is proportional to the classical energy, $B/F$ stands for bosonic/fermionic, $n$ is the mode number, and $i$ is
some label. For example, if we are dealing with frequencies
computed from the algebraic curve, then they will be labeled by a pair of
integers called a polarization, as explained in section 2.2. Although this
formula works well for string solutions in $AdS_{5}\times S^{5}$, it gives a
one-loop correction which disagrees with the all-loop Bethe ansatz when
applied to the folded-spinning string in $AdS_{4}\times CP^{3}$ \cite{Krishnan:2008zs, McLoughlin:2008ms, Alday:2008ut}. In~\cite{Gromov:2008fy} Gromov and Mikhaylov subsequently proposed the following formula for
computing one-loop corrections in $AdS_{4}\times CP^{3}$:
\begin{equation}
\delta E_{1-loop,{\rm GM}}=\lim_{N\rightarrow \infty }\frac{1}{2\kappa}%
\sum_{n=-N}^{N}K_{n},\,\,\,K_{n}=\left\{
\begin{array}{c}
\omega ^{H}(n)+\omega^{L}(n/2)\,\,\,n\in even \\
\omega^{H}(n)\,\,\,n\in odd%
\end{array}%
\right. ,  \label{eq:vic}
\end{equation}%
where $\omega _{n}^{L}$/$\omega _{n}^{H}$ are referred to as heavy/light
frequencies and are defined in eqs.~\ref{eq:light} and \ref{eq:heavy}.
For later convenience, we note that eq.~\ref{eq:bad1} can be written in
terms of heavy and light frequencies as follows:
\begin{equation}
\delta E_{1-loop,old}=\lim_{N\rightarrow \infty }\frac{1}{2\kappa}\sum_{-N}^{N}\left(
\omega _{L}(n)+\omega _{H}(n)\right) {\normalcolor.}  \label{eq:badsum}
\end{equation}%
In the large-$\kappa$ limit, eq.~\ref{eq:vic} can be approximated by the following integral:
\begin{equation}
\delta E_{1-loop}\approx \lim_{N\rightarrow \infty }\frac{1}{2\kappa}\int_{-N }^{N }\left( \omega
_{H}(n)+\frac{1}{2}\omega _{L}(n/2)\right) dn{\normalcolor.}
\label{eq:intvic}
\end{equation}%
In ~\cite{Gromov:2008fy} it was shown that Eq.~(\ref{eq:intvic}) gives a one-loop correction which agrees with the all-loop
Bethe Ansatz when applied to the spectrum of the folded spinning string.

An alternative prescription for computing one-loop corrections, which we refer to as the new prescription, was proposed in \cite{Bandres:2009kw}:
\begin{equation}
\delta E_{1-loop,new}=\lim_{N\rightarrow \infty }\frac{1}{2\kappa}\sum_{-N}^{N}\left(2\omega_{H}(2n)+\omega_{L}(n)\right){%
\normalcolor.}  \label{eq:goodsum}
\end{equation}%
This sum can be motivated by the observation in section 2.2.3.1 that only heavy fluctuations with even mode numbers correspond to composite excitations of the spin chain. Alternatively, from the frequencies in Table~\ref{tabBMN}, one sees that heavy fluctuations with mode number $2n$ can be thought of as threshold bound states of two light fluctuations with mode number $n$ \footnote{Note that the heavy modes are only threshold bound states at infinite 't Hooft coupling. At finite 't Hooft coupling, each heavy fluctuation becomes more massive than two light fluctuations, and therefore becomes unstable \cite{Zarembo:2009au}. Although the heavy fluctuations cannot be asymptotic states, they can still appear on internal lines in Feynman diagrams and therefore should be included in the path integral.}. This suggests that only heavy modes with even mode number should contribute to the one-loop correction. The formula for the one-loop correction should therefore have the form
\[\delta E_{1-loop,new}=\lim_{N\rightarrow \infty }\frac{1}{2\kappa}\sum_{-N}^{N}\left(A\omega_{H}(2n)+B\omega_{L}(n)\right){\normalcolor .}\]
The coefficients $A$ and $B$ can then be fixed uniquely by requiring that the integral approximation to this formula reduces to eq.~\ref{eq:intvic} in the large-$\kappa$ limit, ensuring that this summation prescription gives a one-loop correction to the folded spinning string energy which agrees with the all-loop Bethe Ansatz. One then finds that $A=2$ and $B=1$.

One virtue of the summation prescription in eq.~\ref{eq:goodsum} compared to the one in eq.~\ref{eq:vic} is that it gives more well-defined results for one-loop corrections. For example, consider the case where
$\omega_{L}(n)=-2\omega_{H}(n)=C$, where $C$ is some constant (we will see shortly that the AC frequencies for the point-particle have this form). In this case,
eq.~\ref{eq:vic} does not have a well-defined $N\rightarrow\infty$ limit;
in particular the sum alternates between $\pm C/(4\kappa)$ depending on whether $N$ is
even or odd. On the other hand, eq.~\ref{eq:goodsum} vanishes for all $N$.

In the next two subsections, we will use the summation prescriptions in eqs. \ref{eq:badsum} and \ref{eq:goodsum} to compute the one-loop corrections to the energies of the point-particle and spinning string solutions.

\subsubsection{Point-Particle}

Using eqs.~\ref{eq:light} and \ref{eq:heavy}, we see that $\omega_{H}$
and $\omega_{L}$ are constants for both the world-sheet and algebraic curve
spectra. In particular, for the world-sheet spectrum we find that $%
\omega_{H}(n)=\omega_{L}(n)=0$. As a result, both the standard summation
prescription in eq.~\ref{eq:badsum} and the new summation prescription in
eq.~\ref{eq:goodsum} give a vanishing one-loop correction to the energy.
On the other hand, for the algebraic curve spectrum we find that $\omega_{H}(n)=\kappa$
and $\omega_{L}(n)=-2\kappa$. For these values of $\omega_{H}$ and $\omega_{L}$,
the new summation prescription gives a vanishing one-loop correction but the
standard summation prescription gives a linear divergence:
\begin{equation*}
\delta E_{1-loop,old}=\lim_{N\rightarrow\infty}-(N+1/2){\normalcolor .}
\end{equation*}
Thus we find that both summation prescriptions are consistent with
supersymmetry if we use the spectrum computed from the world-sheet, but only
the new summation is consistent with supersymmetry if we use the spectrum
computed from the algebraic curve.

\subsubsection{Spinning String}

For the spinning string, $\omega _{H}(n)$ and $\omega _{L}(n)$ defined in
eqs.~\ref{eq:light} and \ref{eq:heavy} are nontrivial:
\begin{eqnarray*}
\omega _{H}^{WS}(n) &=&3\omega _{n}^{A}+\omega _{n}^{C}-4\omega _{n}^{F},\,\,
\\
\,\omega _{L}^{WS}(n) &=&2\omega _{n}^{C_{+}}+2\omega _{n}^{C_{-}}-2\omega
_{2n}^{A}, \\
\omega _{H}^{AC}(n) &=&3\omega _{n}^{A}+\omega _{n+m}^{C}-2\omega
_{n}^{F}-2\omega _{n+m}^{F}+2\mathcal{J},\,\, \\
\omega _{L}^{AC}(n) &=&2\omega _{n}^{C_{+}}+2\omega _{n+m}^{C_{-}}-2\omega
_{2n}^{A}-4\mathcal{J},
\end{eqnarray*}%
where WS stands for world-sheet, AC stands for algebraic curve, and we are using the notation in Table~\ref{freqT}.

To compute the one-loop correction, we must evaluate an infinite
sum of the form
\begin{equation}
\delta E_{1-loop}=\sum_{n=-\infty}^{\infty}\Omega\left(\mathcal{J},n,m\right){\normalcolor .}
\label{sum1}
\end{equation}
Note that the frequency $\Omega$ in this equation should not be confused with the off-shell frequencies defined in section 2.2.2. Since we have two summation prescriptions (the old one in eq.~\ref{eq:badsum} and the new one in eq.~\ref{eq:goodsum}) and two sets of frequencies (world-sheet and algebraic curve) there are four choices for $\Omega\left(\mathcal{J},n,m\right)$:
\begin{equation}
\begin{array}{l}
\Omega _{old,WS} =\frac{1}{2\kappa}\left(\omega_{H}^{WS}(n)+\omega_{L}^{WS}(n)\right),\,\,\, \\
\Omega _{new,WS} =\frac{1}{2\kappa}\left(2\omega_{H}^{WS}(2n)+\omega_{L}^{WS}(n)\right), \\
\Omega _{old,AC} =\frac{1}{2\kappa}\left(\omega_{H}^{AC}(n)+\omega_{L}^{AC}(n)\right), \\
\Omega _{new,AC} =\frac{1}{2\kappa}\left(2\omega_{H}^{AC}(2n)+\omega_{L}^{AC}(n)\right),
\end{array}
\label{summands}
\end{equation}
where $old$/$new$ refers to the summation prescription.

To gain further insight, let's look at the summands in eq.~\ref{summands} in two limits: the large-$n$ limit and the large-$\mathcal{J}$ limit. By looking at the large-$n$ limit, we will learn about the convergence properties of the one-loop corrections, and by looking at the large-$\mathcal{J}$ limit and evaluating the sums over $n$ using $\zeta$-function regularization, we will be able to compute the $\mathcal{J}^{-2n}$ contributions to the one-loop corrections. These are referred to as the analytic terms. There are also terms proportional to $\mathcal{J}^{-2n+1}$, which are referred to as the non-analytic terms, and exponentially suppressed terms, i.e., terms that scale like $e^{-\mathcal{J}}$. These terms are sub-dominant compared to the analytic terms in the large-$\mathcal{J}$ limit.

\subsubsection*{Large-$n$ Limit}

Note that in all four cases $\Omega\left(\mathcal{J},-n,m\right)=\Omega\left(\mathcal{J},n,-m\right)$, so the one-loop correction in eq.~\ref{sum1} can be written as
\begin{equation}
\delta E_{1-loop}=\Omega\left(\mathcal{J},0,m\right)+\sum_{n=1}^{\infty}\left(\Omega\left(\mathcal{J},n,m\right)+\Omega\left(\mathcal{J},n,-m\right)\right)
\label{sum2}.
\end{equation}
The large-$n$ limit of $\Omega\left(\mathcal{J},n,m\right)+\Omega\left(\mathcal{J},n,-m\right)$ for the four choices of $\Omega\left(\mathcal{J},n,m\right)$ is summarized in Table~ \ref{largen}.
\begin{table}[tb]
\caption{Large-$n$ limit of  $\Omega\left(\mathcal{J},n,m\right)+\Omega\left(\mathcal{J},n,-m\right)$ for the old summation (where $\Omega\left(\mathcal{J},n,m\right)=\omega_{H}(n)+\omega_{L}(n)$) and the new summation (where $\Omega\left(\mathcal{J},n,m\right)=2\omega_{H}(2n)+\omega_{L}(n)$) applied to the world-sheet (WS) spectrum and algebraic curve (AC) spectrum}
\label{largen}%
\begin{equation}
\begin{array}{c|c|c}
\toprule & \mathrm{\mathbf{WS}} & \mathrm{\mathbf{AC}} \\
\midrule {\mathbf{Old\,\,Sum}} &
\begin{array}{l}
-\frac{m^{2}\left(5m^{2}/4+3\mathcal{J}^{2}\right)}{2\kappa n^{3}}+\mathcal{O}\left(n^{-5}\right)
\end{array}
&
\begin{array}{l}
-\frac{2\mathcal{J}}{\kappa}-\frac{m^{2}\left(11m^{2}/4+5\mathcal{J}^{2}\right)}{2\kappa n^{3}}+\mathcal{O}\left(n^{-4}\right)
\end{array}
\\
\midrule {\mathbf{New\,\,Sum}} &
\begin{array}{l}
-\frac{m^{4}}{4\kappa n^{3}}+\mathcal{O}\left(n^{-5}\right)
\end{array}
&
\begin{array}{l}
\frac{m^{2}\left(\mathcal{J}^{2}-5m^{2}/4\right)}{2\kappa n^{3}}+\mathcal{O}\left(n^{-4}\right)
\end{array}
\\
\bottomrule
\end{array}
\notag
\end{equation}%
\end{table}

From this table we see that all one-loop
corrections are free of quadratic and logarithmic divergences because terms of order $n$ and order $1/n$ cancel out in
the large-$n$ limit. At the same time, we find a linear divergence when we apply the old summation prescription to the algebraic curve spectrum, since the summand has a constant term. In all other cases however, the summands are
at most $\mathcal{O}(n^{-3})$, which suggests that the one-loop corrections are
convergent. Hence we find that both summation prescriptions give finite one-loop corrections when applied to the world-sheet spectrum, but only the new summation prescription gives a finite result when applied to the algebraic
curve spectrum. This is the same thing we found for the point-particle. The
new feature of the spinning string is that the finite one-loop correction is
nonzero and therefore provides a nontrivial prediction to be compared with the dual gauge theory.

\subsubsection*{Large-$\mathcal{J}$ Limit}
In the previous section we found that when $\Omega\left(\mathcal{J},n,m\right)=\Omega _{old,AC}\left(\mathcal{J},n,m\right)$, the one-loop correction is divergent, but for the other three cases in eq.~\ref{summands}, it is convergent. This means we have three possible predictions for the one-loop correction, however by expanding the summands in the large-$\mathcal{J}$ limit and evaluating the sums over $n$ at each order of $\mathcal{J}$ using $\zeta$-function regularization, we find that all three cases give the same result. The technique of $\zeta$-function
regularization, which was developed in \cite{SchaferNameki:2005tn}, is convenient for computing the analytic terms in the
large-$\mathcal{J}$ expansion of one-loop corrections but does not capture non-analytic and exponentially suppressed terms. The latter terms can be computed using the techniques described in \cite{SchaferNameki:2006gk}. We now describe this procedure in more detail.

If we expand the summand in the large-$\mathcal{J}$ limit, only even powers of $\mathcal{J}$ appear:
\begin{equation}
\sum_{n=-\infty}^{\infty}\Omega\left(\mathcal{J},n,m\right)=\sum_{k=1}^{\infty}\mathcal{J}^{-2k}\sum_{n=-\infty}^{\infty}\Omega_{k}(n,m){\normalcolor .}\label{eq:j1}\end{equation}
For each power of $\mathcal{J}$, the sum over $n$ can be written
as follows
\begin{equation}
\sum_{n=-\infty}^{\infty}\Omega_{k}(n,m)=\Omega_{k}(0,m)+\sum_{n=1}^{\infty}\left(\Omega_{k}(n,m)+\Omega_{k}(n,-m)\right){\normalcolor .}\label{eq:j2}\end{equation}
If we expand $\Omega_{k}(n,m)$ in the limit $n\rightarrow\infty$,
we find that it splits into two pieces:

\[
\Omega_{k}(n,m)=\sum_{j=-1}^{2k}c_{k,j}(m)n^{j}+\tilde{\Omega}_{k}(n,m)\]
where $\tilde{\Omega}_{k}(n,m)$ is $\mathcal{O}\left(n^{-2}\right)$. We will refer to $\tilde{\Omega}_{k}(n,m)$
as the finite piece because it converges when summed over $n$, and
$\sum_{j=-1}^{2k}c_{k,j}(m)n^{j}$ as the divergent piece because
it diverges when summed over $n$. Furthermore, we find that $\tilde{\Omega}_{k}(n,m)=\tilde{\Omega}_{k}(n,-m)$
and $c_{k,j}(m)\propto m^{2k-j}$. Hence, the odd powers of $n$ cancel
out of the divergent piece when we add $\Omega_{k}(n,m)$ to $\Omega_{k}(n,-m)$
and we get \[
\Omega_{k}(n,m)+\Omega_{k}(n,-m)=2\left[\sum_{j=0}^{k}c_{k,2j}(m)n^{2j}+\tilde{\Omega}_{k}(n,m)\right]{\normalcolor .}\]
Noting that $\zeta(0)=-1/2$ and $\zeta\left(2j\right)=0$ when $j$ is a positive integer,
we see that only the constant term in the divergent piece contributes
if we evaluate the sum over $n$ using $\zeta$-function regularization:
\[
\sum_{n=1}^{\infty}\left(\Omega_{k}(n,m)+\Omega_{k}(n,-m)\right)\rightarrow-c_{k,0}+2\sum_{n=1}^{\infty}\tilde{\Omega}_{k}(n,m){\normalcolor .}\]
Combining this with eqs.~\ref{eq:j1} and \ref{eq:j2} then gives

\[
\delta E_{1-loop}=\sum_{k=1}^{\infty}\mathcal{J}^{-2k}\left[\Omega_{k}(0,m)-c_{k,0}+2\sum_{n=1}^{\infty}\tilde{\Omega}_{k}(n,m)\right]{\normalcolor .}\]

Using the procedure described above, we obtain a single prediction
for the one-loop correction to the energy of the spinning string:
\begin{eqnarray}
\delta E_{1-loop} &=&\frac{1}{2\mathcal{J}^{2}}\left[ m^{2}/4+\sum_{n=1}^{%
\infty }\left( n\left( \sqrt{n^{2}-m^{2}}-n\right) +m^{2}/2\right) \right]
\label{eq:predicold} \\
&&-\frac{1}{8\mathcal{J}^{4}}\left[ 3m^{4}/16+\sum_{n=1}^{\infty }\left(
\begin{array}{c}
3m^{4}/8-n^{4} \\
+n\sqrt{n^{2}-m^{2}}\left( m^{2}/2+n^{2}\right)  \notag
\end{array}%
\right) \right] +\mathcal{O}\left( \frac{1}{\mathcal{J}^{6}}\right){%
\normalcolor. \notag}
\end{eqnarray}%
Recalling that $\mathcal{J}=\frac{J}{\sqrt{2\pi^{2}\lambda}}$ and making the replacement $\lambda\rightarrow2\lambda^{2}$ in eq.~\ref{eq:predicold} then gives a prediction for the $1/J$ correction to the anomalous
dimension of the gauge theory operator in eq. \ref{NB}:
\begin{equation}
\delta D=\left(\frac{\pi^{2}\lambda^{2}m^2}{J}+...\right)+\frac{1}{J}\left(\frac{2a\pi^{2}\lambda^{2}}{J}+...\right)\label{prediction}\end{equation}
where
\begin{eqnarray*}
a &=&m^2/4+\sum_{n=1}^{\infty }\left( n\left( \sqrt{n^{2}-m^2}-n\right) +m^2/2\right) . \\
\end{eqnarray*}%
The first term in eq.~\ref{prediction} came from expanding the classical dispersion relation for the spinning string to first order in the parameter $\lambda/J^2$ and then making the replacement $\lambda\rightarrow2\lambda^{2}$. Note that this result agrees with the gauge theory computation in eq. \ref{AD}.

%% file: amplitude.tex
 \chapter{Scattering Amplitudes}    		
\label{amplitude}

\section{On-Shell Superspace}

The study of on-shell scattering amplitudes in $\mathcal{N}=4$ sYM has shown that many symmetries, structures, and dualities can be discovered by parameterizing the scattering amplitudes using supertwistors. Furthermore, amplitudes take a much simpler form and can be computed much more efficiently when parametrized in terms of these variables.  Supertwistors are objects which transform in the fundamental representation of the superconformal group. For a three-dimensional superconformal field theory, a supertwistor takes the following form:

$$\zeta^{\mathcal{M}}=\left(\begin{array}{c}\xi^{a } \\ \eta^{\bf{I}} \end{array}\right)$$
where $\xi^{a}$ is a bosonic twistor with four components which trasforms in the fundamental representation of the conformal group $SO(3,2)=Sp(4)$, and $\eta^{\bf{I}}$ is a fermionic twistor which transforms in the fundamental representation of the R-symmetry group.  For a theory with $\mathcal{N}=6$ supersymmetry like the ABJM theory, the index $\bf{I}$ runs from 1 to 6. Note that the twistors are real and self-conjugate, i.e., they satisfy the following canonical commutation relations:
\eq
[\xi^a,\xi^b]=\Omega^{ab}\;\;\{\eta^{\bf{I}},\eta^{\bf{J}}\}=\delta^{\bf{IJ}},
\label{conjugation}
\eqe
where $\Omega^{ab}=-\Omega^{ba}$ is the invariant tensor of $Sp(4)$. The $Sp(4)$ twistors can be broken into two-component objects
$$\xi^{a}=\left(\begin{array}{c}\lambda^\alpha \\ \mu_\beta\end{array}\right),\quad[\lambda^\alpha,\mu_\beta]=\delta^\alpha_\beta,\quad \alpha=1,2 $$
where the $\lambda$'s are $SL(2,R)=SO(2,1)$ spinors used for the bi-spinor representation of massless momenta:
$$p^\mu(\sigma_\mu)^{\alpha\beta}= \lambda^\alpha\lambda^\beta.$$
We will take all momenta to be outgoing. As a result, the spinor $\l^\a$ is real for particles that are physically outgoing ($p^0>0$) and purely imaginary for particles that are physically incoming ($p^0<0$).
Since the right-hand side is symmetric in the spinor indices, the left-hand side is indeed a three-component object, as expected. Note that the twistors of three-dimensional Minkowski space have a geometric interpretation similar to the twistors of four-dimensional Minkowski space. In particular, a point in twistor space corresponds to a null ray in Minkowski space and a point in Minkowski space corresponds to a line in twistor space \cite{Penrose:1986ca}. This is described in greater detail in Appendix C.

Since supertwistors of the ABJM theory are self-conjugate, the scattering amplitudes can be parameterized using the half-supertwistor variables $(\lambda^\alpha,\eta^A)$, where $A$ runs from one to three. The three $\eta^A$ variables are linear combinations of the six $\eta^{\bf{I}}$ variables such that the $\eta^A$'s anticommute. These variables are referred to as the on-shell space. In terms of these variables, the generators of the superconformal group  $OSp(6|4)$  can be represented as follows:
$$ p^{\alpha\beta}=\lambda^\alpha\lambda^\beta$$
$$q^{A\alpha}=\lambda^\alpha\eta^A,\;\;q^\alpha_{A}=\lambda^\alpha\frac{\partial}{\partial \eta^A}$$
$$ m^\alpha\,_\beta=\lambda^\alpha\frac{\partial}{\partial \lambda^\beta}-\delta^\alpha_\beta\frac{1}{2}\lambda^\gamma\frac{\partial}{\partial \lambda^\gamma},\;\;d=\frac{1}{2}\lambda^\gamma\frac{\partial}{\partial \lambda^\gamma}+\frac{1}{2}$$
$$ r^{AB}=\eta^A\eta^B,\;\;r^A\,_B=\eta^A\frac{\partial}{\partial \eta^B}-\delta^A_B\frac{1}{2},\;\;r_{AB}=\frac{\partial}{\partial \eta^A}\frac{\partial}{\partial \eta^B}$$
$$s^{A}_\alpha=\eta^A\frac{\partial}{\partial \lambda^\alpha},\;\;s_{\alpha A}=\frac{\partial}{\partial \lambda_\alpha}\frac{\partial}{\partial \eta^A}$$
\begin{equation}
k_{\alpha\beta}=\frac{\partial}{\partial \lambda^\alpha}\frac{\partial}{\partial \lambda^\beta}.
\label{confalg}
\end{equation}
The constant appearing in the dilatation operator counts the engineering dimension of the scalar field, which is $1/2$. From the R-symmetry generator $r^A_{\,\,\,\,\,B}$, we see that an $n$-point superamplitude must have fermionic degree $3n/2$. This is in contrast to the chiral formulation of $\mathcal{N}=4$ sYM, where the fermionic degree is related to the MHV-degree of the amplitude rather than the number of external legs.

The on-shell superfields of the ABJM theory take the form \cite{Bargheer:2010hn}
\be
&&\Phi = \phi^4 + \eta^A \psi_A + \thalf \e_{ABC} \eta^A\eta^B \phi^C
+ \tfrac{1}{6} \e_{ABC} \eta^A \eta^B \eta^C \psi_4 \,, \\
&&\bar{\Phi} = \psi^4 + \eta^A \phi_A +
\thalf \e_{ABC} \eta^A\eta^B \psi^C
+ \tfrac{1}{6} \e_{ABC} \eta^A \eta^B \eta^C \phi_4.
\label{sfield}
\ee
The matter fields appearing in the superfield expansion have $SU(4)$ R-symmetry indices, so it is easy to match them with the matter fields in the ABJM Lagrangian, which is reviewed in Appendix A. Note that the fields $\phi^I$ and $\psi^I$ , $I=1,...,4$, are in the fundamental representation of the R-symmetry group $SU(4)$, and their adjoints $\phi_I$ and $\psi_I$ are in the anti-fundamental representation. The gauge fields do not appear in the expansion of the on-shell superfields because Chern-Simons fields have no on-shell degrees of freedom.

Equipped with the on-shell superspace, we now introduce the building blocks for the construction of superconformal amplitudes. First we define the supermomentum delta function:
\eqa
\nonumber\delta^6(Q)&=&\delta\left(Q^{1\alpha}\right)\delta\left(Q_{\alpha}^{1}\right)\delta\left(Q^{2\beta}\right)\delta\left(Q_{\beta}^{2}\right)\delta\left(Q^{3\gamma}\right)\delta\left(Q_{\gamma}^{3}\right)
\eqae
where
$$Q^{A\alpha}=\sum_{i=1}^{n}q_{i}^{A \alpha},\,\,\,\,\,q_i^{A \alpha}=\lambda_i^{\alpha}\eta^{A}_i$$
and the summation is over all external lines, which are labeled by $i$. The Lorentz-invariant objects are then:
$$\delta^3(P)\delta^6(Q)$$
$$\epsilon_{\alpha \beta} \lambda_i^\alpha\lambda_j^{\beta}=\langle ij\rangle$$
$$p_i\cdot p_j=-\frac{1}{2}\langle ij\rangle^2$$
 \begin{equation}
q_l^{A \alpha}\lambda_{j\alpha}=\eta_l^{A}\langle lj\rangle,
 \label{blocks}
\end{equation}
where $\epsilon_{\alpha \beta}$ and $\epsilon^{\alpha \beta}$ are antisymmetric and $\epsilon_{12}=\epsilon^{21}=1$.

Given these building blocks, it is not possible to construct  a dilatation-invariant amplitude with an odd number of external legs. This can be seen by noting that the first term in the dilatation operator ($d$) in eq. \ref{confalg} counts the mass dimension while the second term gives a constant factor of $n/2$, where $n$ is the number of external legs. When the number of external legs is odd, it is not possible to cancel the constant term because all of the building blocks have integer mass dimension. It follows that all odd-point on-shell tree-level amplitudes must vanish. This can also be understood from the Lagrangian since any off-shell amplitude with an odd number of external legs will have at least one gauge field as an external leg. Since the gauge fields have no propagating degrees of freedom, all on-shell amplitudes with an odd number of external legs vanish.

The first nontrivial amplitude is the four-point color-ordered superamplitude, which reads
 \eq
 \mathcal{A}_4(1,2,3,4)=\frac{\delta^3(P)\delta^6(Q)}{\langle12\rangle\langle41\rangle}.
\label{4pt}
 \eqe
We describe color-ordering in the ABJM theory in Appendix A. At four-point, the spinor inner products have the following relationships:
\eq
\frac{\langle21\rangle}{\langle34\rangle}=\frac{\langle23\rangle}{\langle14\rangle}=\frac{\langle13\rangle}{\langle42\rangle}=\pm1.
\label{4ptrelat}
\eqe
These relations follow from momentum conservation. Note that the superamplitude encodes the scattering amplitudes of all possible component fields. The component amplitudes can be read off from the superamplitude as the coefficients of monomials of the $\eta$ coordinates. By expanding the four-point superamplitude in the $\eta$ coordinates, one can match the component amplitudes with color-ordered four-point amplitudes computed using Feynman diagrams. 

For example, let's consider the scattering of four scalar fields $\phi_{4},\phi^{4},\phi_{4},\phi^{4}$.
If legs 1 and 3 correspond to $\phi_{4}$ and legs 2 and 4 correspond
to $\phi^{4}$, then this scattering amplitude should be the coefficient
of $\eta_{1}^{1}\eta_{1}^{2}\eta_{1}^{3}\eta_{3}^{1}\eta_{3}^{2}\eta_{3}^{3}$
in the $\eta$-expansion of the superamplitude in eq. \ref{4pt}. This can be
understood by looking at the superfield expansion in eq. \ref{sfield}, from which
we see that the coefficient of $\phi^{4}$ is 1 and the coefficient
of $\phi_{4}$ is $\eta^{1}\eta^{2}\eta^{3}$. Since all the fermionic
coordinates are contained in the supermomentum delta function, this
coefficient can be extracted from the following integral:\[
\int d\eta_{2}^{1}d\eta_{2}^{2}d\eta_{2}^{3}d\eta_{4}^{1}d\eta_{4}^{2}d\eta_{4}^{3}\delta^{6}(Q)=\left\langle 13\right\rangle ^{3}{\normalcolor .}\]
Hence, we find that the color-ordered four-point scalar amplitude
is given by
\begin{equation}
\mathcal{A}_{\phi\phi\phi\phi}\propto\frac{\left\langle 13\right\rangle ^{3}}{\left\langle 12\right\rangle \left\langle 14\right\rangle }\label{eq:4sca}
\end{equation}
where we are leaving out the momentum-conserving delta function.

It is not difficult to verify this result using Feynman diagrams.
The Feynman diagrams contributing the color-ordered amplitude are
illustrated in Fig. \ref{4ptdiag}. Using the color-ordered Feynman rules described
in appendix A, one finds that the amplitude is given by
\[
\mathcal{A}_{\phi\phi\phi\phi}\propto\frac{\epsilon_{\mu\nu\lambda}\left(p_{1}-p_{2}\right)^{\mu}\left(p_{3}-p_{4}\right)^{\nu}\left(p_{1}+p_{2}\right)^{\lambda}}{\left(p_{1}+p_{2}\right)^{2}}+\frac{\epsilon_{\mu\nu\lambda}\left(p_{1}-p_{4}\right)^{\mu}\left(p_{3}-p_{2}\right)^{\nu}\left(p_{1}+p_{4}\right)^{\lambda}}{\left(p_{1}+p_{4}\right)^{2}}{\normalcolor .}\]
To compare this with eq. \ref{eq:4sca}, we must write it in terms
of the spinor variables of the on-shell superspace. Note that the
numerator in the first term can be written as follows: \[
\epsilon_{\mu\nu\lambda}\left(p_{1}-p_{2}\right)^{\mu}\left(p_{3}-p_{4}\right)^{\nu}\left(p_{1}+p_{2}\right)^{\lambda}=4\epsilon_{\mu\nu\lambda}p_{1}^{\mu}p_{2}^{\lambda}p_{3}^{\nu}=-4\epsilon_{\mu\nu\lambda}p_{1}^{\mu}p_{2}^{\lambda}p_{4}^{\nu}{\normalcolor .}\]
From this, we see that the numerator is totally antisymmetric under
exchange of the four external particle labels. Furthermore, it is
not difficult to see that the numerator in the second term is equal
to the numerator in the first term. It follows that the numerators
are equal to $\left\langle 12\right\rangle \left\langle 23\right\rangle \left\langle 31\right\rangle $
(up to a numerical factor), which is totally antisymmetric. For example, exchanging $1$ with
$3$ gives $\left\langle 32\right\rangle \left\langle 21\right\rangle \left\langle 13\right\rangle =\left(-\left\langle 23\right\rangle \right)\left(-\left\langle 12\right\rangle \right)\left(-\left\langle 31\right\rangle \right)=-\left\langle 12\right\rangle \left\langle 23\right\rangle \left\langle 31\right\rangle $.
Hence, we find that
\[
\mathcal{A}_{\phi\phi\phi\phi}\propto\left(\frac{1}{s}+\frac{1}{t}\right)\left\langle 12\right\rangle \left\langle 23\right\rangle \left\langle 31\right\rangle =-\frac{u}{st}\left\langle 12\right\rangle \left\langle 23\right\rangle \left\langle 31\right\rangle \]
where $s,t,u$ are the usual Mandelstam variables, which satisfy $s+t+u=0$.
From eq. \ref{blocks}, we see that $s=-\left\langle 12\right\rangle ^{2}$, $t=-\left\langle 14\right\rangle ^{2}$,
$u=-\left\langle 13\right\rangle ^{2}$. If we plug these relations
into the equation above, we obtain eq. \ref{eq:4sca}. This confirms
that the scalar component of the four-point superamplitude in eq. \ref{4pt}
matches the four-point scalar amplitude of the ABJM theory. Since
all the other four-point amplitudes are related by supersymmetry,
this implies that eq. \ref{4pt} is in fact the four-point superamplitude of
the ABJM theory.
\begin{figure}
\begin{center}
\includegraphics[scale=0.25]{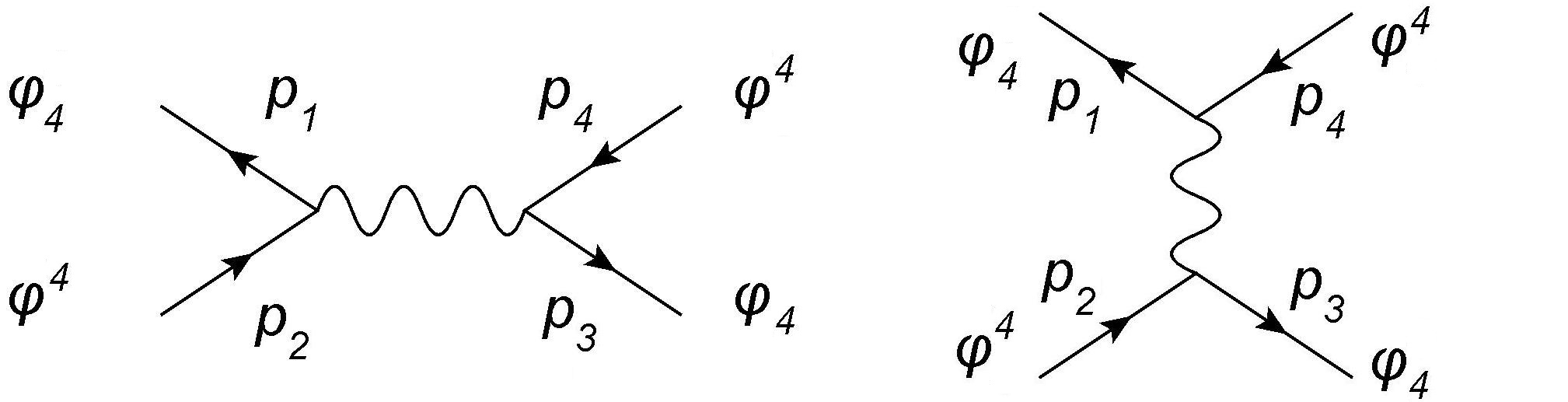}
\caption{Feynman diagrams contributing to the four-point color-ordered scalar amplitude. We take all mometa to be outgoing.
 }
\label{4ptdiag}
\end{center}
\end{figure}

Let us conclude this subsection with some general comments about the superamplitudes of the ABJM theory. It is customary to introduce a collective notation $ \Lambda = ( \lambda , \eta) $. Following the convention of ref.~\cite{Lee:2010du}, the superamplitudes can be written as functions of  $\L_{i}$, where the index $i$ labels the external particles and $\L_{\rm odd/even}$ is associated with the $\bar{\Phi}$/$\Phi$ multiplet. The superamplitudes are invariant up to a sign under $\Lambda_i \rightarrow -\Lambda_i$, which we refer to as an orientation reversal:
\be
{ \cal A}_n ( \Lambda_1, \cdots, - \Lambda_i , \cdots, \Lambda_n )  = (-1)^{i} { \cal A}_n ( \Lambda_1, \ldots , \Lambda_i , \cdots, \Lambda_n )  \,.
\label{super-l-parity}
\ee
Furthermore, since the $\Phi / \bar{\Phi}$ multiplets are bosonic/fermionic, the superamplitudes are invariant up to a sign after cyclic permutation by two sites \cite{Bargheer:2010hn}:
\be
\mathcal{A}_{2k}\left(\Lambda_{3},...,\Lambda_{2k},\Lambda_{1},\Lambda_{2}\right)=\left(-1\right)^{k-1}\mathcal{A}_{2k}\left(\Lambda_{1},...,\Lambda_{2k}\right).
\label{cycle}
\ee

\section{Dual Superconformal Symmetry}

In this section, we will construct a dual superspace and use it to define dual superconformal symmetry in the ABJM theory. We will find that the dual superspace contains three bosonic coordinates $x^{\alpha \beta}=x^{\beta \alpha}$, six fermionic coordinates $\theta^{A\alpha}$, and three Grassmann-even coordinates $y^{AB}=-y^{BA}$. The Grassmann-even coordinates are a new feature of the ABJM theory and arise because the superspace is non-chiral. We will then show that the four-point amplitude has dual superconformal symmetry. In section 3.5, we will explain how to extend dual superconformal symmetry beyond the four-point superamplitude.
\subsection{Dual Superspace}

In order to define dual superconformal symmetry, we must first define a dual space. Since the ABJM theory lives in 3D Minkowski space and has $\mathcal{N}=6$ supersymmetry, we will assume that the dual space has three bosonic coordinates and six fermionic coordinates. These coordinates are related to the on-shell superspace as follows:
\eqa
\nonumber x_i^{\alpha\beta}-x_{i+1}^{\alpha\beta}=p^{\alpha\beta}_i=\lambda^\alpha_i\lambda^\beta_i\\
\theta^{A\alpha}_i-\theta^{A\alpha}_{i+1}=q_i^{\alpha A}=\lambda_i^\alpha\eta_i^A,
\label{hyper}
\eqae
where $x_{n+1}= x_1,\;\theta_{n+1}= \theta_1$. In words, the displacement between points in the dual superspace correspond to the external supermomenta of the scattering amplitudes. In these new coordinates, supermomentum conservation is automatically satisfied. Note that $(x,\theta)$ should not be identified with the usual (super)space-time, since they would have incorrect mass dimensions. Eqs. \ref{hyper} define a hyperplane in the full space $(x,\theta,\lambda,\eta)$. The amplitudes have support on this hyperplane. One can translate from the dual coordinates back to the on-shell space via
\eqa
\nonumber&&x_i^{\alpha\beta}=x_1^{\alpha\beta}-\sum^{i-1}_{k=1}\lambda^\alpha_k\lambda^\beta_k,\\
&&\theta^{A\alpha}_i=\theta^{A\alpha}_1-\sum^{i-1}_{k=1}\lambda^\alpha_k\eta_k^A.
\label{translation}
\eqae
Note that $x_1^{\alpha\beta}$ and $\theta^{A\alpha}_1$ parameterize the ambiguity that arises from the fact that eqs. \ref{hyper} are invariant under a constant shift in the dual coordinates. Furthermore, the hyperplane equations lead to the following relationships:
\eqa
\nonumber &&(x_{i,i+1})^{\alpha\beta}\lambda_{i\beta}=0,\quad \lambda_i^\alpha=\frac{(x_{i,i+1})^{\alpha\beta}\lambda_{i+1\beta}
}{\sqrt{-x^2_{i,i+2}}},\\
&&\theta^{A\alpha}_{i,i+1}\lambda_{i\alpha}=0,\quad\quad \eta_i^A=\;\frac{\theta^{A\alpha}_{i,i+1}\lambda_{i+1\alpha}}{\sqrt{-x^2_{i,i+2}}},
\label{trans}
\eqae
where $x_{ij}=x_i-x_j$ and $\theta_{ij}=\theta_i -\theta_j$. In obtaining these relationships, we noted that
\begin{equation}
{\langle i,i+1\rangle}^2=-x_{i,i+2}^2
\label{i2x}
\end{equation}
which is easy to prove using eqs.~\ref{hyper} and \ref{blocks}. Given the $x$'s and $\theta$'s, one can obtain all the $\lambda$'s and $\eta$'s. In particular, after fixing $x_1$, the $\lambda$ coordinates can be determined using the first relation in eq.~\ref{hyper}. After solving for the $\lambda$ coordinates, the $\eta$ coordinates can then be determined using the last relation in eq.~\ref{trans}.

Having defined a dual superspace, let's try to construct $\mathcal{N}=6$ dual supersymmetry generators. By analogy with the definition of dual supersymmetry in $\mathcal{N}=4$ sYM, there is a simple ansatz for the dual supersymmetry generators in the ABJM theory:
\eqa
\nonumber Q_{A\alpha}&=&\sum^n_{i=1}\frac{\partial}{\partial \theta_i^{A\alpha}},\\
\nonumber Q^A_{\alpha}&=&\sum^n_{i=1}\left( \theta_i^{A\beta}\frac{\partial}{\partial x_i^{\alpha\beta}}+\frac{1}{2}\eta_i^A\frac{\partial}{\partial\lambda_i^\alpha} \right).\\
\label{firstdef}
\eqae
If we follow what was done for $\mathcal{N}=4$ super Yang-Mills, however, we immediately encounter a difficulty: half of the supercharges are inconsistent with the constraints in eq.~\ref{hyper}. In particular, the second supercharge violates the $\theta$-space constraint in eq.~\ref{hyper}:
\eq
 Q^A_{\alpha} (\theta_i-\theta_{i+1})^{B\beta} \neq  Q^A_{\alpha}(\lambda_i^\beta\eta_i^B).
\eqe
With a little thought, one can see that there are no terms which can be added to $Q^A_{\alpha}$ to cancel this ``anomaly.'' 

We can solve this problem by introducing three Grassmann-even coordinates, $y^{AB}=-y^{BA}$, which are related to the on-shell twistor space as follows:
\eq
y_i^{AB}-y_{i+1}^{AB}=\eta_i^A\eta_i^B.
\label{seconddef}
\eqe
In addition to introducing new coordinates, we alter the second supercharge as follows:
\eq
Q^{*A}_{\alpha}=\sum^n_{i=1} \left(\theta_i^{A\beta}\frac{\partial}{\partial x_i^{\alpha\beta}}+\frac{1}{2}\eta_i^A\frac{\partial}{\partial\lambda_i^\alpha}+\frac{1}{2}y_i^{AB}\frac{\partial}{\partial \theta^{B\alpha}_i} \right).
\eqe
Now it is straightforward to see that the hyperplane constraints are preserved
 \eq
 Q^{*A}_{\alpha} (\theta_i-\theta_{i+1})^{B\beta}=Q^{*A}_{\alpha}(\lambda_i^\beta\eta_i^B)=\frac{1}{2}\delta_\alpha^\beta\eta_i^A\eta_i^B.
\eqe
Furthermore, the $y$-space constraint is also preserved, so no additional coordinates are needed.

\begin{figure}
\begin{center}
\includegraphics[scale=0.52]{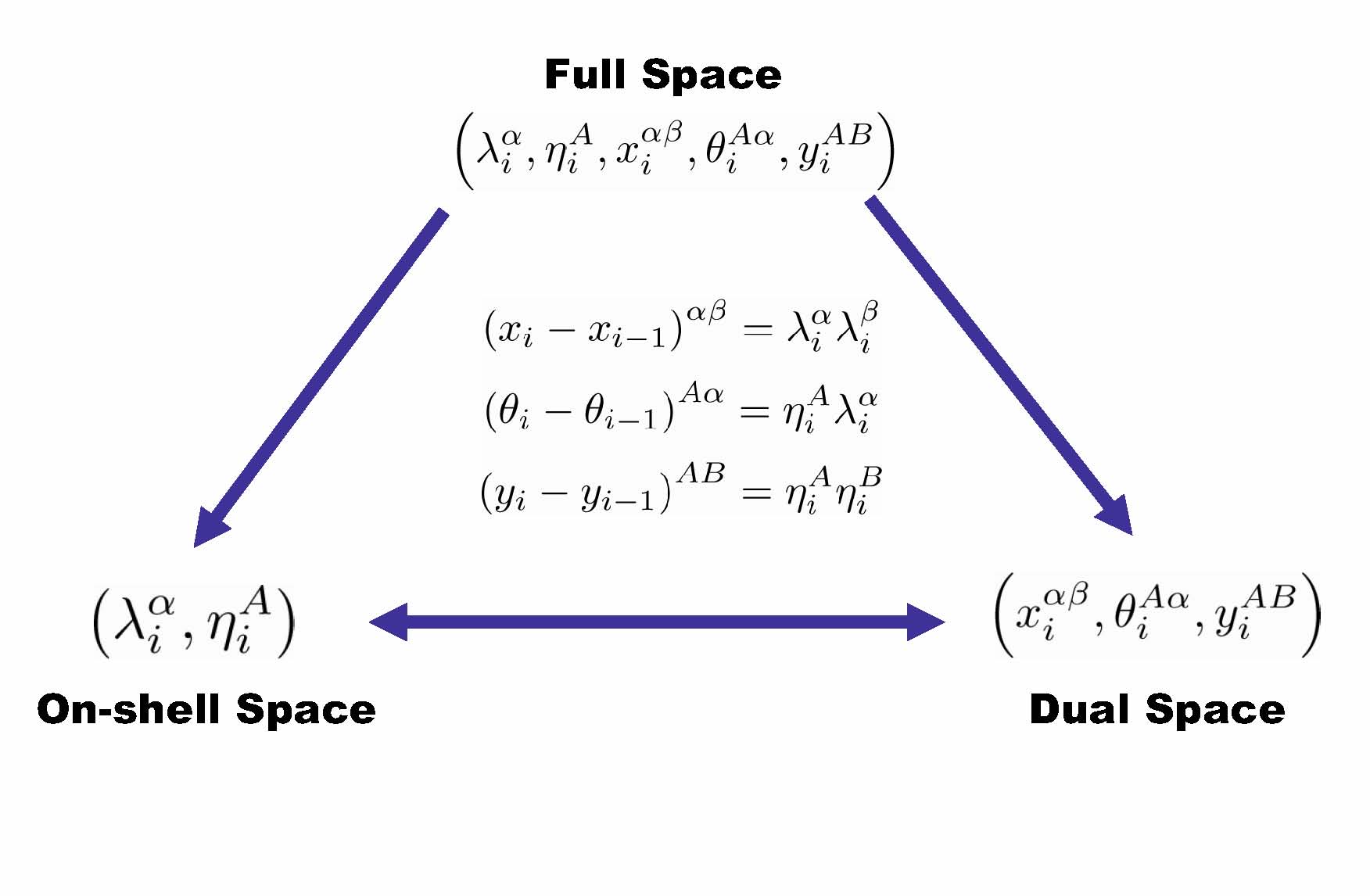}
\caption{Summary of the superspace for $\mathcal{N}=6$ Chern-Simons theory. The full superspace can reduced to the on-shell superspace or the dual superspace using the three hyperplane constraints.}
\label{dualspace}
\end{center}
\end{figure}

In summary, the dual superspace contains three bosonic coordinates, six fermionic coordinates, and three Grassmann-even coordinates. The coordinates of the dual superspace are related to the coordinates of the on-shell superspace via eqs. \ref{hyper}, which can be viewed as defining hyperplanes in the full superspace. The $\mathcal{N}=6$ superspace is summarized in Fig. \ref{dualspace}.

Since the R-symmetry of the field theory corresponds to the isometries of $CP^3$, eq.~\ref{seconddef} suggests that the $y^{AB}$ coordinates should be associated with three commuting Killing vectors in $CP^3$. The fact that type IIB string theory on $AdS_5 \times S^5$ is self-dual after T-dualizing directions corresponding to the dual superspace of $\mathcal{N}=4$ super Yang-Mills suggests that type IIA string theory on $AdS_4 \times CP^3$ should be self-dual if one performs T-dualities along the translational directions of $AdS_4$, three-directions in $CP^3$, and six fermionic directions. Although several attempts have been made to demonstrate this, none have succeeded.

\subsection{Dual Inversion Properties}

In this subsection, we will deduce the transformation properties of the $\mathcal{N}=6$ superspace under dual conformal transformations. The dual translations and dual inversion of the $x$ coordinates are defined in the usual way:
\eq
P_{\alpha\beta}=\sum^n_{i=1}\frac{\partial}{\partial x_i^{\alpha\beta}},\;I[x_i^{\alpha\beta}]=\frac{x_i^{\alpha\beta}}{x_i^2},
\label{dconf}
\eqe
and dual inversion of the $\theta$ coordinates can be defined as follows:
\eq
I[\theta_i^{A\alpha}]=\frac{x_i^{\alpha\beta}\theta^A_{i\beta}}{x_i^2}.
\label{delta}
\eqe

The dual inversion properties of the on-shell superspace can be deduced by requiring compatibility of the above transformation rules with eq. \ref{trans}. From the first line of eq.~\ref{trans}, one can deduce that
\eq
I[(x_{i,i+1})^{\alpha\beta}\lambda_{i\beta}]=0\Longrightarrow I[\lambda_{i\beta}]=\alpha_i(x_{i+1})_{\beta\gamma}\lambda_i^\gamma,
\eqe
where $\alpha_i$ is a proportionality constant. This constant can be fixed through the compatibility of the second relationship in the first line of eq.~\ref{trans} with inversion:
\eq
I[\lambda^\alpha_i]=I\left[\frac{(x_{i,i+1})^{\alpha\beta}\lambda_{i+1\beta}}{\langle i,i+1\rangle}\right]\Longrightarrow \alpha^2_{i}=\frac{1}{(x_{i+1})^2(x_i)^2}.
\eqe
Thus we arrive at
\eq
 I[\lambda_{i\beta}]=\frac{(x_{i})_{\beta\gamma}\lambda_i^\gamma}{\sqrt{(x_{i+1})^2(x_i)^2}}.
\label{lambda}
\eqe
Similarly, for the fermionic variable $\eta$, one has:
\eq
I[\eta_i^A]=I\left[\frac{\theta^{A\alpha}_{i,i+1}\lambda_{i+1\alpha}}{\langle i,i+1\rangle}\right]\Longrightarrow I[\eta_i^A]=-\sqrt{\frac{x_i^2}{x^2_{i+1}}}\left[\eta_i^A+ \frac{x_i^{\alpha\beta}\theta_{i\beta}^A\lambda_{i\alpha}}{x_i^2}\right].
\label{eta}
\eqe
By applying this formula to the hyperplane constraint in eq. \ref{seconddef}, we can deduce the dual inversion formula for the $y$ coordinates:
\eq
I\left[y_{i}^{AB}\right]=\frac{\theta_{i}^{A\alpha}\theta_{i}^{B\beta}x_{i\alpha\beta}}{x_{i}^{2}}+y_i^{AB}.
\eqe

\subsection{Dual Conformal Boost Generator}
In this subsection, we will prove that the four-point amplitude has dual superconformal symmetry. We will then derive an expression for the dual conformal boost generator, which will be useful for demonstrating that the amplitudes have Yangian symmetry.

We will first show that the four-point tree-level superamplitude of the ABJM theory is dual conformal covariant. This is easiest to demonstrate if we write the four-point amplitude in terms of the ($x,\theta$) coordinates of the dual superspace. Note that the $y$ coordinates are auxiliary in the sense that given $x$ and $\theta$, the $y$'s can be determined using the hyperplane constraints in eqs. \ref{hyper} and \ref{seconddef}. As a result, it is possible to express the amplitudes in terms of $(x,\theta)$.

Prior to the identification $x_{n+1}= x_1,\;\theta_{n+1}= \theta_1$ for an $n$-point amplitude, the hyperplane constraints imply that
$$\sum^n_{i=1}\lambda^\alpha_i\lambda^\beta_i=x^{\alpha\beta}_1-x^{\alpha\beta}_{n+1},\;\;\;\sum^n_{i=1}\lambda^\alpha_i\eta^A_i=\theta^{\alpha A}_1-\theta^{\alpha A}_{n+1}.$$
In the dual space, the (super)momentum delta functions are therefore given by
\eq
\delta^3 \left(  \sum_{i=1}^n p_i  \right)=\delta^3(x_1-x_{n+1}),\quad\delta^6 \left( \sum_{i=1}^4\eta^A_{i}\lambda^\alpha_i  \right)=\delta^6(\theta_1-\theta_{n+1} ).
\eqe
The inversion properties of the delta functions follow from the definitions $\int d^3x_1\delta^3(x_1-x_5)=1$ and $I[\int d^3x_1]=\int\frac{d^3x_1}{x^6_1}$. In particular,
\eq
I[\delta^3(x_1-x_5)]=x_1^6\delta^3(x_1-x_{n+1}),\quad I[\delta^6(\theta_1-\theta_5)]=x_1^{-6}\delta^6(\theta_1-\theta_{n+1}),
\eqe
where the inversion property of the fermionic delta function is derived from eq.~\ref{delta} on the support of $\delta^3(x_1-x_{n+1})$. Note that in three dimensions, it is only for $\mathcal{N}$=6 dual supersymmetry that the momentum and supermomentum delta functions combine to give an invariant under dual inversion.

Using eq. \ref{i2x}, we therefore find that the four-point amplitude in eq. \ref{4pt} can be written as follows:
\eq
A^{ABJM}_{4}=\frac{\delta^3(x_1-x_5)\delta^6(\theta_1-\theta_5)}{\sqrt{x_{1,3}^2 x_{2,4}^2}}.
\eqe
Furthermore, using eq. \ref{dconf}, we find that the four-point amplitude transforms as follows under dual inversion:
\eq
I[A^{ABJM}_{4}]=\sqrt{x^2_1x^2_2x^2_3x^2_4}A^{ABJM}_4.
\eqe
From this, we see that the four-point tree amplitude transforms covariantly under a dual conformal boost:
\eq
K^{\alpha\beta}A^{ABJM}_{4}=IP^{\alpha\beta}IA^{ABJM}_{4}=-\frac{1}{2}\sum^4_{i=1}x^{\alpha\beta}_iA^{ABJM}_{4},
\eqe
where the dual conformal boost is defined in the usual way, notably a dual inversion-translation-inversion.  Therefore, under the redefined generator
\eq
\tilde{K}^{\alpha\beta}=K^{\alpha\beta}+\frac{1}{2} \left(\sum^n_{i=1} x^{\alpha\beta}_i \right),
\label{kdef}
\eqe
we have $\tilde{K}^{\alpha\beta}A_4=0$.

Note that the dual supersymmetry generators either vanish or reduce to ordinary superconformal symmetry generators when restricted to the on-shell space. Hence, if the dual conformal boost generator is defined according to eq. \ref{kdef}, this implies that the four-point superamplitude has $OSp(6|4)$ dual superconformal symmetry (since all the other dual superconformal generators can be obtained by commuting the dual superconformal boost generator with the dual supersymmetry generators). More generally, if the $n$-point amplitude of the ABJM theory transforms covariantly under dual inversion,
\eq
I[A_n]=\sqrt{x_1^2...x_n^2}A_n
\eqe
this implies that all the tree-level amplitudes of the ABJM theory have dual superconformal symmetry. We will prove this in section 3.5.

Let's derive an explicit expression for the dual conformal boost generator in the full $\mathcal{N}=6$ superspace. In order to do so, let's first compute how the dual coordinates $x$ and $\theta$ transform under a dual inversion-translation-inversion. For the $x$ coordinates, we have
\eqa
\nonumber IP_{\gamma\delta}I \left[x_i^{\alpha\beta} \right] &=&\sum^n_{j=1}I\frac{\partial}{\partial x_j^{\gamma\delta}}\frac{x_i^{\alpha\beta}}{x_i^2}\\
\nonumber&=&I\left[\frac{1}{2}\frac{\delta^{(\alpha}_\gamma\delta^{\beta)}_\delta}{x_i^2}+\frac{x_{i\gamma\delta}x_i^{\alpha\beta}}{x^4_i}\right]\\
\nonumber&=&\frac{1}{2}x_i^2\delta^{(\alpha}_\gamma\delta^{\beta)}_\delta+x_{i\gamma\delta}x_i^{\alpha\beta},
\eqae
where $A^{(\alpha\beta)}\equiv A^{\alpha\beta}+A^{\beta\alpha}$. Using $x_{i\gamma\delta}x_i^{\alpha\beta}=\frac{1}{2}x_{i\gamma}\,^{(\alpha}x_{i\delta}\,^{\beta)}-\frac{1}{2}x_i^2\delta^{(\alpha}_\gamma\delta^{\beta)}_\delta$, we see that
\eq
IP_{\gamma\delta}I \left[x_i^{\alpha\beta} \right] = \frac{1}{2}x_{i\gamma}\,^{(\alpha}x_{i\delta}\,^{\beta)}.
\eqe
Similarly, for the $\theta$ coordinates we have
\eqa
\nonumber
IP_{\gamma\delta}I \left[\theta_{i}^{A\alpha} \right] &=&I\left[\sum_{j=1}^{n}\frac{\partial}{\partial x_{j}^{\gamma\delta}}\frac{x_{i}^{\alpha\beta}}{x_{i}^{2}}\theta_{i\beta}^{A}\right]
\\
\nonumber &=&I\left[\left(\frac{1}{2}\delta_{(\gamma}^{\alpha}\delta_{\delta)}^{\beta}\frac{1}{x_{i}^{2}}+\frac{x_{i\gamma\delta}x_{i}^{\alpha\beta}}{x_{i}^{4}}\right)\theta_{i\beta}^{A}\right]
\\
\nonumber &=&-\frac{1}{2}\delta_{(\gamma}^{\alpha}x_{i\delta)\omega}\theta_{i}^{A\omega}+x_{i\gamma\delta}\theta_{i}^{A\alpha}.
\eqae
Using $\frac{1}{2}x_{i\,\,\,(\gamma}^{\alpha}\theta_{i\delta)}^{A}=-\frac{1}{2}\delta_{(\gamma}^{\alpha}x_{i\delta)\omega}\theta_{i}^{A\omega}+x_{i\gamma\delta}\theta_{i}^{A\alpha}$
we see that
\eq
IP_{\gamma\delta}I \left[ \theta_{i}^{A\alpha} \right]=\frac{1}{2}x_{i\,\,\,(\gamma}^{\alpha}\theta_{i\delta)}^{A}.
\eqe

Hence, when acting in $(x,\theta)$ space, the dual conformal boost generator is given by
$$
K^{\alpha\beta}=\sum^n_{i=1} x_i^{\alpha\gamma}x_i^{\beta\delta}\frac{\partial}{\partial x_i^{\gamma\delta}}+\frac{1}{2}x^{\gamma(\alpha}_{i}\theta^{A\beta)}_i\frac{\partial}{\partial \theta^{A\gamma}_i}+\frac{1}{2}\sum^n_{i=1} x^{\alpha\beta}_i$$
where the first term generates dual conformal boosts of $x$, the second term generates dual conformal boosts of $\theta$, and the third term follows from eq. \ref{kdef}. We can extend this definition to the on-shell superspace by adding terms so that it commutes with the hyperplane constraints in eqs. \ref{hyper} and \ref{seconddef} modulo constraints \cite{Drummond:2009fd}.
Doing so gives
\eqa
\nonumber K^{\alpha\beta}&=&\sum^n_{i=1} x_i^{\alpha\gamma}x_i^{\beta\delta}\frac{\partial}{\partial x^{\gamma\delta}}+\frac{1}{2}x^{\gamma(\alpha}_{i}\theta^{A\beta)}_i\frac{\partial}{\partial \theta^{A\gamma}_i}+\frac{1}{4}\left(x^{\gamma(\alpha}_{i}\lambda^{\beta)}_i\frac{\partial}{\partial \lambda^\gamma_i}+x^{\gamma(\alpha}_{i+1}\lambda^{\beta)}_i\frac{\partial}{\partial \lambda^\gamma_i}\right)\\
&&+\frac{1}{4}\left(\theta^{B(\alpha}_{i}\lambda^{\beta)}_i\frac{\partial}{\partial \eta^B_i}+\theta^{B(\alpha}_{i+1}\lambda^{\beta)}_i\frac{\partial}{\partial \eta^B_i}\right)+\frac{1}{2}\sum_{i=1}^{n}\theta_i^{A(\alpha}\theta_i^{B\beta)}\frac{\partial}{\partial y_i^{AB}}\\
&&+\frac{1}{2}\sum^n_{i=1} x^{\alpha\beta}_i.
\label{k}
\eqae
Indeed, one can check that eq. \ref{k} preserves eqs. \ref{hyper} and \ref{seconddef}. Alternatively, we can deduce the action of the dual conformal boost generator on all the coordinates of the $\mathcal{N}=6$ superspace by using the dual inversion properties derived in section 3.2.2.

\section{Yangian Symmetry}

In \cite{Bargheer:2010hn}, the authors showed that the four- and six-point tree-level superamplitudes of the ABJM theory have Yangian symmetry. The Yangian algebra is generated by a set of level-zero and level-one generators $J^{(0)a}$ and $J^{(1)a}$ satisfying
\eq
[J^{(0)}_a,\;J^{(0)}_b\}=f_{ab}\,^cJ^{(0)}_c,\;\;[J^{(1)}_a,\;J^{(0)}_b\}=f_{ab}\,^cJ^{(1)}_c,
\label{Yangian}
\eqe
where $f_{ab}\,^c$ are the structure of the superconformal group. The indices can be raised and lowered using the metric of the superconformal group.  For $OSp(6|4)$, these are computed in Appendix F of \cite{Bargheer:2010hn}. The level-zero generators are identified with the superconformal generators and the level-one generators are given by a sum of bi-local products of single-site superconformal generators \cite{Dolan:2003uh, Dolan:2004ps}:
\eq
J^{(1)}_a=f_a\,^{bc}\sum_{1\leq i<j\leq n}J^{(0)}_{ib}J^{(0)}_{jc}.
\eqe
The commutation relations in eq. \ref{Yangian} lead to two sets of Jacobi identities. There is a third set of deformed Jacobi identities which involve two level-1 generators, which are known as the Serre relations.

In $\mathcal{N}=4$ sYM, Yangian symmetry of the tree-level amplitudes follows from combining superconformal symmetry with dual superconformal symmetry. In this subsection, we will demonstrate that this is also true for the ABJM theory. In particular,  we will demonstrate that the dual conformal boost generator is equivalent to a level-one Yangian generator when acting on on-shell amplitudes. Since all the other level-1 Yangian generators can be obtained from the algebra in eq. \ref {Yangian}, it follows that dual superconformal symmetry plus ordinary superconformal symmetry implies Yangian symmetry of the on-shell amplitudes.

Since the amplitudes can be written purely in terms of the on-shell superspace variables $(\lambda,\eta)$, we only consider the part of the dual conformal boost generator which acts on this space, which is given by
\[
K^{\alpha\beta}|_{os}=  \left[\sum^n_{i=1} \frac{1}{4}\left(x^{\gamma(\alpha}_{i}\lambda^{\beta)}_i\frac{\partial}{\partial \lambda^\gamma_i}+x^{\gamma(\alpha}_{i+1}\lambda^{\beta)}_i\frac{\partial}{\partial \lambda^\gamma_i}\right)+\frac{1}{4}\left(\theta^{B(\alpha}_{i}\lambda^{\beta)}_i\frac{\partial}{\partial \eta^B_i}+\theta^{B(\alpha}_{i+1}\lambda^{\beta)}_i\frac{\partial}{\partial \eta^B_i}\right)+\frac{1}{2}x^{\alpha\beta}_i\right],
\]
where $|_{os}$ means that we are restricting to the on-shell space. Next, we trade $x$ and $\theta$ for $\lambda$ and $\eta$ using eq. \ref{translation}. In this way, all of the dual coordinates can be replaced except $x_1$ and $\theta_1$, however the terms containing these variables take the form
\eqa
&&\sum^n_{i=1}\frac{1}{2}x^{\gamma(\alpha}_{1}\lambda^{\beta)}_i\frac{\partial}{\partial \lambda^\gamma_i}+\frac{1}{2}\theta^{B(\alpha}_{1}\lambda^{\beta)}_i\frac{\partial}{\partial \eta^B_i}+\frac{n}{2}x^{\alpha\beta}_1\\
\nonumber&=&x^{\gamma(\alpha}_{1}\frac{1}{2}\left(m^{\beta)}\,_\gamma+\delta^{\beta)}\,_\gamma d\right)+\frac{1}{2}\theta^{B(\alpha}_{1}q^{\beta)}_B.
\eqae
Since $m,q,d$ are the usual Lorentz, supersymmetry, and dilatation generators (under which the amplitudes are invariant), these terms vanish on the amplitudes. The remaining terms, which we denote as $\tilde{K}^{'\alpha\beta}$, are
\eqa
\nonumber\tilde{K}^{'\alpha\beta}&=&-\sum^n_{i=1}\left[ \frac{1}{4}\left(\sum^{i-1}_{k=1}\lambda_k^{\gamma}\lambda_{k}^{(\alpha}\lambda_i^{\beta)}\frac{\partial}{\partial \lambda^\gamma_i}+\sum^{i}_{k=1}\lambda_k^{\gamma}\lambda_{k}^{(\alpha}\lambda_i^{\beta)}\frac{\partial}{\partial \lambda^\gamma_i}\right)\right.\\
\nonumber&&\left.+\frac{1}{4}\left(\sum^{i-1}_{k=1}\lambda_k^{(\alpha}\eta^B_k\lambda^{\beta)}_i\frac{\partial}{\partial \eta^B_i}+\sum^{i}_{k=1}\lambda_k^{(\alpha}\eta^B_k\lambda^{\beta)}_i\frac{\partial}{\partial \eta^B_i}\right)+\frac{1}{2}\sum^{i-1}_{k=1}\lambda_k^{\alpha}\lambda^{\beta}_k\right].\\
\eqae
In order to relate this object to a level-one generator, we will write it in terms of the ordinary superconformal generators given in eq. \ref{confalg}:

\[
\tilde{K}^{'\alpha\beta}=-\sum_{i=1}^{n}\left[\frac{1}{2}\sum_{k=1}^{i-1}p_{k}^{\gamma(\alpha}(m_{i}^{\beta)}\,_{\gamma}+\delta^{\beta)}\,_{\gamma}(d_{i}-1/2))+\frac{1}{4}p_{i}^{\gamma(\alpha}(m_{i}^{\beta)}\,_{\gamma}+\delta^{\beta)}\,_{\gamma}(d_{i}-1/2))\right.\]
\[
\left.+\frac{1}{2}\sum_{k=1}^{i-1}q_{k}^{B(\alpha}q_{iB}^{\beta)}+\frac{1}{4}q_{i}^{B(\alpha}q_{iB}^{\beta)}+\frac{1}{2}\sum_{k=1}^{i-1}p_{k}^{\alpha\beta}\right]\]

\[
=-\frac{1}{2}\sum_{k<i}^{n}\left[p_{k}^{\gamma(\alpha}(m_{i}^{\beta)}\,_{\gamma}+\delta^{\beta)}\,_{\gamma}d_{i})+q_{k}^{B(\alpha}q_{iB}^{\beta)}\right]\]
\[
-\frac{1}{4}\sum_{i=1}^{n}\left[p_{i}^{\gamma(\alpha}(m_{i}^{\beta)}\,_{\gamma}+\delta^{\beta)}\,_{\gamma}d_{i})+q_{i}^{B(\alpha}q_{iB}^{\beta)}\right]+\frac{1}{4}p^{\alpha\beta}.\]

At this point, it's convenient to add the following term (which vanishes on amplitudes):
\eq
\Delta\tilde{K}^{'\alpha\beta}=\frac{1}{4}\sum_{k=1}^n\left[p_k^{\gamma(\alpha}(m^{\beta)}\,_{\gamma}+\delta^{\beta)}\,_{\gamma}d)+q_k^{B(\alpha }q^{\beta)}_{B}\right]-\frac{1}{4}p^{\alpha\beta}.
\eqe
We finally arrive at
\eqa
&&\tilde{K}^{'\alpha\beta}+\Delta\tilde{K}^{'\alpha\beta}=-\frac{1}{4}\sum^n_{k<i}\left[ p_k^{\gamma(\alpha}(m_i^{\beta)}\,_{\gamma}+\delta^{\beta)}\,_{\gamma}d_i)+q_k^{B(\alpha }q^{\beta)}_{iB}-(i\leftrightarrow k)\right]\\
\nonumber&=&-\frac{1}{4}\sum^n_{k<i}\left[ (m_i^{(\alpha}\,_{\gamma}+\delta^{(\alpha}\,_{\gamma}d_i)p_k^{\gamma\beta)}-q_i^{B(\alpha }q^{\beta)}_{kB}-(i\leftrightarrow k)\right],
\eqae
which is indeed the level-one generator $J^{(1)\alpha\beta}$ obtained in \cite{Bargheer:2010hn}.

While the dual supersymmetry generators either vanish or reduce to ordinary superconformal generators when restricted to the on-shell space, the dual conformal boost generator implies a new constraint on the amplitudes because it is equivalent to a level-1 Yangian generator when restricted to the on-shell space. By commuting the dual conformal boost generator with the dual supersymmetry generators, it is not difficult to deduce that the dual generators $S^{A \alpha}$ and $R^{AB}$ are also nontrivial, while the remaining dual superconformal symmetry generators either vanish or reduce to ordinary superconformal symmetry generators when restricted to the on-shell space. Furthermore, it is possible to match $S^{A \alpha}$ and $R^{AB}$ with level-1 Yangian generators with the help of the $y$ coordinates \cite{Huang:2010qy}. Since the level-$n$ Yangian generators can be obtained by taking $n-1$ commutators of level-1 Yangian generators, they do not impose additional constraints on the tree-level amplitudes when $n>1$.

\section{Recursion Relations}

The computation of amplitudes can be drastically simplified using the BCFW recursion relations, which relate higher-point tree-level amplitudes to lower-point tree-level amplitudes. In the BCFW approach, one analytically continues the external momenta of an on-shell amplitude into the complex plane in such a way that the amplitude remains on-shell and becomes a rational function in the complex plane whose only singularities arise from internal propagators becoming on-shell. The recursion relation then follows from the fact that a rational function can be reconstructed from the residues of its poles, provided that the function vanishes at infinity. The residues in this case are simply products of lower-point on-shell amplitudes \cite{Britto:2005fq}.

In the simplest version of the BCFW recursion relation, one shifts two external momenta by a complex parameter $z$. For legs $i$ and $j$, the shift is given by
\be
p_i\rightarrow p_i+zq,\;\; p_j\rightarrow p_j-zq\,,
\ee
for some vector $q$. In order for the external legs to remain on-shell, the shift vector $q$ must satisfy
\be
q\cdot p_i=q\cdot p_j=q^2=0\,.
\label{Ddimshift}
\ee
Although these constraints admit nontrivial solutions in $D>3$ dimensions, for $D=3$ the only solution is $q=0$.

The analysis above can be translated into spinor language. Under a general two-line shift in three-dimensions, $(\lambda_i,\lambda_j) \rightarrow (\hat{\lambda}_i(z),\hat{\lambda}_j(z))$, momentum conservation implies that
\be
\lambda_i\lambda_i+\lambda_j\lambda_j=\hat{\lambda}_i(z) \hat{\lambda}_i(z)+\hat{\lambda}_j(z) \hat{\lambda}_j(z)\,.
\ee
Assuming that the shift is a linear transformation of $(\l_i,\l_j)$, one can write
\begin{align}
\begin{pmatrix} \hat \lambda_i(z)  \\ \hat \lambda_{j} (z)  \end{pmatrix} = R(z)  \begin{pmatrix} \lambda_i \\ \lambda_{j}
\end{pmatrix} \,.
\label{R}
\end{align}
Then momentum conservation implies that
\be
R^T(z)R(z)=I \,,
\label{Rcondition}
\ee
i.e., the shift is an element of O$(2,\IC)$.

If we assume that $R(z)$ in eq. \ref{R} is linear in $z$ and reduces to the unit matrix when $z=0$, then  it can be parameterized as
\eq
R(z)=\left(\begin{array}{cc}1+a_{11}z & a_{12}z \\ a_{21}z & 1+a_{22}z\end{array}\right)\,.
\eqe
It is not difficult to see, however,  that eq. \ref{Rcondition} constrains all of $a$'s to be zero. A similar analysis shows that eq. \ref{Rcondition} also constrains the three-line shift to be trivial, i.e., if we take $R(z)$ to be an O$(3,\IC)$ matrix that depends linearly on $z$, then eq. \ref{Rcondition} constrains it to be the identity. The first non-trivial solution which depends linearly on $z$ appears for a four-line shift. It is possible to define a linear two-line shift in a three-dimensional field theory with massive fields. In this case, the kinematics is essentially four-dimensional. Scattering amplitudes in mass-deformed superconformal Chern-Simons theories were considered in \cite{Agarwal:2008pu}.

In order to construct a two-line shift without introducing a mass deformation, we must relax the assumption that  $R(z)$ is linear in $z$. In the next subsection, we will construct such a matrix and use it to derive a recursion relation for tree-level amplitudes.

\subsection{Two-Line Shift}

We parameterize the $R(z)$ matrix by
\begin{equation}
R(z) = \begin{pmatrix} \frac{z+z^{-1}}{2} & - \frac{ z - z^{-1} }{ 2 i }   \\ \frac{ z- z^{-1} }{2 i } & \frac{ z+ z^{-1} }{ 2} \end{pmatrix} .
\label{Rmat}
\end{equation}
Note that this matrix reduces to the unit matrix when $z=1$. Also note that for the ABJM theory, this deformation has a $z \rightarrow 1/z$ symmetry since $R(1/z)$ is related to $R(z)$ by an orientation reversal:
\[
R(1/z)=\left(\begin{array}{cc}
1 & 0\\
0 & -1\end{array}\right)R(z)\left(\begin{array}{cc}
1 & 0\\
0 & -1\end{array}\right),
\]
under which the superamplitudes are invariant up to a sign (see eq. \ref{super-l-parity}).
Without loss of generality, let's suppose the deformation acts on legs 1 and $l$. Since we are interested in a recursion relation for superamplitudes,
we also need to consider super-momentum conservation,
\eq
\sum^n_{i=1}q^{\alpha I}_i=\sum^n_{i=1}\lambda^\alpha_i\eta_i^I=0 \ .
\eqe
The conservation of both momentum and super-momentum can be maintained if we also deform the fermionic coordinates $\eta_1$ and $\eta_l$. Thus we define the super-shift as:
\eqa
\begin{pmatrix} \hat \lambda_1(z)  \\ \hat \lambda_{l} (z)  \end{pmatrix} = R(z)  \begin{pmatrix} \lambda_1 \\ \lambda_{l} \end{pmatrix},\; \begin{pmatrix} \hat \eta_1(z)  \\ \hat \eta_{l} (z)  \end{pmatrix} = R(z)  \begin{pmatrix} \eta_1 \\ \eta_{l} \end{pmatrix}
\,, \label{deformation}
\eqae
or
\eq
\begin{pmatrix} \hat \Lambda_1(z)  \\ \hat \Lambda_{l} (z)  \end{pmatrix} = {R}(z)  \begin{pmatrix} \Lambda_1 \\ \Lambda_{l} \end{pmatrix} \
\label{supershift}
\eqe
where we are using the collective notation $\Lambda=(\lambda,\eta)$.

We will now derive the recursion relation that follows from this supershift. In the following discussion, we will factor out the momentum-conserving delta function from the amplitudes:
\be
\CA( \Lambda_1, \cdots, \Lambda_{2k}) = A( \Lambda_1, \cdots, \Lambda_{2k}) \delta^{3}(P).
\ee
After applying the deformation to an on-shell amplitude, it acquires poles in the complex parameter $z$. Near these poles, the amplitude factorizes into two lower-point on-shell amplitudes connected by a propagator. We will denote the two lower-point amplitudes by $A_{L}$ and $A_{R}$. The poles of the amplitude in this factorization channel are therefore given by the roots of $\hat{p}_{f}^2(z)=0$, where $f$ labels the factorization channel and $\hat{p}_f(z)$ is the momentum in the propagator. If the amplitudes vanish at $z=0$ and $z=\infty$, it follows that undeformed amplitude $A_n(z=1)$ is given by
\begin{equation}
A_n(z=1)=\frac{-1}{2\pi i}\sum_{f,j}\int d^{3}\eta\oint_{z= z_{j,f}}\frac{dz}{z-1}\frac{A_{L}(z,\eta)A_{R}(z,i\eta)}{\hat{p}_{f}(z)^{2}},
\label{recursion}
\end{equation}
where the index $j$ labels the roots of $\hat{p}_{f}^2(z)=0$, which we denote as $ z_{j,f}$, and the integration variable $\eta$ is allocated to the internal propagator of each channel. Integrating over $\eta$ is equivalent to summing over all fields which can appear in the propagator. For simplicity, we have suppressed the dependence on the other variables of the on-shell superspace. This is the recursion relation for three-dimensional superconformal field theories. This recursion relation is applicable to the ABJM theory, because the superamplitudes vanish as $z$ goes to zero and infinity, as we will explain in the next subsection. Furthermore, this recursion relation was used to compute six- and eight-point component amplitudes in the ABJM theory and the results agree with Feynman diagram calculations and calculations based on the Grassmannian integral formula, which will be described in section 3.6~\cite{Gang:2010gy}.

For any channel, $\hat{p}_{f}^{2}(z)$ in eq. \ref{recursion} has the form $a_{f}z^{-2}+b_{f}+c_{f}z^{2}$, so the roots are obtained by solving a quadratic equation. To see this, note that the deformed external momenta are
\[
\hat{ p}_1 = \hat{ \lambda}_1 \hat{ \lambda}_1  =  \half ( p_1 + p_{l} ) +  z^2 q +  z^{-2} \tilde q,
\]
\begin{equation}
\hat p_{l}  =\hat \lambda_{l} \hat \lambda_{l} =  \half ( p_1 + p_{ l} ) -   z^2 q -  z^{-2} \tilde q,
\label{shiftedq}
\end{equation}
where $q$ and $ \tilde q$ are given by
\begin{align}
q^{ \alpha \beta} = \frac{1}{4} ( \lambda_1 + i  \lambda_{ l} )^{ \alpha} ( \lambda_1 + i \lambda_{ l} )^{ \beta} , \quad \tilde q^{ \alpha \beta} = \frac{1}{4} ( \lambda_1 - i \lambda_{ l} )^{ \alpha} ( \lambda_1 - i \lambda_{l} )^{\beta}  .
\end{align}
Since $q^2=\tilde{q}^2=0$, it follows that $\hat{p}_{f}^{2}(z)$ has the form claimed above.

After a little algebra, eq.\ref{recursion} can be written more explicitly as follows:
\begin{align}
A  (z=1) =
&  \sum_f    \int d^3 \eta  \frac{1}{ p_f^2 } \big(   H(z_{1,f} , z_{2,f} ) A_L ( z_{1,f}; \eta ) A_R (z_{1,f} ; i\eta)+(z_{1,f} \leftrightarrow z_{2,f})  \big) \,,\label{fact-limit}
\end{align}
where $p_f=\hat{p}(z=1)$, $\{\pm z_{1,f}, \pm z_{2,f} \}$ are the roots of $\hat{p}_{f}^{2}(z)=0$, and
\begin{align}
H(a,b)
& = \begin{cases}  \frac{a^2 (b^2 -1) }{a^2 -b^2} , & ( l \mbox{ odd}), \\
 \frac{a (b^2 -1)}{a^2 - b^2}  , &  ( l \mbox{ even}).
\end{cases}
\label{function F}
\end{align}
In obtaining the above equation, we used the fact that
$$A_L (-z) A_R(-z) = (-1)^{l+1} A_L (z) A_R (z), $$
which follows from eq. \ref{super-l-parity}.

Since $ H(a,b) A_L(a) A_R (a)$ is invariant under both $ a \to -a $ and $b \to -b$, it must be a function of $a^2$ and $b^2 $. Furthermore, since the integrand in eq. \ref{fact-limit} is symmetric under $z_{1,f} \leftrightarrow z_{2,f}$, it can be written as a rational function of $(z_{1,f})^2 + (z_{2,f})^2 = - b_f/c_f$ and $(z_{1,f})^2 (z_{2,f})^2 = a_f/c_f$. Hence, the final result is free from any square-root factors that $ \{  \pm z_{1,f},  \pm z_{2, f} \} $ may contain.

\subsection{Large-$z$ Behavior}

The recursion relation derived in the previous subsection is justified only when  $\mathcal{A}(z)$  vanishes as $z \to 0, \infty$. For the ABJM theory, $A(1/z)$ is the same as $A(z)$ up to a sign because exchanging $z$ with $1/z$ in eq. \ref{Rmat} acts as an orientation reversal, so it suffices focus on the behavior as $z \rightarrow \infty$. Since the proof of good large-$z$ behavior is somewhat technical, we will just summarize the main ideas in the analysis. The basic strategy is to identify component amplitudes which have the same large-$z$ behavior as the superamplitudes, and use background field techniques to show that these amplitudes vanish sufficiently quickly as $z \rightarrow \infty$.

To identify which component amplitudes have the same large-$z$ behavior as the superamplitudes, consider shifting legs $i$ and $j$ of the superamplitude, and expand the superamplitude in $(\eta_i,\eta_j)$:
\be
\CA = \CA^{(0,0)} + \CA^{(1,0)}_I \eta_i^I + \CA^{(0,1)}_I \eta_j^I
+\ldots + \CA^{(3,3)} (\eta_i)^3 (\eta_j)^3 \,.
\ee
Each sub-amplitude $\CA^{(m,n)}$ depends on all $\l$'s and
all $\eta$'s except $(\eta_i,\eta_j)$.
After the super-shift, $(\L_i,\L_j)\goto (\hat{\L}_i(z),\hat{\L}_j(z))$,
the superamplitude becomes
\begin{align}
\CA(z) &= \CA^{(0,0)}(z) +
\CA^{(1,0)}_I(z) \hat{\eta}_i^I(z) + \CA^{(0,1)}_I(z) \hat{\eta}_j^I(z)
+\cdots
 \\
&= \tilde{\CA}^{(0,0)}(z) +
\tilde{\CA}^{(1,0)}_I(z) \eta_i^I + \tilde{\CA}^{(0,1)}_I(z) \eta_j^I +
\cdots \,.
\end{align}
On the first line, the $z$-dependence of $\CA^{m,n}(z)$ is entirely due to the shift $(\l_i, \l_j)\goto (\hat{\l}_i(z), \hat{\l}_j(z))$. Hence, these sub-amplitudes have the same large-$z$ behavior as the component amplitudes under the bosonic shift. On the second line, the expansion variables are the undeformed $(\eta_i,\eta_j)$, so the relation between $\tilde{\CA}^{(m,n)}(z)$ and $\CA^{(m,n)}(z)$ follows from the shift property of $(\eta_i,\eta_j)$. In particular, $\tilde{\CA}^{(0,0)}= \CA^{(0,0)}$ and $\tilde{\CA}^{(3,3)}=\CA^{(3,3)}$. Furthermore, using the supersymmetric Ward identity, one can show that all the $\tilde{\CA}^{(m,n)}(z)$ have the same large-$z$ behavior. It follows that the component amplitudes contained in  $\CA^{(3,3)}$ have the same large-$z$ behavior as the superamplitude. Note that $\CA^{(3,3)}$ contains the component amplitudes whose shifted legs are $\phi_4$ or $\psi_4$ fields, since these are the lowest-weighted components in the superfield expansion in eq. \ref{sfield}. Hence, the large-$z$ behavior of the superamplitude is the same as the large-$z$ behavior of component amplitudes of the form  $\langle\cdots\  \psi_4(\hat{\lambda}_i(z)) \cdots \phi_4(\hat{\lambda}_j(z)) \cdots\rangle$, for example.

\begin{figure}
\begin{center}
\includegraphics[scale=0.75]{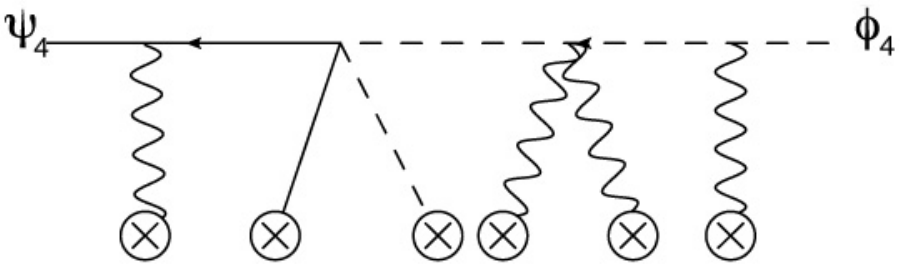}
\caption{Diagrammatic representation of a background field computation
 }
\label{Background}
\end{center}
\end{figure}

It is convenient to analyze the large-$z$ behavior of these amplitudes using a background field formulation which describes hard particles scattering through a soft background~\cite{ahk}.  If we take the shifted legs to be adjacent, the amplitudes under consideration reduce to diagrams of the form depicted in Fig. \ref{Background}, where the horizontal line represents fields whose momenta scale like $z^2$ in the large-$z$ limit, and the crosses represent soft background fields.
In the background field formalism, there are two kinds of gauge symmetry, notably the gauge symmetry of the background and the gauge symmetry of the fluctuations.
From eq. \ref{shiftedq}, we see that in the large-$z$ limit, the shifted momenta scale like $z^2 q$, where $q$ is a null vector. It is therefore convenient to use the background gauge symmetry to impose the following conditions on the background gauge fields:
\eq
q \cdot A^a=q\cdot \hat{A}^a=0.
\label{backchoice}
\eqe
After gauge-fixing the background, we still have the gauge symmetry of the fluctuations (under which the background is inert). We can therefore use the Faddeev-Popov procedure to introduce the following gauge-fixing terms for the gauge-field fluctuations:
\eq
\mathcal{L}_{gf}=tr\left[\frac{1}{\xi_1}(\partial \cdot a)^2-\frac{1}{\xi_2}(\partial\cdot \hat{a})^2\right]\,
\eqe
where $a$ and $\hat{a}$ represent the fluctuations of the gauge fields and $\xi_1$, $\xi_2$ are gauge-fixing parameters.
Note that these gauge-fixing terms don't  preserve the background gauge symmetry. On the other hand, the background gauge symmetry is already broken by eq. \ref{backchoice}, so there's no need to choose gauge-fixing terms which preserve the background gauge symmetry.

Using the approach described above, it can be shown that the tree-level superamplitudes scale like $1/z$ as $z\rightarrow \infty$, which is sufficient for the recursion relation to be applicable \cite{Gang:2010gy}.

\section{Dual Symmetry of All Tree Amplitudes}

While an on-shell recursion relation provides an efficient way to construct higher-point amplitudes from lower-point amplitudes, it also serves as a convenient tool for analyzing the symmetries of amplitudes. If one can demonstrate that the recursion relation preserves a symmetry of the lowest-point tree-level amplitude, it follows by induction that the symmetry holds for all tree-level amplitudes. Indeed, the dual superconformal symmetry of tree-level amplitudes in maximal sYM defined in four~\cite{bht}, six~\cite{Dennen:2010dh}, and ten dimensions~\cite{CaronHuot:2010rj} follows from the fact that the BCFW recursion relations preserve dual superconformal symmetry. In this section, we will use the recursion relation described in section 3.4 to demonstrate that the dual superconformal symmetry of the four-point superamplitude can be extended to all on-shell tree-level superamplitudes in the ABJM theory. This boils down to showing that the amplitudes are covariant under dual inversion.

Note that the dual coordinates are defined up to an overall translation in the dual superspace. We are therefore free to choose the origin of the dual superspace such that
\begin{equation}
x_{1}=-x_{n},\,\,\,\theta_{1}=q_1,
\label{choice}
\end{equation}
where $q_i=\lambda_{i}\eta_{i}$, and we are considering an $n$-point amplitude. For this choice, eq. \ref{hyper} implies that
\begin{align}
&x_1 = \frac{p_1 - p_{n}}2 , \; x_2 = - x_{ n} =  - \frac{p_1 +p_{n}}2 , \ldots , x_{n} = \frac{p_1 +p_{n}}2, \nonumber
\\
&\theta_1 = q_1 , \; \theta_2 = 0, \ldots, \theta_{n} = q_1 +q_{n}. \label{x-lambda relation}
\end{align}
This choice is convenient because if we apply the deformation in eq \ref{supershift} to legs 1 and $n$, this only shifts $(x_1,\theta_1)$:
\be
\hat{x}_1 (z) = \frac{\hat p_1 (z) - \hat p_{n}(z)}2, \; \hat{\theta}_1 (z) = \hat {q}_1 (z),
\\
\hat x_{i} (z) = x_i, \; \hat \theta_i (z) = \theta_i , \; \textrm{for $i>1$.}
\ee
Furthermore, the magnitude of $x_1$ is invariant under the shift, i.e., $\hat{x}_1(z)^2=x_1^2$. This can be seen by noting that
\[
\hat{x}_{1}^{2}\propto\left(\hat{p}_{1}-\hat{p}_{n}\right)^{2}\propto\hat{p}_{1}\cdot\hat{p}_{n}\propto\left(\hat{p}_{1}+\hat{p}_{n}\right)^{2}=\left(p_{1}+p_{n}\right)^{2}\propto x_{1}^{2}.
\]

We would now like to show that when the on-shell amplitudes are written in terms of the dual superspace coordinates $(x,\theta)$, they transform covariantly under dual inversion:
\be
&&I[\tilde{A}_{2k}] = \sqrt {\prod_{i=1}^{2k} x^2_i} \tilde{A}_{2k},
\label{dual conformal invariance}
\ee
where 
\[
\mathcal{A}_{2k} = \tilde{A}_{2k}\delta^3(P)\delta^6(Q). 
\]
In section 3.2.3, we argued that eq. \ref{dual conformal invariance} implies that the amplitudes have dual superconformal symmetry.

\begin{figure}
\begin{center}
\includegraphics[scale=0.8]{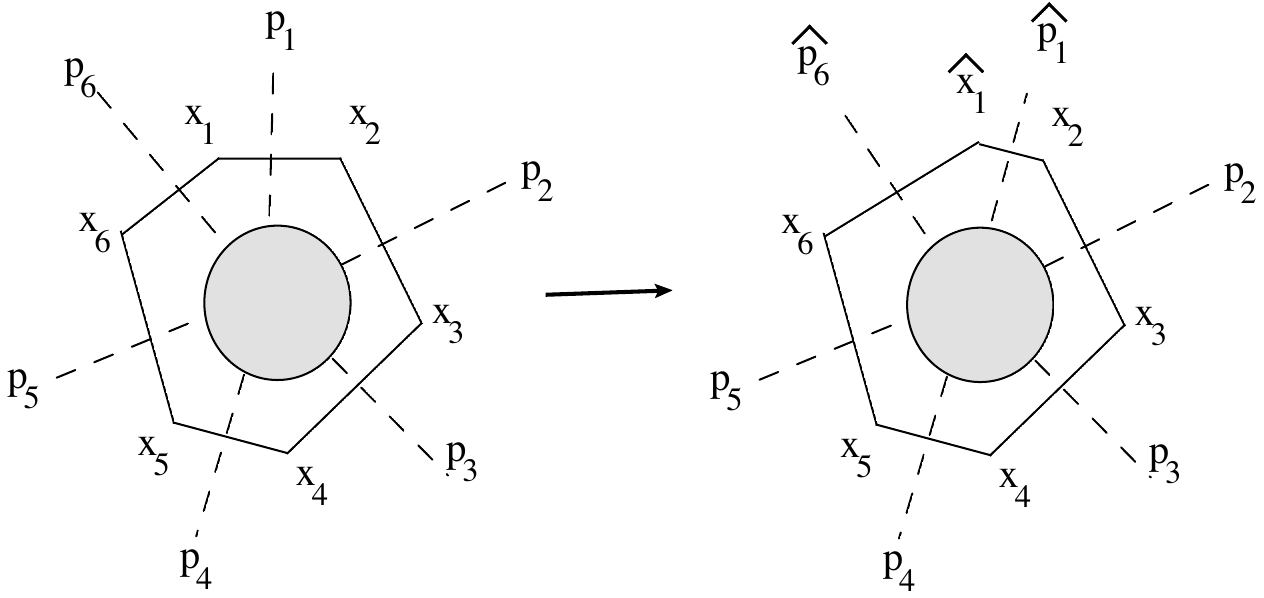}
\caption{The recursion in dual coordinates for $2k=6$. Note that the amplitude is deformed by shifting $x_1$ to $\hat{x}_{1}(z)$. The shift creates poles in the $p_{123}$ channel. The residues of these poles are proportional to the product of four-point tree amplitudes at shifted kinematics.   }
\label{bcfwdual}
\end{center}
\end{figure}

The proof is based on induction. Assuming that eq. \ref{dual conformal invariance} is satisfied by all amplitudes with fewer than $2k$ external legs, we will use the recursion relation defined in section 3.4.1 to show that $\tilde{A}_{2k}$ inverts the same way. In particular, the recursion relation in eq. \ref{fact-limit} implies that
\be
\tilde{A}_{2k} = \sum_f \int \frac{d^3 \eta}{p_f^2} [\delta^6(Q_R)H(z_{1,f},z_{2,f})\tilde{A}_L (z_{1,f};\eta) \tilde{A}_R (z_{1,f};i\eta) +(z_{1,f}\leftrightarrow z_{2,f})].
\label{BCFW for dualconformal}
\ee
In obtaining this expression, we noted that $\delta^6(Q_L)\delta^6(Q_R) = \delta^6(Q_{2k})\delta^6 (Q_R)$.
Note that since we chose the origin of the dual space according to eq. \ref{choice}, shifting legs 1 and $2k$ corresponds to shifting the dual coordinate $x_1$ and $\theta_1$. We illustrate this in Fig. \ref{bcfwdual} for the case $2k=6$.

Let's focus on a factorization channel where $\tilde{A}_L$ is a $(j+1)$-point amplitude and $\tilde{A}_{R}$ is a $(2k-j+1)$-point amplitude. When written in terms of dual coordinates,
\be
\tilde{A}_L (z) = \tilde{A}_L (\hat{x}_1(z),x_2,...,x_{j+1}), \; \tilde{A}_R (z) = \tilde{A}_R (x_{j+1},x_{j+2},\ldots, x_{2k},\hat x_1 (z)),
\ee
where we suppress the $\theta$ coordinates for simplicity. We will now deduce the inversion property of $\tilde{A}_{2k}$ by applying a dual inversion to eq. \ref{BCFW for dualconformal}. By assumption, the functions $\tilde{A}_L$ and $\tilde{A}_R$ invert as
\be
&&I[\tilde{A}_L (\hat x_1 (z), \ldots, x_{j+1})] = \sqrt{\hat x_1^2(z)\ldots x_{j+1}^2} \tilde{A}_L  (\hat x_1 (z), \ldots, x_{j+1}), \nonumber
\\
&&I[\tilde{A}_R (x_{j+1}, \ldots, \hat x_1 (z))] = \sqrt{x_{j+1}^2 \dots \hat x_1^2(z)} \tilde{A}_R (\hat x_1(z),\ldots, x_{2k}). \label{inversion bcfw 1}
\ee
The inversion of the propagator term in eq.\ref{BCFW for dualconformal}  is given by
\be
I \left[ \frac{1}{p_f^2} \right]  = I  \left[\frac{1}{x_{1,j+1}^2} \right] = \frac{x_1^2 x_{j+1}^2}{x_{1,j+1}^2}
= \frac{x_1^2 x_{j+1}^2}{p_f^2  }.  \label{inversion bcfw 2}
\ee
Furthermore, a simple calculation shows that
$$
\int d^3 \eta \delta^6 (Q_R) = \int d^3 \eta \delta^6 (q^{I \a}_{j+1}+ \ldots + \hat q^{I\a}_{2k}- \eta^I  \hat \l^\a_{f})
= \int d^3 \eta \delta^6 (\theta^{I\a}_{j+1} - \hat \theta^{I\a}_{1} - \eta^I \hat{\l}^\a_{f})
$$
$$
= \frac{1}{6} \epsilon_{IJK} (\hat \theta_1 - \theta_{j+1})^{I \a}\hat{\l}_{f,\a}(\hat \theta_1 - \theta_{j+1})^{J \b}\hat{\l}_{f,\b} (\hat \theta_1 - \theta_{j+1})^{K \g}\hat{\l}_{f,\g}\,.
$$
Using the dual inversion rules in section 3.2.2, we find that
\begin{align}
I \left[ \int d^3 \eta \delta^6(Q_R) \right] = \frac{1}{\hat x_1^2(z) x^2_{j+1}\sqrt{\hat x_1^2 (z) x^2_{j+1}}} \int d^3 \eta \delta^6(Q_R). \label{inversion bcfw 3}
\end{align}
The only remaining piece to invert in eq.~\ref{BCFW for dualconformal} is $H(z_1,z_2)$. From the observation that
\begin{align}
I [ \hat p_f (z)^2] & 
= \frac{\hat p_f (z)^2}{x_1^2 x_{j+1}^2}, \label{inversion-pf}
\end{align}
one can easily see that
$
\hat p_f (z)^2 =0 $ is equivalent to $ I [ \hat p_f (z)^2] =0 $ so
\be
I[H(z_1, z_2)]= H(z_1,z_2). \label{inversion bcfw 4}
\ee
Combining eqs. \ref{BCFW for dualconformal},  \ref{inversion bcfw 1}, \ref{inversion bcfw 2}, \ref{inversion bcfw 3}, and \ref{inversion bcfw 4}, and recalling that $\hat x_1^2 = x_1^2$ implies that
\be
I [\tilde{A}_{2k}] =\sqrt{\prod_{i=1}^{2k} x_i^2 } \tilde{A}_{2k}. \label{inversion-amp}
\ee

Hence, if $\tilde{A}_{l}$ inverts according to eq. \ref{dual conformal invariance} for all $l<n$, then so does $\tilde{A}_{n}$. Since $\tilde{A}_4$ satisfies this property (as shown in section 3.2.3), this completes the proof that all tree-level ABJM amplitudes are dual superconformal invariant. Given that the tree-level amplitudes have dual superconformal symmetry, one can use generalized unitarity methods \cite{Bern:1994zx,Bern:1994cg} to show that the integrands of the cut-constructible loop amplitudes have this symmetry as well \cite{Gang:2010gy}.

\section{Grassmannian Integral Formula}
In this section, we will briefly describe the matrix integral formula proposed in \cite{Lee:2010du}, which is conjectured to generate all tree-level amplitudes of the ABJM theory. For the $2k$-point superamplitude, the integral is given by
\be
\CL_{2k}(\L) =
\int  \frac{d^{k\times 2k} C}{{\rm vol}[{\rm GL}(k)]}\frac{\d^{k(k+1)/2}(C\cdot  C^T)\, \d^{2k|3k}(C\cdot \L)}
{M_1 M_2 \cdots M_{k-1} M_k}
\,.
\label{3d-master}
\ee
The integration variable $C$ is a $(k\times 2k)$ matrix. The dot products denote $(C\cdot C^T)_{mn} = C_{mi}C_{ni}$, $(C\cdot \L)_m = C_{mi}\L_i$.
Note that the index $m$ runs from $1$ to $k$ and the index $i$ labels the external particles and runs from $1$ to $2k$. The $i$-th minor, $M_i$, of $C$ is defined as
\be
M_i = \e^{m_1 \cdots m_k} C_{m_1 (i)} C_{m_2 (i+1)} \cdots C_{m_k (i+k-1)} \,.
\ee
The delta functions can be written more explicitly as
\[
\delta^{2k|3k}(C\cdot\Lambda)=\Pi_{m=1}^{k}\delta^{2}\left(C_{m}\cdot\vec{\lambda}\right)\delta^{3}\left(C_{m}\cdot\vec{\eta}\right)
\]
where we are organizing the spinors $\lambda_\i^\alpha$ into two $2k$-dimensional vectors $\vec{\lambda}^{\alpha}$, $\alpha=1,2$, and we are organizing the fermionic variables $\eta^A_i$ into three $2k$-dimensional vectors $\vec{\eta}^A$, $A=1,2,3$. We see that the integral has fermionic degree $3k$, which is required by superconformal symmetry, as explained in section 3.1. Furthermore, ref.~\cite{Lee:2010du} demonstrated that this formula has Yangian symmetry and the same cyclic symmetry as the scattering amplitudes of the ABJM theory (see eq. \ref{cycle}).

Note that the integrand in eq. \ref{3d-master} has a $GL(k)$-symmetry $C \sim g C$, where $g \in GL(k)$. This symmetry can be used to remove $k^2$ degrees of freedom from the $k \times 2k$ matrix $C$. Furthermore, the bosonic delta function $\delta(C \cdot \lambda)$ removes $2k$ degrees of freedom and the orthogonality constraint $\delta (C\cdot C^T)$ removes $k(k+1)/2$ degrees of freedom. The net number of integration variables is therefore given by
\[
2k^{2}-k^{2}-2k-\frac{k(k+1)}{2}+3=\frac{(k-2)(k-3)}{2}
\]
where the $+3$ appears because three of the bosonic delta functions ultimately become momentum-conserving delta functions. Hence, no integration is required to obtain the four- and six-point tree-level amplitudes from eq. \ref{3d-master}.

If we think of the matrix $C$ as a set of $k$ vectors in $\mathbb{C}^{2k}$, then eq. \ref{3d-master} can be interpreted as an integral over $k$-planes in $\mathbb{C}^{2k}$, which are subject to an equivalence relation $C \sim g C$, where $g \in GL(k)$, and an orthogonality constraint coming from $\delta (C\cdot C^T)$. This space is referred to as the orthogonal Grassmannian, $OG(k,2k)$.

As an example, let's reproduce the four-point superamplitude using the Grassmannian integral formula. In ref. \cite{Gang:2010gy}, the integral formula was also used to compute six- and eight-point component amplitudes.
For the case $k=2$, one can use the GL$(2)$ gauge symmetry to set
\be
C = \begin{pmatrix}
c_{21} & 1 & c_{23} & 0 \\
c_{41} & 0 & c_{43} & 1
\end{pmatrix} \,.
\ee
The bosonic delta function $\delta (C \cdot \lambda)$ then implies that
\be
c_{21}\lambda_{1}+\lambda_{2}+c_{23}\lambda_{3}=0,
\label{bosconst1}
\ee
\be
c_{41}\lambda_{1}+c_{43}\lambda_{3}+\lambda_{4}=0.
\label{bosconst2}
\ee
Using the Schouten identity
\[
\lambda_{i}\left\langle jk\right\rangle +\lambda_{j}\left\langle ki\right\rangle +\lambda_{k}\left\langle ij\right\rangle =0,
\]
we see that eqs. \ref{bosconst1} and \ref{bosconst2} are solved by
\be
c \equiv
\left
(\begin{array}{cc}
c_{21} & c_{23}\\
c_{41} & c_{43}\end{array}\right)=\frac{1}{\left\langle 31\right\rangle }\left(\begin{array}{cc}
\left\langle 23\right\rangle  & \left\langle 21\right\rangle \\
\left\langle 43\right\rangle  & \left\langle 14\right\rangle \end{array}\right)\equiv c_{0}.
\ee
Hence, the delta function $\delta (C \cdot \lambda)$ reduces to
\be
\delta^{4}\left(C\cdot\vec{\lambda}\right)=\frac{1}{\left\langle 13\right\rangle ^{2}}\delta^{4}\left(c-c_{0}\right).
\label{4pt-1}
\ee
Inserting this solution into the remaining delta functions and the minors, and using eq. \ref{4ptrelat}, one finds that
\be
\d^{3}(C\cdot C^T) =  \frac{\langle 13 \rangle^6}{\langle 24 \rangle^3} \d^3(P) \,,
\;\;
\d^6(C\cdot \eta) = \frac{\langle 24 \rangle^3}{\langle 13 \rangle^6}\d^6(Q) \,,
\;\;
\frac{1}{M_1 M_2}= \frac{\langle 13 \rangle^2}{\langle 1 4 \rangle \langle 34 \rangle} \,.
\label{4pt-2}
\ee
Combining eq.\eqref{4pt-1} and eq.\eqref{4pt-2}, we see that the integral over $C$ trivially gives the four-point superamplitude in eq. \ref{4pt}. 

%% file: conclusion.tex
\chapter{Conclusion}    		
\label{conclusion}

In this thesis, we have presented evidence that the superconformal
Chern-Simons theory discovered by ABJM is integrable in the planar
limit. In a certain limit, this theory is dual to type IIA string
theory on $AdS_{4}\times CP^{3}$ and therefore represents a new example
of the $AdS/CFT$ correspondence. Although $\mathcal{N}=4$ sYM and the ABJM theory have some common
features, notably they are both superconformal field theories which
have string theory duals, the ABJM theory exhibits many new properties.
As a result, the techniques for computing anomalous dimensions and
scattering amplitudes that were developed for $\mathcal{N}=4$ sYM
cannot be trivially applied to the ABJM theory. Below we list some
of the main differences between $\mathcal{N}=4$ sYM and the ABJM theory:
\begin{itemize}
\item While $\mathcal{N}=4$ sYM lives in four dimensions, the ABJM theory
lives in three dimensions. This has several important implications:

\begin{itemize}
\item Conformal symmetry in three dimensions allows for Chern-Simons gauge
fields but not Yang-Mills gauge fields, since the latter would introduce
a dimensionful parameter. Since Chern-Simons gauge fields do not have
dynamical degrees of freedom, this implies that only matter fields
can appear on external lines of the on-shell amplitudes.
\item Chirality is not defined in three dimensions. Whereas there are two
types of spinors in four dimensions (which have opposite chirality),
there is only one type of spinor in three dimensions. As a result,
the position twistor of the ABJM theory is self-conjugate, as described in section 3.1. Nevertheless,
as we describe in Appendix C, twistors in three dimensions have the same geometric interpretation
as twistors in four dimensions in the sense that a point in twistor
space corresponds to a null ray in Minkowski space and a point in
Minkowski space corresponds to a line in twistor space.
\item Since the superspace of the ABJM theory is non-chiral, this implies
that the dual superspace of the ABJM theory should include three Grassmann-even
coordinates which carry $R$-symmetry indices, in addition to three
bosonic and six fermionic coordinates (see section 3.2.1 for more details). Such coordinates do not
arise in the dual superspace of $\mathcal{N}=4$ sYM (unless one uses a non-chiral superspace \cite{Huang:2011um}). This definition of the dual superspace suggests that
type IIA string theory on $AdS_{4}\times CP^{3}$ should be self-dual
after performing T-duality transformations along the three translational
directions of $AdS_{4}$, three directions in $CP^{3}$, and six fermionic
directions. All attempts to demonstrate this however, have encountered
singularities \cite{Adam:2009kt,Adam:2010hh,Dekel:2011qw,Bakhmatov:2010fp}.
\item Helicity is not defined in three dimensions (except by dimensional
reduction from a four-dimensional theory \cite{Chiou:2005jn}, but three-dimensional Chern-Simons
theories do not arise by dimensional reduction). Hence, while the
on-shell color-ordered superamplitudes of $\mathcal{N}=4$ sYM are labeled by the
number of external lines and MHV-degree, the on-shell color-ordered superamplitudes
of the ABJM theory are only labeled by the number of external lines. Formally, the $2k$-point superamplitude of the ABJM theory is similar to the $2k$-point superamplitude of $\mathcal{N}=4$ sYM with MHV-degree $k$. Whereas the generating function for tree-level amplitudes in $\mathcal{N}=4$ sYM takes the form of an integral over the Grassmannian $G(k,n)$, the analogous formula for the ABJM theory is an integral over the orthogonal Grassmannian, $OG(k,2k)$. This is described in section 3.6.
\item The BCFW recursion relation, which involves linearly deforming the momenta
of two external lines of an on-shell amplitude by a complex parameter,
only works in four and higher dimensions. In order
to define a deformation of two external momenta in three dimensions
which preserves the on-shell properties of the amplitudes and conformal
symmetry, one must allow the deformation to be nonlinear. This is described in greater detail in section 3.4.
\end{itemize}
\item While $\mathcal{N}=4$ sYM has maximal supersymmetry, the ABJM theory
only has $3/4$ maximal supersymmetry.

\begin{itemize}
\item As a result, many quantities, such as the magnon dispersion relation
and the all-loop Bethe Ansatz, are dependent on an interpolating function
$h(\lambda)$ which has different asymptotics at weak and strong coupling.
This is in contrast to $\mathcal{N}=4$ sYM, where $h(\lambda)\propto\sqrt{\lambda}$
for all values of the coupling. So far, $h(\lambda)$ is only known
at very small and very large coupling. It has been suggested that
the value of $h(\lambda)$ may be scheme-dependent in the ABJM theory \cite{McLoughlin:2008he}.
If this is the case, then $h(\lambda)$ should be eliminated in favor
of a physical observable which has been computed in the same scheme.
\item Another implication of less-than-maximal supersymmetry is that the radius of
the string theory background receives $\mathcal{O}(1/\sqrt{\lambda})$ corrections, although
this effect only becomes relevant at two loops in the string theory sigma model \cite{Bergman:2009zh}.

\end{itemize}
\item While the matter fields of $\mathcal{N}=4$ sYM are in the adjoint
representation of the gauge group $U(N)$, the matter fields in the
ABJM theory are in the bifundamental representation of the gauge group
$U(N)\times U(N)$.

\begin{itemize}
\item It still possible to define color-ordering and double-line notation
for Feynman diagrams. In the double-line notation, there there are two types of lines; one for each $U(N)$ of the gauge group. See Appendix A for more details.
\item As explained in section 2.1, gauge invariant operators in the ABJM theory must contain an even
number of fields which alternate between the representations $(N,\bar{N})$
and $(\bar{N},N)$. If these operators are thought of as spin chains, then
there are two types of elementary excitations: those residing on the
even sites and those residing on the odd sites. In total, there are
eight elementary excitations which transform in the representation
$(2|2)_{even} \bigoplus (2|2)_{odd}$. This is in contrast to the $\mathcal{N}=4$
sYM spin chain, which has sixteen elementary excitations which transform in the representation $(2|2)_{L} \bigotimes (2|2)_{R}$.
\item In the string theory dual to $\mathcal{N}=4$ sYM, the sixteen transverse
fluctuations about a generic classical solution are dual to the sixteen
elementary excitations of the spin chain. On the other hand, in the
string theory dual to the ABJM theory, only half of the transverse
fluctuations are dual to to the elementary excitations of the spin
chain. These are referred to as {}``light'' fluctuations. The other
eight fluctuations are referred to as {}``heavy.'' As a result, care must be taken when computing one-loop corrections
to the energy of classical string solutions, which involves adding
up the fluctuation frequencies. In section 2.2.4, we describe various summation prescriptions and argue that the appropriate prescription for type IIA string theory in $AdS_4 \times CP^3$ treats heavy and light frequencies on unequal footing.

\end{itemize}
\item Whereas the $AdS_{5}\times S^{5}$ superspace is a supercoset, the
$AdS_{4}\times CP^{3}$ superspace is not  \cite{Gomis:2008jt,Grassi:2009yj}. As a result, the string
theory dual to the ABJM theory cannot be fully described by a coset
sigma model. It is possible to describe a subsector of the string
theory using the $OSp(6|4)/SO(3,1)\times U(3)$ coset sigma model,
but $\kappa$-symmetry breaks down in this model when the superstring
has trivial support in $CP^{3}$. Although integrability has been
demonstrated in this subsector of the theory and one other subsector \cite{Stefanski:2008ik,Arutyunov:2008if,Sorokin:2010wn},
proving integrability of the full superstring theory will probably
require new techniques.
\end{itemize}

Despite the many new features exhibited by the ABJM theory, a great
deal of progress has been made in computing it's anomalous dimensions
at weak coupling via a Bethe Ansatz and at strong coupling using string
theory. Furthermore, as we described in section 3.5, the tree-level amplitudes enjoy dual superconformal symmetry.

Regarding the anomalous dimensions of the ABJM theory, there are still a number of open questions:
\begin{itemize}
\item
As mentioned above, it would be useful to compute the interpolating function $h(\lambda)$ which appears in the asymptotic Bethe Ansatz equations and the Thermodynamic Bethe Ansatz and to determine if this function is scheme-dependent. One way to approach this is to compute higher-loop corrections to the magnon dispersion relation in the gauge theory and string theory. The interpolating function was computed to fourth order at weak coupling in \cite{Leoni:2010tb}. Beyond one-loop in the string theory sigma model, one must take into account the full $AdS_4 \times CP^3$ superspace, whose radius receives $\mathcal{O}(1/\sqrt{\lambda})$ corrections.
\item
It is important to understand the interpretation of the heavy fluctuations which appear in type IIA string theory on $AdS_4 \times CP^3$. This question is closely related to which summation prescription should be used when computing one-loop corrections to classical string energies. In \cite{Bandres:2009kw}, it was argued that only heavy fluctuations with even mode number should be included since the resulting one-loop corrections are generically finite, well-defined, and match gauge theory calculations. It would be interesting to further test various summation prescriptions by computing exponentially suppressed finite-size corrections to classical string energies and comparing them to gauge theory calculations. The first steps in this direction were taken in \cite{Bombardelli:2008qd,Abbott:2010yb,Ahn:2010eg}, which computed finite-size corrections for giant magnons in $AdS_4 \times CP^3$, and in \cite{Astolfi:2011ju}, which computed finite-size corrections to the magnon dispersion relation.
\item
Although the all-loop $AdS_4/CFT_3$ Bethe Ansatz is highly constrained by symmetry and satisfies various nontrivial consistency conditions, it contains three dressing phases which are not uniquely fixed by crossing symmetry \cite{AllLoop,Ahn:2008aa}. The first hint that the all-loop $AdS_5/CFT_4$ Bethe Ansatz should include a dressing phase came from including curvature corrections to the plane-wave limit of the Green-Schwarz action and comparing the string spectrum to perturbative calculations in the gauge theory \cite{Callan:2003xr,Callan:2004uv}. It would therefore be interesting to perform a similar calculation in the $AdS_4/CFT_3$ context. The first steps in this direction were taken in \cite{Astolfi:2009qh,Astolfi:2011ju}.
\item
A Lax connection has been constructed for the $OSp(6|4)/SO(3,1)\times U(3)$ coset sigma model, as well as a subsector of the string theory that is not reachable by the coset sigma model. Furthermore, the Lax connection of the full string theory was shown to be flat to at least second order in the fermionic fields \cite{Sorokin:2010wn}. Although this strongly suggests that the full superstring theory is integrable, this example illustrates the need to develop a method to construct a Lax connection when the target space of the string theory is not a supercoset.
\end{itemize}

There are also many open questions regarding the scattering amplitudes of the ABJM theory:

\begin{itemize}
\item
Using unitarity methods, it has been shown that the integrands of the loop amplitudes have dual superconformal symmetry, but due to infrared singularities, the loop integrals are-ill defined without a regulator. Since a regulator will generally render the symmetry anomalous, it would be interesting to see if one can define a regulator and modify the dual superconformal symmetry generators in such a way that the regulator becomes symmetry-preserving.  This was done for $\mathcal{N}=4$ sYM by considering the dual symmetry to be five-dimensional, with the extra dimension giving rise to a massive regulator~\cite{Alday:2009zm,henn1,henn2}.
\item
After learning how to compute loop integrals in the ABJM theory, it would be interesting to investigate a possible amplitude/Wilson loop duality by computing the two-loop correction to the four-point superamplitude and comparing it to the four-cusp null-polygonal Wilson loop computed in \cite{Henn:2010ps}. For $\mathcal{N}=4$ sYM, the amplitude/Wilson loop duality has recently been extended beyond MHV amplitudes and conjectured to include correlators as well. It would therefore be interesting to see if the four-point superamplitude of the ABJM theory can be matched with a correlator \cite{Bianchi:2011rn}.
\item
It would be interesting to investigate if the recursion relation developed for the ABJM theory is applicable to other superconformal Chern-Simons theories, such as the BLG theory \cite{BLG,Bagger:2007vi,Gustavsson:2007vu}. In order for this recursion relation to be applicable, the superamplitudes must vanish when the complex deformation parameter $z$ goes to zero and infinity. Although various component amplitudes in the ABJM theory have bad large-$z$ behavior, there is enough supersymmetry to ensure that the superamplitudes have good large-$z$ behavior. This may not be true for theories with less supersymmetry.
\item
In $\mathcal{N}=4$ sYM, dual superconformal symmetry is a consequence of the fact that type IIB string theory on $AdS_5 \times S^5$ is self-dual after performing T-duality transformations along isometries corresponding to the dual superspace of $\mathcal{N}=4$ sYM. As mentioned above, a similar analysis for type IIA string theory on $AdS_4 \times CP^3$ encounters singularities. It would be interesting to determine if there is some combination of T-duality transformations under which the string theory is self-dual.
\end{itemize}

The study of integrability of $\mathcal{N}=6$ superconformal Chern-Simons theories is still in its early stages, so there are many other open questions. Ultimately, this subject should continue to provide new perspectives on integrability and the $AdS/CFT$ correspondence. 

%% file: abjm.tex
\chapter{Review of the ABJM Theory}					

The ABJM theory is a three-dimensional superconformal Chern-Simons gauge theory with $\mathcal{N}=6$ supersymmetry. The field content consists of four complex scalars $\phi^I$, four Dirac fermions $\psi_I$, their adjoints $\phi_I=\phi^{I\dagger}$, $\psi^I=\psi_I^\dagger$, and two gauge fields, $A_\mu$ and $\hat{A}_\mu$. Note that the index $I$ is an $SU(4)$ R-symmetry index. The gauge group is $ U(N) \times U(N)$, where $A_{\mu}$ and $\hat{A}_{\mu}$ are the associated gauge fields. The fields $\phi^I$ and $\psi_I$ transform in the $(\bar{N},N)$ representation of the gauge group and their adjoints transform in the $(N,\bar{N})$ representation.

The Lagrangian for the ABJM theory is given by \cite{Benna:2008zy,scs6,Bandres:2008ry}:
\[
\mathcal{L}=\mathcal{L}_{2}+\mathcal{L}_{CS}+\mathcal{L}_{4}+\mathcal{L}_{6},
\]
\[
\mathcal{L}_{2}={\normalcolor tr}\left(D_{\mu}\phi^{I}D_{\mu}\phi_{I}+i\bar{\psi}_{I}\gamma^{\mu}D_{\mu}\psi^{I}\right),
\]
\[
\mathcal{L}_{CS}= \epsilon^{\mu\nu\lambda}{\normalcolor tr}\left(\frac{1}{2}A_{\mu}\partial_{\nu}A_{\lambda}+\frac{i}{3}gA_{\mu}A_{\nu}A_{\lambda}-\frac{1}{2}\hat{A}_{\mu}\partial_{\nu}\hat{A}_{\lambda}-\frac{i}{3}g\hat{A}_{\mu}\hat{A}_{\nu}\hat{A}_{\lambda}\right),
\]
\[
\mathcal{L}_{4}=ig^{2}\epsilon^{IJKL}{\normalcolor tr}\left(\bar{\psi}_{I}\phi_{J}\psi_{K}\phi_{L}\right)-ig^{2}\epsilon_{IJKL}{\normalcolor tr}\left(\bar{\psi}^{I}\phi^{J}\psi^{K}\phi^{L}\right)+ig^{2}{\normalcolor tr}\left(\bar{\psi}^{I}\psi_{I}\phi_{J}\phi^{J}\right)
\]
\[
-2ig^{2}{\normalcolor tr}\left(\bar{\psi}^{J}\psi_{I}\phi_{J}\phi^{I}\right)-ig^{2}{\normalcolor tr}\left(\bar{\psi}_{I}\psi^{I}\phi^{J}\phi_{J}\right)+2ig^{2}{\normalcolor tr}\left(\bar{\psi}_{I}\psi^{J}\phi^{I}\phi_{J}\right) \,,
\]
\[
\mathcal{L}_{6}=\frac{g^{4}}{3}tr\left[\phi^{I}\phi_{I}\phi^{J}\phi_{J}\phi^{K}\phi_{K}+\phi_{I}\phi^{I}\phi_{J}\phi^{J}\phi_{K}\phi^{K}\right.
\]
\[
\left.+4\phi_{I}\phi^{J}\phi_{K}\phi^{I}\phi_{J}\phi^{K}-6\phi^{I}\phi_{J}\phi^{J}\phi_{I}\phi^{K}\phi_{K}\right]\,,
\]
where $\bar{\psi}_I= \psi_I^T \gamma^0$, i.e., we take the transpose of fermion rather than the conjugate transpose, and
\[
D_{\mu}\phi_{I}=\partial_{\mu}\phi_{I}+ig\left(A_{\mu}\phi_{I}-\phi_{I}\hat{A}_{\mu}\right) \,,
\;\; D_{\mu}\psi^{I}=\partial_{\mu}\psi^{I}+ig\left(A_{\mu}\psi^{I}-\psi^{I}\hat{A}_{\mu}\right)\,.
\]
Note that $g=\sqrt{2\pi/k}$, where $k$ is an integer known as the Chern-Simons level. The coupling constant can absorbed into the normalization of the fields. In particular, if we re-scale the fields by $1/g$, the coupling constant only appears as an overall factor of $1/g^2=k/2\pi$ multiplying the Lagrangian. Also note
that the first two terms in $\mathcal{L}_{4}$ break the R-symmetry
from $SO(8)$ to $SU(4)$ because they contain $\epsilon^{IJKL}$.

The fields can be expanded in terms of matrices as follows:
\[
A_{\mu}=A_{\mu}^{a}T^{a}\,,\;\;\; \hat{A}_{\mu}=\hat{A}_{\mu}^{a}T^{a} \,,
\qquad
\phi_{I}=\phi_{I}^{a}\tilde{T}^{a}\,,
\;\;\;
\psi^{I}=\psi^{Ia}\tilde{T}^{a} \,,
\]
where $T^{a}$, $a=1,...,N^{2}-1$, are hermitian generators of ${\rm SU}(N)$
satisfying ${\normalcolor tr}\left(T^{a}T^{b}\right)=\delta^{ab}$,
$T^{N^{2}}$ is $1/\sqrt{N}$ times an $N\times N$ matrix, $\tilde{T}^{a}=\frac{i}{\sqrt{2}}T^{a}$
for $a=1,...,N^{2}-1$, and $\tilde{T}^{N^{2}}=\frac{1}{\sqrt{2}}T^{N^{2}}$.
Hence, at the level of matrix representations, the main difference
between the gauge fields and the matter fields is that the matter
fields are not hermitian. It should be emphasized that the matter fields
carry two different indices, one for each $U(N)$. Using these conventions, the definition of color-ordering in the ABJM theory is similar to the definition of color ordering  in Yang-Mills theories. In particular, a color-dressed amplitude can be expressed as a sum of color-ordered amplitudes which are multiplied by a trace of matrices:
\[
\hat{A}_{n}\left(\bar{\Phi}_{1},\Phi_{2},\bar{\Phi}_{3},...\Phi_{2k}\right)=\sum_{\sigma}sgn(\sigma) Tr\left(\tilde{T}_{\sigma_{1}}\tilde{T}_{\sigma_{2}}^{\dagger}\tilde{T}_{\sigma_{3}}...\tilde{T}_{\sigma_{2k}}^{\dagger}\right)\mathcal{A}_{2k}\left(\Lambda_{\sigma_{1}},...,\Lambda_{\sigma_{2k}}\right),
\]
where the sum is over all permutations, $\sigma$, that mix even and odd sites among themselves modulo cyclic permutations by two sites. A cyclic permutation by two sites is trivial because of the cyclicity of the trace. The function $sgn(\sigma)$ gives $-1$ if $\sigma$ involves an odd permutation of the odd sites, and $+1$ otherwise. This sign arises because the superfields on the odd sites are fermionic.

\begin{figure}
\begin{center}
\includegraphics[scale=0.25]{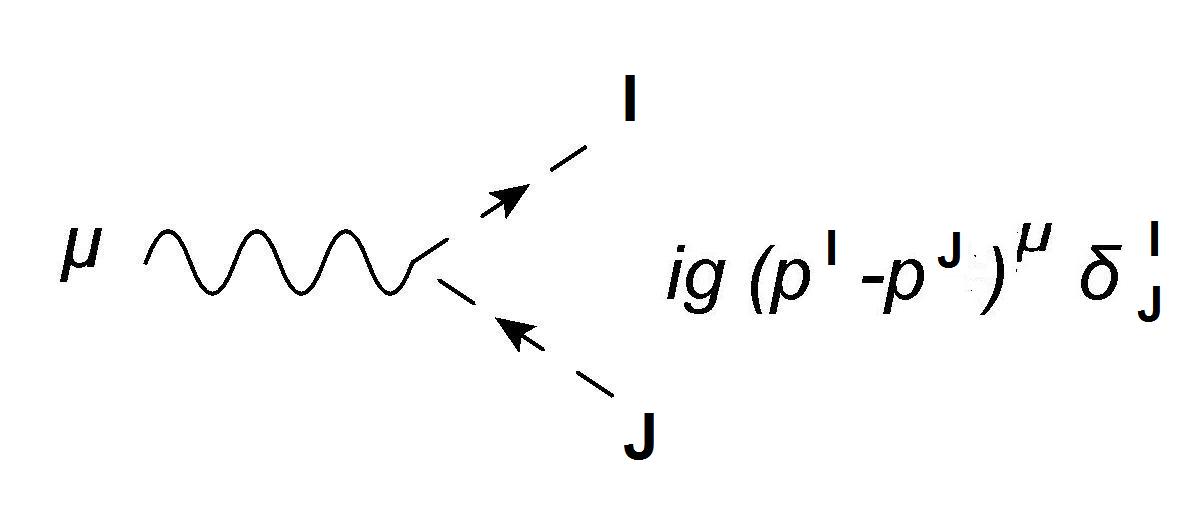}
\caption{Color-ordered cubic vertex.}
\label{3pt}
\end{center}
\end{figure}

\begin{figure}
\begin{center}
\includegraphics[scale=0.25]{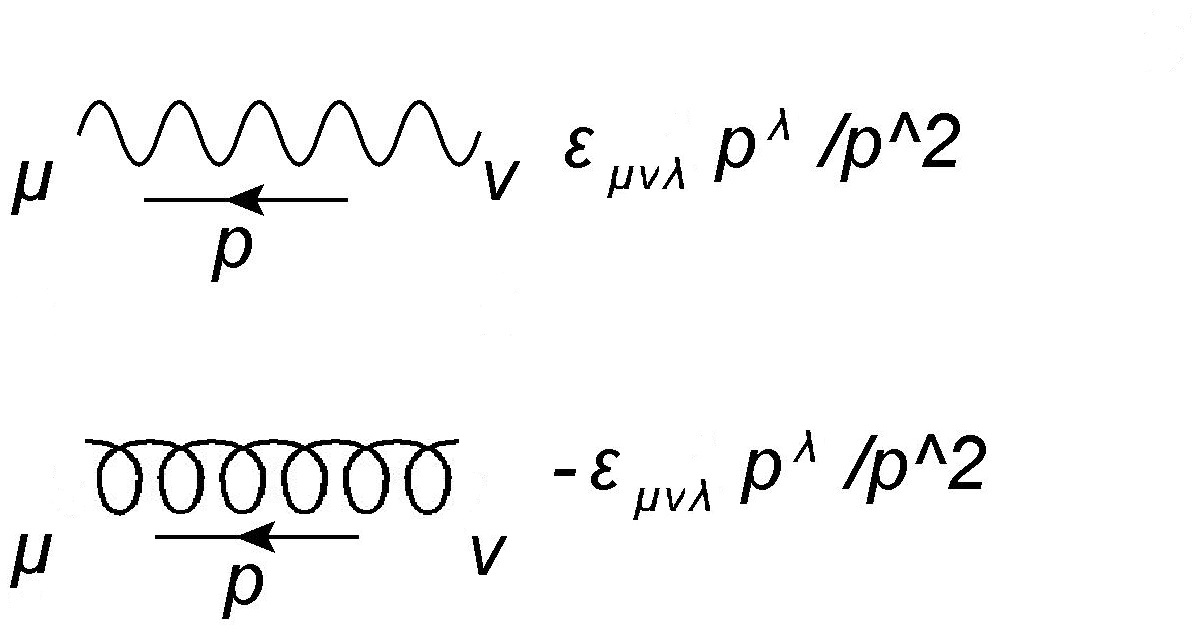}
\caption{Gauge field propagators.}
\label{gagprop}
\end{center}
\end{figure}

In section 3.1, we compute the 4-point color-ordered scalar amplitude using Feynman diagrams. The color-ordered cubic vertex coupling two scalars to a gauge field is depicted in Fig. \ref{3pt}.
Moreover, the gauge field propagators are then given by
 \eq
\pm \frac{\delta^{ab}}{p^{2}}\left(\epsilon_{\mu\nu\lambda}p^{\lambda}+i\xi_i\frac{p_\mu p_\nu}{p^2}\right) \,,
\eqe
where $\pm$ refers to the $A/\hat{A}$ field, $\xi_1,\xi_2$ are gauge-fixing parameters, and $a,b$ are adjoint indices. Taking $\xi_i=0$ gives Landau gauge propagators:
 \eq
\pm\frac{\epsilon_{\mu\nu\lambda}p^{\lambda}}{p^{2}}\delta^{ab} \,.
\eqe
The resulting propagators are depicted in Fig. \ref{gagprop}. A more complete list of color-ordered Feynman rules can be found in \cite{Gang:2010gy}, for example.

\begin{figure}
\begin{center}
\includegraphics[scale=0.22]{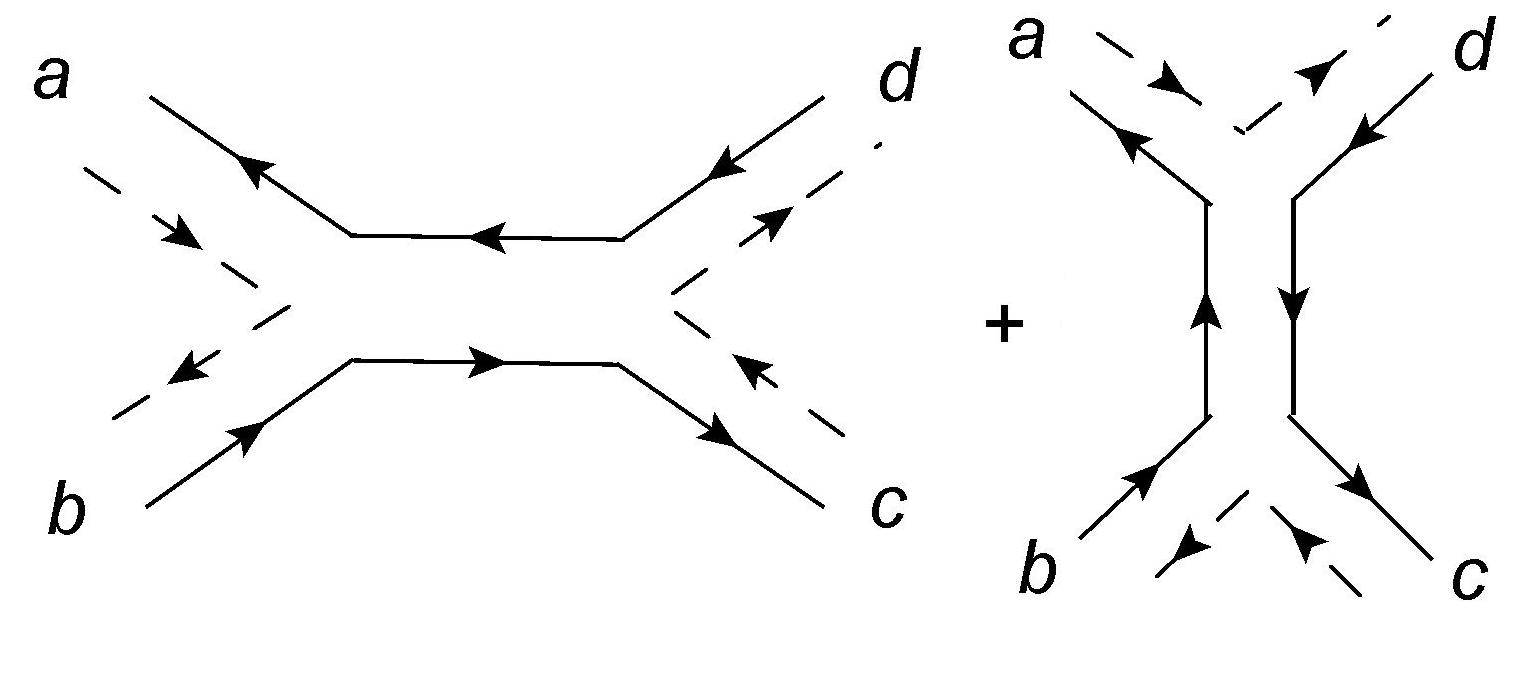}
\caption{Four-point scalar diagrams in double-line notation.}
\label{4ptdoub}
\end{center}
\end{figure}

It is possible to draw Feynman diagrams in the ABJM theory using double-line notation. Since the matter fields transform in the bifundamental representation of the gauge group, there are two types of lines \cite{Huang:2010rn}. For example, when the Feynman diagrams in Fig. \ref{4ptdiag} are written in double-line notation, they are given by Fig. \ref{4ptdoub}. The color-factor associated with these diagrams is $tr\left(\tilde{T}_a\tilde{T}_b^{\dagger}\tilde{T}_c\tilde{T}_d^{\dagger}\right)$.

%% file: IIA.tex
\chapter{Review of Type IIA String Theory on $AdS_4 \times CP^3$}	
The supergravity background of the string theory dual to the ABJM theory consists of the following string frame
metric, dilaton, and Ramond-Ramond fluxes \cite{Aharony:2008ug}:
\begin{eqnarray}
ds^{2} &=&G_{MN}dx^{M}dx^{N}=R^{2}\left( \frac{1}{4}%
ds_{AdS_{4}}^{2}+ds_{CP^{3}}^{2}\right) ,  \label{IIAGeomM} \\
e^{\phi } &=&\frac{R}{k},  \label{Dilaton} \\
\,F_{4} &=&\frac{3}{8}kR^{2}{\normalcolor Vol}_{AdS_{4}},  \label{F4} \\
F_{2} &=&k\mathbf{J},  \label{F2}
\end{eqnarray}
where $R^{2}$ is the radius of
curvature in string units, $\mathbf{J}$ is the Kahler form on $CP^{3}$, and $k$ is an integer corresponding to the level of the dual Chern-Simons theory. The 2-form gauge field in the NS-NS sector is zero.  
Note that the $AdS_{4}$ space has radius $R/2$ while the $CP^{3}$ space has
radius $R$.

The metric for an $AdS_{4}$ space with unit radius in global coordinates $%
(t,\rho ,\theta ,\phi )$ is given by%
\begin{equation}
ds^{2}_{AdS_4}=-\cosh ^{2}\rho \mathrm{d}t^{2}+\mathrm{d}\rho ^{2}+\sinh ^{2}\rho
\left( \mathrm{d}\theta ^{2}+\sin ^{2}\theta \mathrm{d}\phi ^{2}\right) ,
\label{adsmetric}
\end{equation}
where $-\infty <t<\infty ,$ $0\leq \rho <\infty ,$ $0\leq \theta \leq\pi ,$ $%
0\leq \phi <2\pi .$

The embedding coordinates are defined by%
\begin{equation}
n_{1}^{2}+n_{2}^{2}-n_{3}^{2}-n_{4}^{2}-n_{5}^{2}=1,  \label{emb1}
\end{equation}%
and they are related to the global coordinates by%
\begin{equation}
\begin{array}{l}
n_{1}=\cosh \rho \cos t, \\
n_{2}=\cosh \rho \sin t, \\
n_{3}=\sinh \rho \cos \theta \sin \phi , \\
n_{4}=\sinh \rho \sin \theta \sin \phi , \\
n_{5}=\sinh \rho \cos \phi .%
\end{array}
\label{adsmap}
\end{equation}%
Because the global coordinates are not well defined at $\rho =0,$ it is
useful to define Cartesian coordinates $\left( t,\eta _{1},\eta _{2},\eta
_{3}\right)={\bf(0,1,2,3)}$ for which the metric is given by%
\begin{equation}
ds_{AdS_{4}}^{2}=\frac{1}{\left( 1-\eta
^{2}\right) ^{2}}\left[ -\left( 1+\eta ^{2}\right) ^{2}dt^{2}+4d\vec{\eta}%
\cdot d\vec{\eta}\right] {\normalcolor.}  \label{eq:ads4}
\end{equation}%
These coordinates are related to the global coordinates by $\cosh \rho
=(1+\eta ^{2})/(1-\eta ^{2})$. Note that this metric is only valid for $\eta ^{2}=\vec{\eta}\cdot \vec{\eta}%
=\eta _{1}^{2}+\eta _{2}^{2}+\eta _{3}^{2}<1$.

The metric for a unit $CP^{3}$ space is given by
\begin{eqnarray}
ds_{CP^3}^{2} &=&\mathrm{d}\xi ^{2}+\cos ^{2}\xi \sin
^{2}\xi \left( \mathrm{d}\psi +\frac{1}{2}\cos \theta _{1}\mathrm{d}\varphi
_{1}-\frac{1}{2}\cos \theta _{2}\mathrm{d}\varphi _{2}\right) ^{2}
\label{cpmett} \\
&&+\frac{1}{4}\cos ^{2}\xi \left( \mathrm{d}\theta _{1}^{2}+\sin ^{2}\theta
_{1}\mathrm{d}\varphi _{1}^{2}\right) +\frac{1}{4}\sin ^{2}\xi \left(
\mathrm{d}\theta _{2}^{2}+\sin ^{2}\theta _{2}\mathrm{d}\varphi
_{2}^{2}\right) ,  \notag
\end{eqnarray}%
where $0\leq \xi <\pi /2,$ $0 \leq \psi < 2\pi ,$ $0\leq \theta _{i}\leq \pi ,$ and $0\leq \varphi _{i}<2\pi$.

The embedding coordinates ($z^{I}\in {\mathbb{C}}$) are defined by%
\begin{equation}
\sum_{I=1}^{4}\left\vert z^{I}\right\vert ^{2}=1,\qquad z^{I}\sim
e^{i\lambda }z^{I},  \label{emb2}
\end{equation}%
where $\lambda \in {\mathbb{R}}$. The embedding coordinates are related to the global coordinates by
\begin{equation}
\begin{array}{l}
z_{1}=\cos \xi \cos \frac{\theta _{1}}{2}\exp \left( i\frac{\psi +\varphi
_{1}}{2}\right) , \\
z_{2}=\cos \xi \sin \frac{\theta _{1}}{2}\exp \left( i\frac{\psi -\varphi
_{1}}{2}\right) , \\
z_{3}=\sin \xi \cos \frac{\theta _{2}}{2}\exp \left( i\frac{-\psi +\varphi
_{2}}{2}\right) , \\
z_{4}=\sin \xi \sin \frac{\theta _{2}}{2}\exp \left( i\frac{-\psi -\varphi
_{2}}{2}\right) .%
\end{array}
\label{mapping}
\end{equation}%
Note that the $CP^3$ metric can be written in terms of embedding coordinates as follows:
\[
ds_{CP^{3}}^{2}= dz\cdot dz^{\dagger }-\left( z^{\dagger }\cdot dz \right)
\left( z \cdot dz^{\dagger } \right)
\]
where $z \cdot z^{\dagger}=\sum_{I=1}^{4} z^{I} z^{\dagger}_{I}$.

We use the following representation of the 10d Dirac matrices ($%
\left\{ \Gamma ^{A},\Gamma ^{B}\right\} =2\eta ^{AB}$):%
\begin{equation*}
\begin{array}{lll}
\Gamma ^{0}=i\gamma ^{0}\otimes \mathrm{I}\otimes \mathrm{I}\otimes \mathrm{I%
},\quad  & \Gamma ^{1}=i\gamma ^{1}\otimes \mathrm{I}\otimes \mathrm{I}%
\otimes \mathrm{I}\quad  & \Gamma ^{2}=i\gamma ^{2}\otimes \mathrm{I}\otimes
\mathrm{I}\otimes \mathrm{I},\quad  \\
\Gamma ^{3}=i\gamma ^{3}\otimes \mathrm{I}\otimes \mathrm{I}\otimes \mathrm{I%
}, & \Gamma ^{4}=\gamma ^{5}\otimes \sigma _{2}\otimes \mathrm{I}\otimes
\sigma _{1}, & \Gamma ^{5}=\gamma ^{5}\otimes \sigma _{2}\otimes \mathrm{I}%
\otimes \sigma _{3}, \\
\Gamma ^{6}=\gamma ^{5}\otimes \sigma _{1}\otimes \sigma _{2}\otimes \mathrm{%
I}, & \Gamma ^{7}=\gamma ^{5}\otimes \sigma _{3}\otimes \sigma _{2}\otimes
\mathrm{I}, & \Gamma ^{8}=\gamma ^{5}\otimes \mathrm{I}\otimes \sigma
_{1}\otimes \sigma _{2}, \\
\Gamma ^{9}=\gamma ^{5}\otimes \mathrm{I}\otimes \sigma _{3}\otimes \sigma
_{2}, &  &
\end{array}%
\end{equation*}%
where $\mathrm{I}$ is the $2\times 2$ identity matrix, the $\gamma ^{\prime
}s$ are 4d Dirac matrices given by%
\begin{equation*}
\begin{array}{ll}
\gamma ^{0}=\sigma _{1}\otimes \mathrm{I},\qquad  & \gamma ^{1}=i\sigma
_{2}\otimes \sigma _{1}, \\
\gamma ^{2}=i\sigma _{2}\otimes \sigma _{2},\qquad  & \gamma ^{3}=i\sigma
_{2}\otimes \sigma _{3}, \\
\gamma ^{5}=i\gamma ^{0}\gamma ^{1}\gamma ^{2}\gamma ^{3}, &
\end{array}%
\end{equation*}%
and the Pauli matrices are given by
\begin{equation*}
\sigma _{1}=\left[
\begin{array}{cc}
0 & 1 \\
1 & 0%
\end{array}%
\right] ,\qquad \sigma _{2}=\left[
\begin{array}{cc}
0 & -i \\
i & 0%
\end{array}%
\right] ,\qquad \sigma _{3}=\left[
\begin{array}{cc}
1 & 0 \\
0 & -1%
\end{array}%
\right] .
\end{equation*}%
Finally, we define the 10d chirality operator as%
\begin{equation*}
\Gamma_{11}=\Gamma ^{0}\Gamma ^{1}\Gamma ^{2}\Gamma ^{3}\Gamma ^{4}\Gamma
^{5}\Gamma ^{6}\Gamma ^{7}\Gamma ^{8}\Gamma ^{9}.
\end{equation*}

%% file: twistorgeom.tex
\chapter{Twistor Geometry of 3D Minkowski Space}					

In this appendix, we will describe the geometric interpretation of the twistors of three-dimensional Minkowski space. A similar discussion for the twistors of four-dimensional Minkowski space can be found in \cite{Mason:2009qx}. We begin with the conformal compactification of
3D Minkowski space:
\begin{equation}
-T^{2}-V^{2}+X^{2}+Y^{2}+W^{2}=0{\normalcolor .}\label{eq:1}\end{equation}
If we complexify the spacetime, it can also be parameterized by the following five coordinates (which
are defined up to re-scalings):
\[
\tilde{X}_{ab} \propto \left(\begin{array}{cccc}
0 & V+iT & W & X-iY\\
-V-iT & 0 & X+iY & -W\\
-W & -X-iY & 0 & -V+iT\\
-X+iY & W & V-iT & 0\end{array}\right){\normalcolor .}\]
In terms of $\tilde{X},$ eq. \ref{eq:1} can be expressed as follows:\begin{equation}
\epsilon^{abcd}\tilde{X}_{ab}\tilde{X}_{cd}=0\label{eq:4}\end{equation}
where $\epsilon$ is totally antisymmetric. 
Note that $\tilde{X}$ is antisymmetric and traceless with respect
to the invariant tensor of $Sp(4)$:
\begin{equation}
\tilde{X}_{ab}=-\tilde{X}_{ba},\,\,\,\Omega^{ab}\tilde{X}_{ab}=0\label{eq:3}\end{equation}
where $\Omega^{ab}=\left(\begin{array}{cc}
0 & \delta_{\delta}^{\alpha}\\
-\delta_{\gamma}^{\beta} & 0\end{array}\right)$.
The Latin indices $a,b$ run from 1 to 4, and the Greek indices $\alpha,\beta$ run from 1 to 2.  

The most general solution
to eq. \ref{eq:4} is
\begin{equation}
\tilde{X}_{ab} \propto A_{[a}B_{b]}\label{eq:5}\end{equation}
where $A$ and $B$ are defined up to re-scalings by complex parameters,
i.e., $A\sim rA$ and $B\sim tB$ where $r$ and $t$ are complex numbers.
From eq. \ref{eq:3}, we also have\begin{equation}
\Omega^{ab}A_{a}B_{b}=0{\normalcolor .}\label{eq:6}\end{equation}
$A$ and $B$ are referred to as twistors. Since $A$ and $B$ each
have three complex degrees of freedom, when combined with the constraint
in eq. \ref{eq:6}, we find that $\tilde{X}$ in eq. \ref{eq:5} indeed
has five complex degrees of freedom, which confirms that it is the
most general solution to eq. \ref{eq:4}. Since two distinct twistors define a line in twistor space, we see that a point in Minkowski space corresponds to a line in twistor space. Although we are treating the twistors as elements of $CP^3$, they will ultimately be subject to reality constraints.

Given two points in conformally compactified 3D Minkowski space, we
have a natural inner product:
\[
\tilde{X}^{ab}\tilde{Y}_{ab}=\epsilon^{abcd}\tilde{X}_{ab}\tilde{Y}_{cd}{\normalcolor .}\]
Note that this inner product changes under re-scalings of $\tilde{X}$
and $\tilde{Y}$. To fix this, we introduce the infinity twistor:
\[
I_{ab}=\left(\begin{array}{cc}
\epsilon^{\alpha \beta} & 0\\
0 & 0\end{array}\right),\]
where each element in the matrix is a $2 \times 2$ matrix.
Using the infinity twistor, we can define an inner product that's
invariant under rescalings:

\begin{equation}
\frac{\tilde{X}^{ab}\tilde{Y}_{ab}}{I_{cd}\tilde{X}^{cd}I_{ef}\tilde{Y}^{ef}}{\normalcolor .}\label{eq:8}\end{equation}
Furthermore, we can relate points in conformally compactified Minkowski
space to points in Minkowski space as follows:

\begin{equation}
\tilde{X}_{ab}\propto\left(\begin{array}{cc}
\epsilon^{\alpha \beta}x^{2} & -x_{\,\,\, \gamma}^{\alpha}\\
x_{\delta}^{\,\,\, \beta} & \epsilon_{\delta \gamma}\end{array}\right)\label{eq:7}\end{equation}
where we use a proportionality symbol because $\tilde{X}$ is only
defined up to re-scalings. Note that $\Omega^{ab}\tilde{X}_{ab}=-\epsilon^{\alpha \beta}\left(x_{\alpha \beta}+x_{\beta \alpha}\right)=0$
since $x_{\alpha \beta}=x_{\beta \alpha}$. We can confirm that eq. \ref{eq:7} is sensible
by plugging it into eq. \ref{eq:8}. Indeed, this is proportional to the distance between two points in Minkowski space:

\[
\left(x-y\right)^{2} \propto \frac{\tilde{X}^{ab}\tilde{Y}_{ab}}{I_{cd}\tilde{X}^{cd}I_{ef}\tilde{Y}^{ef}}{\normalcolor .}\]

Suppose that we take the twistors $A$ and $B$ in eq. \ref{eq:5}
to be $\left(\mu_{i}^{\alpha},\lambda_{i \beta}\right)$ and $\left(\mu_{j}^{\alpha},\lambda_{j \beta}\right)$.
Then\begin{equation}
\tilde{X}_{ab}\propto\left(\begin{array}{cc}
\mu_{i}^{[\alpha}\mu_{j}^{\beta]} & \frac{1}{2}\left(\mu_{i}^{\alpha}\lambda_{j \gamma}-\lambda_{i \gamma}\mu_{j}^{\alpha}\right)\\
\frac{1}{2}\left(\lambda_{i \delta}\mu_{j}^{\beta}-\mu_{i}^{\beta}\lambda_{j \delta}\right) & \lambda_{i[\delta}\lambda_{j \gamma]}\end{array}\right){\normalcolor .}\label{eq:9}\end{equation}
Combining eq. \ref{eq:7} with eq. \ref{eq:9} gives

\begin{equation}
\left(\begin{array}{cc}
\epsilon^{\alpha \beta}x^{2} & -x_{\,\,\, \gamma}^{\alpha}\\
x_{\delta}^{\,\,\, \beta} & \epsilon_{\delta \gamma}\end{array}\right)=\kappa\left(\begin{array}{cc}
\mu_{i}^{[\alpha}\mu_{j}^{\beta]} & \frac{1}{2}\left(\mu_{i}^{\alpha}\lambda_{j \gamma}-\lambda_{i \gamma}\mu_{j}^{\alpha}\right)\\
\frac{1}{2}\left(\lambda_{i \delta}\mu_{j}^{\beta}-\mu_{i}^{\beta}\lambda_{j \delta}\right) & \lambda_{i[\delta}\lambda_{j \gamma]}\end{array}\right)\label{eq:10}\end{equation}
where $\kappa$ is a proportionality constant. We can determine $\kappa$
by comparing the bottom-right matrix element on each side of eq. \ref{eq:10}.
Noting that $\lambda_{i[\delta}\lambda_{j \gamma]}=-\frac{1}{2}\epsilon_{\delta \gamma}\left\langle ij\right\rangle $,
we readily find that $\kappa=-2/\left\langle ij\right\rangle $. Comparing
the top-right matrix element on each side of eq. \ref{eq:10} gives
\begin{equation}
x^{\alpha \gamma}=\frac{\mu_{i}^{\alpha}\lambda_{j}^{\gamma}-\mu_{j}^{\alpha}\lambda_{i}^{\gamma}}{\left\langle ij\right\rangle }{\normalcolor .}\label{eq:11}\end{equation}
Multiplying each side of eq. \ref{eq:11} by $\lambda_{i \gamma}$ gives

\[
\lambda_{i \gamma}x^{\alpha \gamma}=\mu_{i}^{\alpha}{\normalcolor .}\]
Similarly, multiplying each side of eq. \ref{eq:11} by $\lambda_{j \gamma}$
gives $\lambda_{j \gamma}x^{\alpha \gamma}=\mu_{j}^{\alpha}$. We will refer to this as
the incidence relation or twistor equation. From eq. \ref{eq:6},
we find that the twistors obey the following constraint:

\eq
\Omega^{ab}\tilde{X}_{ab}=\mu_{i}^{\alpha}\lambda_{j \alpha}-\mu_{j}^{\alpha}\lambda_{i \alpha}=0{\normalcolor .}
\label{3dconstraint}
\eqe
Noting that $\mu_{i}^{\alpha}\lambda_{j}^{\gamma}-\lambda_{i}^{\gamma}\mu_{j}^{\alpha}=\mu_{i}^{(\alpha}\lambda_{j}^{\gamma)}-\lambda_{i}^{(\gamma}\mu_{j}^{\alpha)}+\mu_{i}^{[\alpha}\lambda_{j}^{\gamma]}-\lambda_{i}^{[\gamma}\mu_{j}^{\alpha]}=\mu_{i}^{(\alpha}\lambda_{j}^{\gamma)}-\lambda_{i}^{(\gamma}\mu_{j}^{\alpha)}-\frac{1}{2}\epsilon^{\alpha \gamma}\left(\mu_{i}^{\beta}\lambda_{j \beta}-\lambda_{i \beta}\mu_{j}^{\beta}\right)=\mu_{i}^{(\alpha}\lambda_{j}^{\gamma)}-\lambda_{i}^{(\gamma}\mu_{j}^{\alpha)}$,
we see that eq. \ref{eq:11} can be written as follows:\[
x^{\alpha \gamma}=\frac{\mu_{i}^{(\alpha}\lambda_{j}^{\gamma)}-\lambda_{i}^{(\gamma}\mu_{j}^{\alpha)}}{\left\langle ij\right\rangle },\]
which makes it manifest that $x^{\alpha \gamma}=x^{\gamma \alpha}$.

Consider two points in Minkowski space $x_i$ and $x_j$ whose lines in twistor space intersect. At the point of intersection, we have
\[\mu^\alpha= x_i^{\alpha \gamma} \lambda_\gamma,\,\,\,\,\,\,\,\mu^\alpha= x_j^{\alpha \gamma} \lambda_\gamma.\]
If we take take the difference of these two equations we have
\[ \left( x_i -x_j \right)^{\alpha \gamma} \lambda_\gamma=0, \]
which implies that
\[ \det \left( x_i -x_j \right)=\left ( x_i -x_j \right)^2 =0 \]
since $ x_i -x_j $ has a zero eigenvalue. Hence, $x_i$ and $x_j$ are null separated.

In summary, a point in Minkowski space defines a line in twistor space and a point in twistor space defines a null ray in Minkowski space. The essential difference between the twistors of four-dimensional Minkowski space and the twistors of three-dimensional Minkowski space, is that the latter must obey an additional constraint given in eq. \ref{3dconstraint}.  

%% file: bib.tex
\bibliographystyle{utcaps}
\bibliography{main_doc}